\documentclass[aps,prfluids,groupedaddress]{revtex4-2}

\usepackage{bm}
\usepackage{amsmath}
\usepackage{amssymb}
\usepackage{graphicx}
\usepackage{url}
\usepackage{xcolor}

\newcommand{\Reyn}{\mathrm{Re}}

\newcommand{\rmd}{\mathrm{d}}
\newcommand{\sym}{\mathrm{sym}}

\newcommand{\be}{\bm{e}}
\newcommand{\bff}{\bm{f}}
\newcommand{\bg}{\bm{g}}

\newcommand{\bn}{\bm{n}}
\newcommand{\bp}{\bm{p}}
\newcommand{\bq}{\bm{q}}
\newcommand{\br}{\bm{r}}
\newcommand{\bu}{\bm{u}}
\newcommand{\bv}{\bm{v}}
\newcommand{\bx}{\bm{x}}

\newcommand{\bchi}{\bm{\chi}}
\newcommand{\bmu}{\bm{\mu}}
\newcommand{\bomega}{\bm{\omega}}

\newcommand{\btheta}{\bm{\theta}}
\newcommand{\blambda}{\bm{\lambda}}

\newcommand{\ba}{\bm{a}}

\newcommand{\bE}{\bm{E}}
\newcommand{\bF}{\bm{F}}

\newcommand{\bH}{\bm{H}}

\newcommand{\bN}{\bm{N}}

\newcommand{\bQ}{\bm{Q}}
\newcommand{\bR}{\bm{R}}

\newcommand{\bT}{\bm{T}}
\newcommand{\bU}{\bm{U}}
\newcommand{\bV}{\bm{V}}

\newcommand{\bX}{\bm{X}}
\newcommand{\bTheta}{\bm{\Theta}}
\newcommand{\bOmega}{\bm{\Omega}}

\newcommand{\tI}{\bm{\mathit{1}}}
\newcommand{\ta}{\bm{{a}}}
\newcommand{\tb}{\bm{{b}}}
\newcommand{\tc}{\bm{{c}}}
\newcommand{\tB}{\bm{{B}}}
\newcommand{\tC}{\bm{{C}}}
\newcommand{\tE}{\bm{{E}}}
\newcommand{\tEd}{\bm{\dot {E}}}
\newcommand{\tF}{\bm{{F}}}

\newcommand{\tJ}{\bm{J}}
\newcommand{\tK}{\bm{{K}}}
\newcommand{\tM}{\bm{{M}}}

\newcommand{\tPi}{\bm{{\Pi}}}

\newcommand{\tR}{\bm{{R}}}
\newcommand{\tS}{\bm{{S}}}

\newcommand{\tsigma}{\bm{{\sigma}}}

\newcommand{\bad}{\bm{\dot a}}

\newcommand{\bqd}{\bm{\dot q}}

\newcommand{\bxd}{\bm{\dot \chi}}
\newcommand{\bthetad}{\bm{\dot \theta}}
\newcommand{\bEd}{\bm{{\dot E}}}

\newcommand{\bQd}{\bm{{\dot Q}}}
\newcommand{\bRd}{\bm{{\dot R}}}
\newcommand{\bXd}{\bm{{\dot X}}}
\newcommand{\bThetad}{\bm{\dot \Theta}}

\newcommand{\hTheta}{\bm{\hat \Theta}}
\newcommand{\hg}{\bm{\hat g}}

\newcommand{\mRot}{\bm{\mathcal{R}}}

\DeclareMathOperator{\cotan}{cotan}

\begin{document}

\title{Soft Mobility Theory}

\author{Christophe Eloy}
\affiliation{Aix-Marseille Univ, CNRS, Centrale Med, IRPHE, Marseille, France}
\email[]{celoy@centrale-med.fr}
\affiliation{Department of Applied Mathematics and Theoretical Physics, University of Cambridge, Cambridge, UK}

\begin{abstract}
Predicting how a deformable body moves and deforms in a viscous flow underlies problems ranging from microorganism locomotion to soft microrobotics, yet existing frameworks are either problem-specific or ill-suited to inverse design. We propose a soft mobility theory. Applying the principle of virtual power to an assembly of hydrodynamically interacting spheres connected by springs and rods yields a saddle-point problem, whose explicit solution is a configuration-dependent ordinary differential equation for the generalized coordinates of the body. This soft mobility equation extends classical rigid-body mobility theory in that the elastic, body-force, and flow-coupling tensors all depend explicitly on the instantaneous deformation. It is not restricted to sphere assemblies: the same equation follows from the continuum problem of a hyperelastic body in a background Stokes flow, of which the assembly is a Galerkin restriction. We compute the mobility from the Rotne--Prager--Yamakawa approximation and validate the framework on canonical problems spanning rigid and flexible bodies in quiescent and shear flows. An open-source JAX implementation makes entire simulations end-to-end differentiable. This allows efficient gradient-based inverse design: as proofs of concept, we recover the asymptotic optimum of a three-sphere swimmer and design a soft gyrotactic ``surfer'' that exploits passive deformation to ascend faster than its rigid counterpart in a Taylor--Green flow.
\end{abstract}

\maketitle

\section{Introduction\label{sec.intro}}

Microorganisms rely on their deformability to interact with the surrounding fluid~\cite{Visser2001,Wheeler2019}. 
This deformation can be active when microswimmers propel themselves through their slender flagella and cilia~\cite{Lauga2009,Lauga2016,Lauga2020,Guasto2012,Wheeler2019}. It can also be passive, like red blood cells undergoing large shape changes as they flow through capillaries~\cite{BarthesBiesel2016,Farutin2013}, or diatoms using flexible structures such as mucilage stalks and fine spines to adapt their sinking velocity~\cite{Smayda1974,Wiseman1981,Padisak2003}. 
Across these systems, deformability is not incidental but functional, often conferring sensing, locomotion, or feeding advantages that a rigid body could not achieve.

In artificial systems, flexible fibers have been shown to buckle and tumble in quiescent, shear, and turbulent flows~\cite{Wiggins1998,Brouzet2014,Du-Roure2019,Zuk2021}, and a growing class of soft microrobots is being engineered for biomedical applications, such as in-situ sensing, drug delivery, and microsurgery~\cite{Sitti2021,Yasa2023,Tsang2020b,Elnaggar2024,Dabbagh2022,Mo2023,Jiang2022}.
Even microplastics deform and disperse in the ocean in ways that depend on their shape and flexibility~\cite{Sutherland2023}. 

These problems share a common physical structure: a deformable body whose shape evolves under the combined action of active, elastic, viscous, and body forces. 
This broad class of fluid--structure interactions is conventionally referred to as \emph{elastohydrodynamics}. 
We adopt here the more specific term \emph{soft mobility}, by which we mean the transport and deformation of a small deformable body immersed in a background flow, in direct analogy with the classical rigid-body mobility theory~\cite{Brenner1963,Happel2012,Kim2013}. 

Soft mobility encompasses both the \emph{forward} problem of predicting how a given soft body moves and deforms in a prescribed flow, and the \emph{inverse} problem of designing soft bodies to perform a target function: efficient locomotion, flow sensing, clustering, or autonomous navigation in complex environments. 
The latter is increasingly central to soft robotics~\cite{Kim2013a,Yasa2023} and embodied intelligence~\cite{Pfeifer2006}, and underlies the related notions of morphological computation~\cite{Pfeifer2014,Hauser2011,Hauser2014,Nakajima2015} and physical intelligence~\cite{Sitti2021,Hauser2023}, in which sensing, control, and decision-making are partly offloaded from a centralized processor to the mechanical response of the body itself.
Solving this inverse problem at low Reynolds number requires a specific approach, which we aim to develop in the present paper.

The natural starting point is the classical mobility theory developed in the 1960s for rigid bodies in Stokes flow~\cite{Brenner1963,Happel2012,Kim2013}. 
Owing to the linearity of the Stokes equations, the six-component velocity of a rigid body depends linearly on the applied force and torque and on the local background velocity, vorticity, and rate-of-strain. 
The mobility tensors entering this relation depend only on the body's shape and incorporate all the hydrodynamic complexity of the Stokes equations into a finite set of geometric coefficients. 
These tensors can be computed with different methods: boundary integral methods~\cite{Pozrikidis1992} that exploit Lorentz reciprocal theorem~\cite{Happel2012}, multiblob methods that discretize body surfaces with ``blobs''~\cite{Balboa-Usabiaga2017,Balboa-Usabiaga2022,Broms2023}, or slender-body theory for fibers~\cite{Tornberg2004,Nazockdast2017,Maxian2022}.
This formulation has recently been revived for complex-shaped macromolecules~\cite{Garcia-de-la-Torre1981,Zuk2014,Zuk2018} and reinterpreted as a low-dimensional dynamical system describing rigid-body sinking trajectories~\cite{Gonzalez2004,Witten2020}.
Its appeal, from the standpoint of the inverse problem, is the explicit, shape-dependent linear relation it provides between velocities and loads~\cite{Moreau2022}.

For $N$ rigid bodies, the same linear relationship extends to a $6N\times 6N$ grand mobility tensor encoding both single-body mobility and many-body hydrodynamic interactions. 
This tensor is at the core of Stokesian dynamics, which aims to simulate a large ensemble of rigid bodies interacting in a flow~\cite{Brady1988,Ichiki2002,Fiore2019,Swan2016}. 
To compute the grand mobility tensor efficiently, multipole expansions are generally used, such as the Rotne--Prager--Yamakawa (RPY) approximation, which provides a closed-form mobility for assemblies of spheres of possibly different radii~\cite{Rotne1969,Yamakawa1970,Wajnryb2013,Zuk2014,Cichocki2021} and has been validated against experiment on biomacromolecules~\cite{Waszkiewicz2024}.
Because these expansions are truncated at low order, they capture the far-field interactions but not the lubrication forces that diverge as bodies come into near contact. Stokesian dynamics therefore supplements them with pairwise lubrication corrections~\cite{Brady1988,Jeffrey1984}.
This approach is now mature for both passive rigid suspensions and active suspensions of rigid swimmers, routinely tackling problems with millions of degrees of freedom~\cite{Delmotte2015b}. 

A natural next step has been to leverage these tools to simulate deformable bodies by treating them as assemblies of rigid
sub-bodies linked by elastic or rigid constraints. 
For instance, bead models or multiblob methods approximate fibers as an ensemble of touching spheres connected by springs~\cite{Delmotte2015,Zuk2021,Fuchter2023}, and general articulated bodies can be simulated in a similar way~\cite{Balboa-Usabiaga2022}. 
At each timestep, these formulations enforce the constraints through a saddle-point linear system, typically solved with preconditioned iterative solvers (GMRES, FMM-accelerated iterations~\cite{Liang2013,Fiore2019}). 
This is efficient for forward simulation, but the implicit solve is costly to differentiate, which makes it ill-suited to shape optimization. 
We show below that the present formulation leads to a saddle-point system of the same structure, and that solving for its constraint loads in closed form, rather than iterating on them, is what yields a differentiable system suited to gradient-based optimization.

Closest in spirit to the present paper is the framework of Solovev and Friedrich~\cite{Solovev2021} (see also the earlier account of a related projection argument by Vilfan and Stark~\cite{Vilfan2009}). 
They explicitly project the deformation of a soft body onto a small set of generalized coordinates using Lagrangian mechanics.
This ``trick'' allows them to compute the dynamics of a shape-changing body without solving the constraints. 
In the conjugate space of generalized coordinates, they balance generalized hydrodynamic forces against active driving forces, which yields an explicit ordinary differential equation for the generalized velocities. 
Their framework has been applied to a range of biological and artificial model swimmers. In particular, they studied cases with elastic energy terms and waveform compliance~\cite{Klindt2016,Klindt2017}, with a uniform background flow~\cite{Klindt2016}, and with hydrodynamic forces evaluated by boundary-element methods, resistive force theory, or assemblies of interacting spheres~\cite{Friedrich2012,Geyer2013}. 
Two differences with the present work should be noted. 
First, their configuration-dependent friction matrix is tabulated on a grid of shapes and interpolated during integration, regardless of the underlying solver, and has no closed analytic form, which makes it difficult to differentiate cleanly for gradient-based optimization. 
Second, their formulation carries no coupling to the gradients of the background flow. 
This rate-of-strain coupling is what drives passive bodies in non-uniform flows, and it is central here.

The goal of this paper is to develop a general ``soft mobility'' framework that simultaneously: 
(1) is derived from the continuum elastohydrodynamic problem of a hyperelastic body in a background Stokes flow (Appendix~\ref{app:continuum}), of which the sphere assembly used throughout is a Galerkin restriction; 
(2) accommodates arbitrary generalized coordinates and yields an ordinary differential equation in which the elastic mobility, the body-force mobility, and the flow-gradient coupling tensor appear in closed form; and 
(3) is constructed so that all tensors are smooth, differentiable functions of the design parameters, enabling end-to-end gradient-based inverse design.

The paper is organized as follows. 
Section~\ref{sec.kinematics} states the assumptions under which the problem is posed and the kinematic description of a sphere assembly. 
Section~\ref{sec.dynamics} derives its dynamics: the generalized force balance, the saddle-point problem, its explicit solution as the soft mobility equation, and the open-source Python implementation~\cite{softmobility}. 
Section~\ref{sec.results} applies the framework to five canonical problems: the gravity-driven sinking of a chiral rigid body; a flexible fiber sinking under gravity, wrapping around a rotating clamp, and tumbling in shear; Jeffery orbits of a rigid and of a soft dumbbell; the optimization of a three-sphere swimmer; and a soft gyrotactic surfer that exploits passive deformation to navigate through a Taylor--Green vortical flow.

\section{Kinematics\label{sec.kinematics}}

We consider assemblies of rigid spheres connected by springs and rods.
Each sphere carries translational and rotational degrees of freedom. The springs endow the assembly with elasticity, and the rods impose rigid links.
We assume that such an assembly is immersed in a background viscous flow, subject to external body forces, such as gravity or magnetic forces, and possibly subject to internal forces or prescribed kinematic deformations.
Our goal is to determine how this sphere assembly translates, rotates, and deforms in response to this flow and to these forces.
The present section sets out the assumptions under which the problem is posed and the kinematic description of the assembly. The dynamics will be derived in \S\ref{sec.dynamics}.

\subsection{Assumptions\label{sec.assumptions}}

\begin{figure}[t]
\begin{center}
    \includegraphics[scale=0.60]{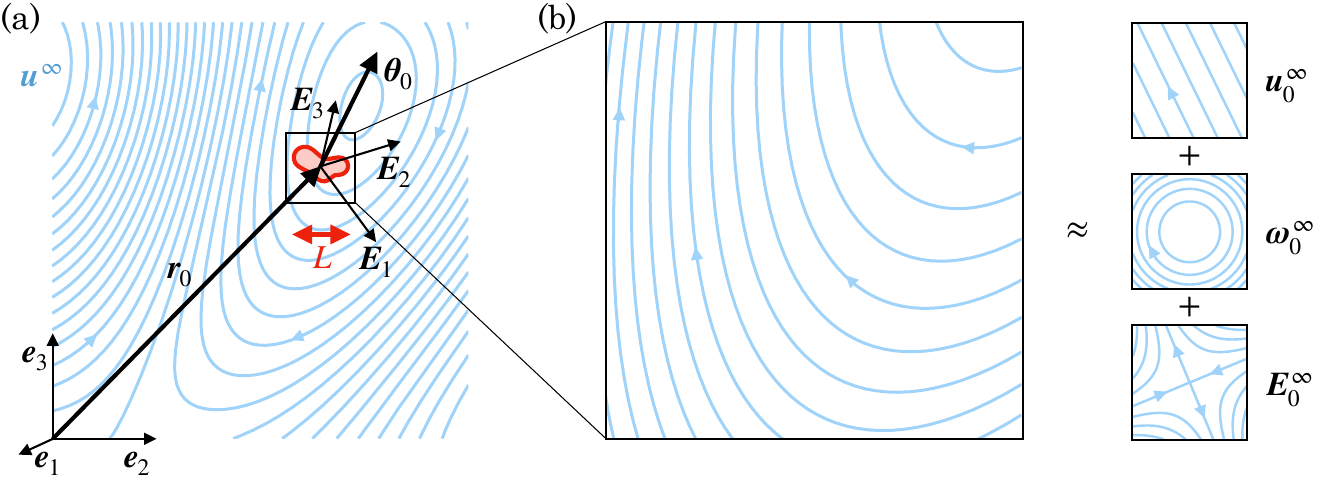}
\caption{A soft body in a non-uniform flow.
(a) The body, of size $L$, is transported and deformed by a background flow varying over a length $\ell \gg L$. Its position and orientation are $\br_0$ and $\btheta_0$. Its deformation is described by the generalized coordinates $\bQ$.
(b) In the neighborhood of $\br_0$, the flow is approximately linear and decomposes into a uniform translation $\bu_0^\infty$, a solid-body rotation of angular velocity $\bomega_0^\infty$, and a pure strain $\tE_0^\infty$, as given by Eq.~(\ref{eq:bg_flow}). 
Streamlines are equally spaced contours of the stream function, so their density measures the flow speed.}
\label{fig:soft_body_in_flow}
\end{center}
\end{figure}

We consider an assembly of characteristic size $L$, immersed in a fluid of dynamic viscosity $\mu$ and kinematic viscosity $\nu$, and subject to a background flow field $\bu^\infty(\br)$.
Two assumptions underlie the entire framework: inertia is negligible, and the flow is locally linear (Fig.~\ref{fig:soft_body_in_flow}).

The first assumption is that inertia is negligible at the scale of the body.
The Reynolds number is $\Reyn = U L/\nu$, with $U$ a characteristic velocity of the body relative to the fluid, and we assume $\Reyn \ll 1$ throughout.
The body is then in quasi-static equilibrium at each instant, and the surrounding fluid obeys the Stokes equations.

The second assumption is that the background flow varies slowly at the scale of the body.
Denoting by $\ell$ the smallest length scale over which $\bu^\infty$ varies appreciably, we assume $L \ll \ell$, so that the flow can be linearized about a reference point $\br_0$ attached to the body,
\begin{equation}\label{eq:bg_flow}
\bu^\infty(\br) \approx \bu_0^\infty + \bomega_0^\infty \times (\br-\br_0)  + \tE_0^\infty \cdot (\br-\br_0),
\end{equation}
with $\bu_0^\infty$ the translational velocity, $\bomega_0^\infty = \frac{1}{2}\nabla\times\bu^\infty$ the angular velocity, and $\tE_0^\infty = \frac{1}{2}(\nabla\bu^\infty + \nabla^\top\bu^\infty)$ the rate-of-strain tensor, all evaluated at $\br_0$.

Neither assumption implies the other.
A particle in a channel whose height is comparable to its own size satisfies $\Reyn \ll 1$ while sampling a parabolic profile whose curvature cannot be discarded, so that Eq.~(\ref{eq:bg_flow}) fails although the flow is a Stokes flow.
Conversely, a body much smaller than $\ell$ may still have $\Reyn = \mathcal{O}(1)$ if it moves fast enough relative to the fluid, in which case Eq.~(\ref{eq:bg_flow}) holds but quasi-static equilibrium does not.

Turbulent environments are the special case in which the two conditions collapse onto one, and they motivate much of what follows.
There, the smallest scale of the flow is the Kolmogorov scale $\eta = (\nu^3/\epsilon)^{1/4}$, so that $\ell = \eta$, where $\epsilon$ is the turbulent dissipation rate.
The Reynolds number can be estimated by taking the Kolmogorov velocity $u_\eta = (\nu\epsilon)^{1/4}$ as the characteristic velocity, which gives $\Reyn = L/\eta$. Alternatively, estimating the characteristic velocity from the local velocity gradient, $U \sim \|\nabla\bu^\infty\| L \sim L/\tau_\eta$ with $\tau_\eta = (\nu/\epsilon)^{1/2}$ the Kolmogorov time scale, gives $\Reyn = (L/\eta)^2$.
Both estimates return the same condition, $L \ll \eta$, which is also the condition $L \ll \ell$ for the linearization.
A body small enough to sit below the Kolmogorov scale therefore sees a linear Stokes flow at each instant, even though the trajectory it follows samples a turbulent flow at large scales.
This is the sense in which the framework applies to turbulent transport, and it is the regime of the navigation problem of \S\ref{sec.surfer}.

\subsection{Conventions\label{sec.notation}}

\begin{figure}[t]
\begin{center}
    \includegraphics[scale=0.60]{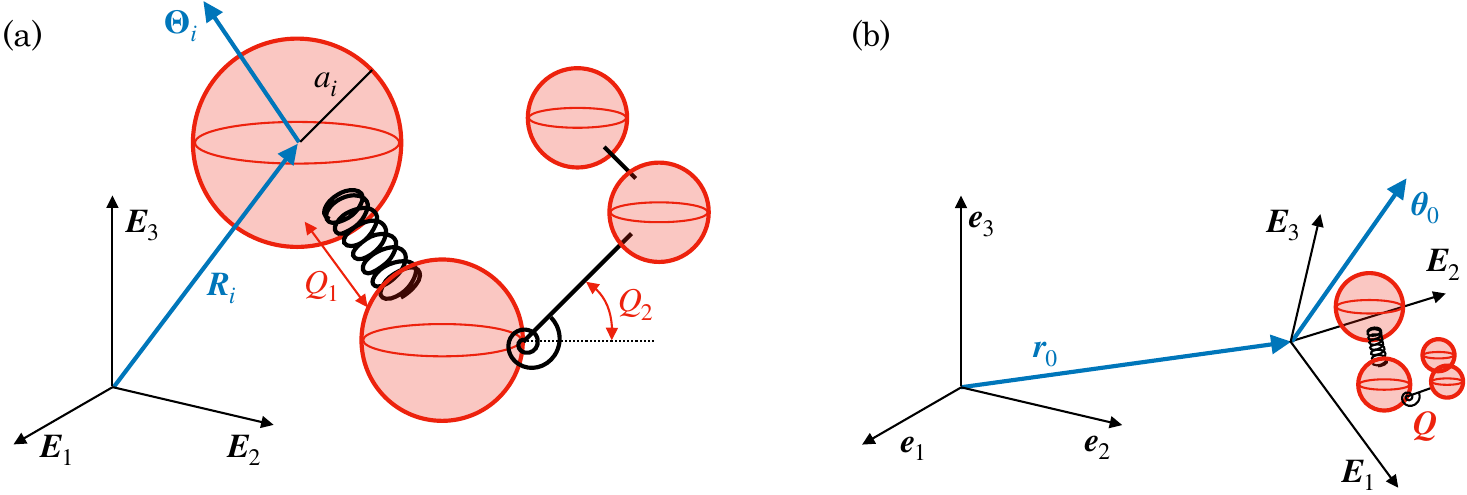}
\caption{Notations for an assembly of spheres.
(a) In the body frame, of basis $(\bE_1,\bE_2,\bE_3)$, the $i$-th sphere has radius $a_i$, position $\bR_i(\bQ,t)$, and Rodrigues orientation $\bTheta_i(\bQ,t)$. Here $N=4$ spheres are described by $N_Q=2$ degrees of freedom $\bQ = [Q_1,Q_2]$.
(b) In the laboratory frame, of basis $(\be_1,\be_2,\be_3)$, the configuration of the assembly is entirely described by the generalized coordinates $\bq = [\br_0, \btheta_0, \bQ]$, with $\br_0$ the position, $\btheta_0$ the orientation of the body frame, and $\bQ$ the degrees of freedom.}
\label{fig:assembly_spheres}
\end{center}
\end{figure}

Two reference frames are used throughout (Fig.~\ref{fig:assembly_spheres}): the laboratory frame and the body frame, which translates and rotates with the assembly. 
The body frame has a position $\br_0$ and orientation $\btheta_0$ in the laboratory frame, where $\btheta_0$ is the Rodrigues vector: the body is rotated by the angle $\theta_0 = \|\btheta_0\|$ about the axis $\hat{\btheta}_0 = \btheta_0/\theta_0$. 
This parameterization is singularity-free for $\theta_0 \in [0,2\pi)$, unlike Euler angles. 
The relation between $\bthetad_0$ and the angular velocity is given by the Bortz equation~\cite{Bortz1971} in Appendix~\ref{app:kinematics}.
 
The case of a symbol records the frame, not the 
basis: uppercase for quantities measured relative to the body frame, lowercase 
for their laboratory-frame counterparts. For instance, $\bR_i$ denotes the position of sphere 
$i$ in the body frame, and $\br_i$ in the laboratory frame. 
Forces and torques are unchanged by the change of frame, so the case convention does not apply to them. 
Operators and tensors are written in uppercase irrespective of frame.

The equations below 
are vectorial and hold in any basis. When vector components are needed, unless specified otherwise, they are taken 
in the body basis $(\bE_1,\bE_2,\bE_3)$. The rotation matrix $\mRot(\btheta_0)$ 
of the Euler--Rodrigues formula~(\ref{eq:Rodrigues_formula}) converts vector components from the body basis to the laboratory basis $(\be_1,\be_2,\be_3)$, and its transpose converts them back [e.g., Eq.~(\ref{eq:rotation_fields})]. 

Positions, velocities, and loads are gathered into six-component vectors, one per sphere (Fig.~\ref{fig:assembly_spheres}).
The $i$-th sphere, of radius $a_i$, has position $\bR_i$ and Rodrigues orientation vector $\bTheta_i$, collected in the six-component position $\bX_i = [\bR_i, \bTheta_i]$.
Its translational and angular velocities $\bU_i$ and $\bOmega_i$ are collected in $\bV_i = [\bU_i, \bOmega_i]$, and the force $\bF_i$ applied at its center together with the associated torque $\bT_i$ are collected in the six-component load $\bff_i = [\bF_i, \bT_i]$.
Stacking over the $N$ spheres produces the grand vectors $\bff = [\bff_1,\dots,\bff_N]$ and $\bV = [\bV_1,\dots,\bV_N]$, both of size $6N$, with $\bv$ the laboratory-frame counterpart of the grand velocity $\bV$.

The deformations of the assembly are described by the degrees of freedom $\bQ$, of size $N_Q$, measured in the body frame: the elongation of each linear spring, the angle of each torsional spring, and so on.
Together with the six-component position $\bx_0 = [\br_0, \btheta_0]$ of the body frame in the laboratory frame, they form the generalized coordinates of the assembly in the Lagrangian mechanics sense,
\begin{equation}\label{eq:generalized_coordinates}
\bq = \left[\bx_0,\, \bQ \right],
\end{equation}
of size $N_q = 6 + N_Q$.
The mixed case in Eq.~(\ref{eq:generalized_coordinates}) is deliberate: $\bx_0$ locates the body frame in the laboratory frame, whereas $\bQ$ measures an internal deformation and is therefore naturally attached to the body frame.
The associated generalized velocity is
\begin{equation}\label{eq:generalized_velocity}
\bp = \left[\bv_0,\, \bQd \right] = \tB(\btheta_0) \cdot \bqd,
\end{equation}
with $\bv_0 = [\bu_0, \bomega_0]$ the six-component velocity of the body frame in 
the laboratory frame. The Bortz matrix $\tB$, given in Appendix~\ref{app:kinematics}, 
is not the identity, since $\bomega_0$ is not the time derivative of $\btheta_0$.

Two conventions attached to this description should be made explicit.
First, the reference point $\br_0$ is an arbitrary point fixed in the body frame.
It need not coincide with the center of mass, nor with the center of resistance, and no result below requires it to do so. 
Second, the degrees of freedom $\bQ$ are assumed to parameterize the deformation minimally and without degeneracy.
All internal constraints, including the rigid rods, are holonomic and are eliminated by this choice, so that the configuration of the assembly is described exactly by $\bq$ with no residual constraint left to enforce.

\subsection{Sphere assembly\label{sec.sphere_kinematics}}

The six-component position $\bX_i$ of sphere $i$ depends on the configuration of 
the assembly only through the degrees of freedom $\bQ$, and, where the kinematics 
is prescribed, on time (Fig.~\ref{fig:assembly_spheres}). Differentiating, the sphere velocities are affine in the 
generalized velocity,
\begin{equation}\label{eq.U_is_J_Qdot}
\bv = \bV_\mathrm{act} + \tJ \cdot \bp = \bV_\mathrm{act} + \tJ \cdot \tB \cdot \bqd,
\end{equation}
with $\tJ = \partial \bv/\partial \bp$ the kinematic Jacobian, of size $6N \times N_q$, 
and $\bV_\mathrm{act}$ the grand active velocity produced by the prescribed deformation (Appendix~\ref{app:kinematics}). 
Their construction from the sphere positions is given in 
Appendix~\ref{app:kinematics}; here only the block structure of the Jacobian is needed,
\begin{equation}\label{eq:J_blocks}
\tJ = \left[\,\tC_U,\; \tJ_\mathrm{sph}\,\right],
\end{equation}
where the $6N\times 6$ block $\tC_U$ transports the rigid-body motion of the body 
frame to the sphere centers, and the $6N\times N_Q$ block $\tJ_\mathrm{sph}$ 
carries the deformation.

Two properties of $\tJ$ matter later.
First, $\tJ(\bQ,t)$ is assembled entirely from the sphere positions and orientations in 
the body frame, whose components are taken in the body basis. It therefore depends 
on the configuration only through the degrees of freedom $\bQ$ and, where part of 
the kinematics is prescribed, explicitly through time. It does not depend on the 
position and orientation $\bx_0$ of the body frame. The reduced tensors of 
\S\ref{sec.dynamics} inherit this property, which is what allows them to be 
evaluated, or tabulated, as functions of the deformation state alone.
Second, because $\bQ$ parameterizes the deformation without degeneracy, $\tJ$ has 
full column rank. The generalized velocity $\bp$ is then in one-to-one 
correspondence with the sphere velocities compatible with the constraints, so that 
$\bv$ is a quasi-velocity of the Lagrangian system. Full rank is also what makes 
$\tJ^\top\cdot\tM_G^{-1}\cdot\tJ$ invertible in \S\ref{sec.dynamics}.

The background flow enters through the same structure. Evaluating the 
linearization~(\ref{eq:bg_flow}) at the sphere centers gives the grand flow 
velocity $\bv^\infty = [\bv^\infty_1,\dots,\bv^\infty_N]$, where $\bv_i^\infty$ 
collects the translational and angular velocities of the undisturbed flow at the center of sphere $i$,
\begin{equation}\label{eq:grandflowvelocity}
\bv^\infty = \tC_U \cdot \bv_0^\infty + \tC_S : \tE_0^\infty,
\end{equation}
with $\bv_0^\infty = [\bu_0^\infty, \bomega_0^\infty]$ the six-component velocity 
of the undisturbed flow at the reference point $\br_0$ and $\tC_S$ a 
$6N\times3\times3$ stresslet coupling tensor. The rate of strain, uniform at the 
scale of the body, is the same at every sphere. The block $\tC_U$ appearing here 
is the same as in Eq.~(\ref{eq:J_blocks}): the rigid-body part of the background 
flow is transported across the assembly exactly as the rigid-body motion of the 
body frame is.

\section{Dynamics\label{sec.dynamics}}

\subsection{Principle of virtual power}

With the kinematic description established, we now derive the equation of motion by applying the principle of virtual power to the sphere assembly. 
We decompose the  grand load as
\begin{equation}\label{eq.force_balance}
\bff_c + \bff_\mathrm{hydro} + \bff_\mathrm{ext} + \bff_K = \mathbf{0},
\end{equation}
where $\bff_c$ are internal constraints, $\bff_\mathrm{hydro}$ are hydrodynamic loads, $\bff_\mathrm{ext}$ are external body loads, and $\bff_K$ are internal elastic restoring loads. The sum is zero because the problem is inertialess, so forces and torques balance on each sphere separately. 

The elastic loads depend on the deformation alone and vanish in the rest
state $\bQ = \bm{0}$, from which $\bQ$ is measured, so that
\begin{equation}\label{eq.elastic_force}
\bff_K = \tC_K(\bQ) \cdot \bQ,
\end{equation}
with $\tC_K$ of size $6N\times N_Q$ distributing the restoring loads over the
spheres. 
The tensor $\tC_K$ is constant for small deformations and depends on $\bQ$ in general,
which is where the geometric nonlinearities enter.
Because these are internal loads, they satisfy $\tC_U^\top\cdot\bff_K = 0$ (Newton's third law). The body loads are linear in an external field $\bH$,
\begin{equation}\label{eq.body_force}
\bff_\mathrm{ext} = \tC_H \cdot \bH,
\end{equation}
with $\tC_H$ of size $6N\times3$ collecting the coupling of each sphere to the
field. For gravity, $\bH$ is the acceleration $\bg$ and sphere $i$ carries its
weight as a force at its center, with $m_i$ the buoyancy-corrected mass of sphere $i$. For magnetic actuation, $\bH$ is the applied
field and sphere $i$ carries a moment $\bmu_i$, which exerts a torque and no
force, so that
\begin{equation}\label{eq:CH_blocks}
(\tC_H)_i =
\begin{bmatrix} m_i\,\tI \\ \mathbf{0} \end{bmatrix}
\quad\text{(gravity)},
\qquad
(\tC_H)_i =
\begin{bmatrix} \mathbf{0} \\ \left[\bmu_i\right]_\times \end{bmatrix}
\quad\text{(magnetic field)}.
\end{equation}

The principle of virtual power allows us to eliminate the unknown constraints. For any virtual velocity $\bv^*$ compatible with the constraints, ideal constraints do no work and hence, their virtual power $P^* = \bff_c \cdot \bv^*$ is zero. 
Since $\bv^* = (\partial \bv/\partial \bp)\cdot \bp^*$ ($\bv^*$ is taken at frozen time to eliminate $\bV_\mathrm{act}$),
and the equality $P^*=0$ must hold for all $\bp^*$, one obtains
\begin{equation}\label{eq.generalized_force_Q}
\tJ^\top \cdot \bff_c = 0,
\end{equation}
where $\tJ = \partial \bv/\partial \bp$ is the same Jacobian as in Eq.~(\ref{eq.U_is_J_Qdot}). This projection, deduced from the principle of virtual power, reduces the $6N$-dimensional load balance to the $N_q$ dimensions of the generalized coordinates while eliminating the constraints $\bff_c$ from the equations of motion.

\subsection{Saddle-point problem\label{sec.saddle}}

The hydrodynamic response of the assembly is carried by the $6N\times6N$ grand mobility matrix $\tM_G$, and the $6N\times 3\times 3$ grand strain-coupling tensor $\tC_G$. Throughout, the subscript $G$ marks grand quantities, of first dimension $6N$. 
We evaluate $\tM_G$ and $\tC_G$ from the Rotne--Prager--Yamakawa approximation~\cite{Rotne1969,Yamakawa1970,Wajnryb2013,Zuk2014}, which provides a closed form for spheres of arbitrary radii, resolving translation, rotation, their hydrodynamic interactions, and their coupling with the background flow.  
The grand mobility equation reads
\begin{equation}\label{eq.mobility_problem_all_spheres}
\bv - \bv^\infty = - \tM_G \cdot \bff_\mathrm{hydro} + \tC_G : \tE_0^\infty.
\end{equation}
It expresses the linear relationship between relative velocities and hydrodynamic loads due to the linearity of the Stokes equations. 

Eliminating $\bff_\mathrm{hydro}$ between Eq.~(\ref{eq.mobility_problem_all_spheres}) and the force 
balance \eqref{eq.force_balance}, substituting the kinematic relations~\eqref{eq.U_is_J_Qdot} and \eqref{eq:grandflowvelocity}, and using the principle of virtual power \eqref{eq.generalized_force_Q} leaves the 
constraint loads $\bff_c$ and the generalized velocity $\bp$ as unknowns of a 
saddle-point system,
\begin{equation}\label{eq:saddle_point}
\begin{pmatrix}
-\tM_G & \tJ\\
\tJ^\top & 0
\end{pmatrix} 
\cdot
\begin{pmatrix}
\bff_c \\
\bp 
\end{pmatrix} 
=
\begin{pmatrix}
\tM_G \cdot \bff + \tC_U\cdot\bv_0^\infty + (\tC_S + \tC_G) : \tE_0^\infty - \bV_\mathrm{act}\\
\mathbf{0}
\end{pmatrix},
\end{equation}
where $\bff = \bff_\mathrm{ext} + \bff_K$ denotes the non-hydrodynamic non-constraint loads.
The first row of this saddle-point problem is the grand mobility equation and the second is the principle of virtual power. The right-hand side collects the three inputs of the problem: the applied loads, the linearized background flow, and any prescribed kinematics.

Equation~(\ref{eq:saddle_point}) has the same block structure as the linear systems 
solved in rigid multiblob methods~\cite{Balboa-Usabiaga2017,Balboa-Usabiaga2022,Broms2023} 
and in Stokesian dynamics~\cite{Brady1988,Swan2016,Fiore2019}, with $\tM_G$ the grand 
mobility, $\tJ$ the geometric or kinematic matrix (denoted $\boldsymbol{\mathcal{K}}$ in Ref.~\cite{Balboa-Usabiaga2017}), and $\bff_c$ the constraints. What differs 
is the range of $\tJ$. In these methods, it spans the rigid-body motions of each body, and the 
constraint it enforces is rigidity. Here, it spans the $N_q = 6 + N_Q$ generalized 
coordinates, so that the admissible deformations lie in its range from the outset. 
The blobs of multiblob methods are, in addition, quadrature nodes discretizing a 
surface, carrying translation only, whereas the spheres here are physical sub-bodies 
of arbitrary radii with six degrees of freedom and their own Fax\'en responses, from 
which $\tC_G$ follows.

So written, Eq.~(\ref{eq:saddle_point}) invites the treatment standard for large 
assemblies: a preconditioned iterative solver, typically GMRES~\cite{Saad1986}, which 
never forms $\tM_G^{-1}$ and costs $\mathcal{O}(N^2)$ per iteration, the number of iterations depending on the conditioning of the system and on the preconditioner. We take the 
opposite route and eliminate $\bff_c$ explicitly. The price is an $\mathcal{O}(N^3)$ 
inversion of $\tM_G$, which confines us to modest $N$. What it buys is a closed-form reduced 
system, interpretable for small $N_Q$ and differentiable with respect to the design 
parameters. We will return to this trade-off in \S\ref{sec.discussion}.

\subsection{Soft mobility equation\label{sec.soft_mobility}}

The saddle-point system~(\ref{eq:saddle_point}) admits an explicit solution. 
Solving the first row for the constraint loads and substituting into the second 
eliminates $\bff_c$, leaving the Schur complement $\tJ^\top\cdot\tM_G^{-1}\cdot\tJ$ 
acting on $\bp$ alone. Since $\tM_G$ is symmetric positive definite and $\tJ$ has 
full column rank, this $N_q\times N_q$ matrix is invertible, and
\begin{equation}\label{eq:schur_solution}
\bp = \tPi \cdot \left[\tM_G \cdot \bff + \tC_U\cdot\bv_0^\infty 
+ (\tC_S + \tC_G) : \tE_0^\infty - \bV_\mathrm{act}\right],
\qquad
\tPi = \left( \tJ^\top \cdot \tM_G^{-1} \cdot \tJ \right)^{-1} 
\cdot \tJ^\top \cdot \tM_G^{-1},
\end{equation}
where $\tPi$, of size $N_q\times 6N$, satisfies $\tPi \cdot \tJ = \tI$. It maps the 
grand space of the spheres  onto the generalized coordinates, retaining only the 
components of the $6N$-dimensional problem compatible with the constraints.

Each term of Eq.~(\ref{eq:schur_solution}) reduces to a tensor on the generalized 
coordinates. The first two define the soft mobility and flow-coupling tensors,
\begin{equation}\label{eq:soft_tensors}
\tM = \left(\tJ^\top \cdot \tM_G^{-1} \cdot \tJ\right)^{-1},
\qquad
\tC_E = \tPi \cdot \left(\tC_S + \tC_G\right),
\end{equation}
of sizes $N_q\times N_q$ and $N_q\times3\times3$, the first term of Eq.~(\ref{eq:schur_solution}) being $\tPi\cdot\tM_G\cdot\bff = \tM\cdot\tJ^\top\cdot\bff$. The third follows from the block partition~(\ref{eq:J_blocks}): $\tPi\cdot\tJ = \tI$ read on its first six columns gives $\tPi\cdot\tC_U = [\tI_6;\,\mathbf{0}]$, so that the background flow enters through the generalized flow velocity $\bp_0^\infty = [\bv_0^\infty, \mathbf{0}]$ alone.

The soft mobility tensor $\tM$ acts on the generalized force $\tJ^\top\cdot\bff$, conjugate to $\bp$. Its symmetry follows from that of $\tM_G$ and expresses the reciprocal theorem on the reduced coordinates. It is also positive definite, as in the framework of Solovev and Friedrich~\cite{Solovev2021}, by the same energy argument as for the grand mobility: for any $\bp\neq\mathbf{0}$, the sphere velocities $\tJ\cdot\bp$ do not vanish, $\tJ$ having full column rank, and $\bp\cdot\tJ^\top\cdot\tM_G^{-1}\cdot\tJ\cdot\bp$ is the power they dissipate in a fluid at rest, which is positive. It extends the classical rigid-body mobility to deformable assemblies, to which it reduces when $N_Q = 0$. The soft mobility equation follows, in terms of this generalized force,
\begin{equation}\label{eq:soft_mobility}
\bp - \bp_0^\infty = \tM \cdot \tJ^\top \cdot \bff + \tC_E : \tE_0^\infty - \tPi \cdot \bV_\mathrm{act}.
\end{equation}
Its left-hand side is the deviation of the generalized velocity from the rigid-body 
part of the background flow: only relative velocities couple, as Galilean invariance 
requires.

Substituting the load decompositions in Eqs.~(\ref{eq.elastic_force}--\ref{eq.body_force}) 
gives the form used throughout the paper,
\begin{equation}\label{eq:soft_mobility_body}
\bp - \bp_0^\infty = 
\left(
\begin{array}{c}
\bu_0 - \bu_0^\infty\\[0.5ex]
\bomega_0 - \bomega_0^\infty\\[0.5ex]
\bQd
\end{array}
\right) =
  \tM_K \cdot \bQ + \tM_H \cdot \bH + \tC_E : \tE_0^\infty - \tPi \cdot \bV_\mathrm{act},
\end{equation}
with $\tM_K = \tM \cdot \tJ^\top \cdot \tC_K$ of size $N_q\times N_Q$ and $\tM_H = \tM \cdot \tJ^\top \cdot \tC_H$ 
of size $N_q\times 3$. All four tensors depend on the deformation state $\bQ$ and, 
when the kinematics is prescribed, explicitly on time, but not on $\bx_0$. 
The four right-hand-side terms admit a physical interpretation: the elastic mobility $\tM_K$ carries the response to the internal elastic loads, the body-force mobility $\tM_H$ the response to the external field $\bH$, the flow-coupling tensor $\tC_E$ is the soft-body analogue of the Bretherton tensor and couples the assembly to the background rate of strain, and $-\tPi\cdot\bV_\mathrm{act}$ projects any prescribed 
kinematics onto the generalized coordinates. 

Nothing in the derivation is specific to spheres: Eq.~(\ref{eq:soft_mobility}) follows from the linearity of the Stokes equations and the linearization of the background flow alone. The same structure therefore holds for any elastic body, the sphere assembly being a Galerkin restriction of that more general problem (see Appendix~\ref{app:continuum}). 

Since $\bp = \tB\cdot\bqd$, Eq.~\eqref{eq:soft_mobility_body} is a first-order 
ordinary differential equation $\bqd = f(\bq, t)$, integrated forward in time from 
given initial conditions. Like every equation above, it holds in any basis. In 
practice we evaluate it in the body basis, in which the sphere positions and 
orientations, and hence the tensors, are naturally expressed. Flow and fields 
prescribed in the laboratory basis are then rotated beforehand,
\begin{equation}\label{eq:rotation_fields}
\bu_0^\infty = \mRot^\top \cdot \left(\bu_0^\infty\right)_\mathrm{lab}, \quad
\tE_0^\infty = \mRot^\top \cdot \left(\tE_0^\infty\right)_\mathrm{lab} \cdot \mRot, \quad
\bH = \mRot^\top \cdot \left(\bH \right)_\mathrm{lab},
\end{equation}
with $\mRot$ given by the Euler--Rodrigues formula~(\ref{eq:Rodrigues_formula}).

\subsection{Python implementation}

The soft mobility framework for sphere assemblies is implemented in \emph{softmobility}, a Python library freely available as open-source code~\cite{softmobility}. The library is organized around four capabilities, each illustrated by a tutorial.

First, a sphere assembly is constructed by declaring its spheres (radii, positions, orientations, forces, torques) symbolically in terms of three classes of variables: degrees of freedom $\bQ$, design parameters $\blambda$ (stiffnesses, rest lengths, radii) targeted by optimization, and external inputs (fields and active controls). A parser imports these spheres from a YAML file and linearizes the symbolic forces in the inputs to build the coupling matrices $\tC_K$ and $\tC_H$ of Eqs.~(\ref{eq.elastic_force}) and (\ref{eq.body_force}).

Second, given the design parameters $\blambda$ and the current deformation state $\bQ$, the library assembles the grand mobility $\tM_G$ and strain coupling $\tC_G$ from the Rotne--Prager--Yamakawa approximation, together with the kinematic Jacobian $\tJ(\bQ,t;\blambda)$, whose explicit time dependence arises only when part of the kinematics is prescribed. 
It then forms the projection $\tPi$ of Eq.~(\ref{eq:schur_solution}) and the reduced tensors $\tM_K$, $\tM_H$ and $\tC_E$ of Eq.~(\ref{eq:soft_mobility_body}), without the user assembling Jacobians or coupling tensors by hand. 
The dense inversion in $\tPi$ dominates the cost, and is what confines the method to modest $N$.

Third, the trajectory of the soft body is integrated forward in time with a fourth-order Runge--Kutta scheme, given a prescribed background flow, external fields (e.g., gravity, magnetic field), and any active kinematics. Anchored bodies (clamped fibers) are handled through the mixed mobility system of Appendix~\ref{app:clamped_anchor}.

Fourth, all numerical operations are implemented in JAX~\cite{jax2018github}, making an entire simulation just-in-time compilable and end-to-end differentiable. This enables gradient-based optimization of shape and stiffness parameters $\blambda$ via the Optax library~\cite{optax2020github}: any scalar objective of the trajectory (e.g., average swim speed) can be maximized or minimized by automatic differentiation through the simulation.

\section{Results\label{sec.results}}

The proposed framework is illustrated on five canonical problems organized along two axes: increasing complexity of the body (rigid $\to$ soft) and of its environment (quiescent fluid $\to$ background flow), followed by design optimization.
These five problems are provided as example notebooks in the \emph{softmobility} library~\cite{softmobility} (all figures below are generated by these notebooks).

\subsection{Sinking of a rigid body}
To test our numerical implementation for non-trivial rigid body dynamics in a quiescent fluid, we first consider the gravity-driven sinking of a rigid four-sphere assembly.
In that case, the degrees of freedom $\bQ$ are absent and the soft mobility equation~(\ref{eq:soft_mobility}) reduces to a rigid mobility equation. We set $a=1$, $mg=1$, and $\mu=1$, so the problem is entirely dimensionless and purely geometrical. 

If the reference point $\br_0$ coincides with the center of mass, such that gravity does not exert a torque on the body, the mobility equation is
\begin{equation}\label{eq:rigid_mobility_body} 
\left(
\begin{array}{c}
	\bu_0 \\[0.5ex]
	\bomega_0 
\end{array}
\right) = \tM \cdot \left(
\begin{array}{c}
	\hg \\[0.5ex]
	\mathbf{0} 
\end{array}
\right),
\end{equation}
with $\hg$ the unit vector along gravity, a fixed vector of the laboratory frame
whose components are here taken in the body basis, and $\tM$ the $6\times6$ symmetric rigid-body mobility matrix, also expressed in the body's basis ($\tM$ is the soft mobility tensor of Eq.~(\ref{eq:soft_mobility}) for $N_Q = 0$, the generalized force $\tJ^\top\cdot\bff_\mathrm{ext} = \tC_U^\top\cdot\bff_\mathrm{ext}$ being the total weight and its torque about $\br_0$). The mobility matrix $\tM$ can be written in terms of three $3\times 3$ blocks   
\begin{equation}
\tM = \begin{pmatrix} \ta & \tb^\top \\ \tb & \tc \end{pmatrix}. 
\end{equation}

Furthermore, the reference point $\br_0$ being arbitrary (\S\ref{sec.notation}), it can be placed at the hydrodynamic center of mobility, where the block $\tb$ is symmetric~\cite{Kim2013} and the dynamics of $\hg$ can be written~\cite{Gonzalez2004}
\begin{equation}\label{eq:Gdot}
\frac{\rmd\hg}{\rmd t} = -\bomega_0 \times \hg = \hg \times \left( \tb \cdot \hg\right).
\end{equation}
Here, ${\rmd\hg}/{\rmd t}$ is the rate of change of the body-basis components of $\hg$. Equation~(\ref{eq:Gdot}) is similar to the classical problem of a rigid body spinning in a vacuum~\cite{Marsden2013} with the difference that the eigenvalues of $\tb$ are not necessarily positive. The solution to Eq.~(\ref{eq:Gdot}) can be expressed in terms of Jacobi elliptic functions of time in the eigenvector basis of $\tb$~\cite{Marsden2013}. 
This analytical solution serves as a benchmark for the code time integration.
It is interesting to note that Eq.~(\ref{eq:Gdot}) has been rederived several times in the literature in different contexts, but few authors point to its analytical solution~\cite{Weinberger1972, Gonzalez2004, Weber2013, Palusa2018, Braverman2020, Witten2020, Miara2024, Huseby2025, Joshi2025}.

\begin{figure}[t]
\begin{center}
    \includegraphics[scale=0.60]{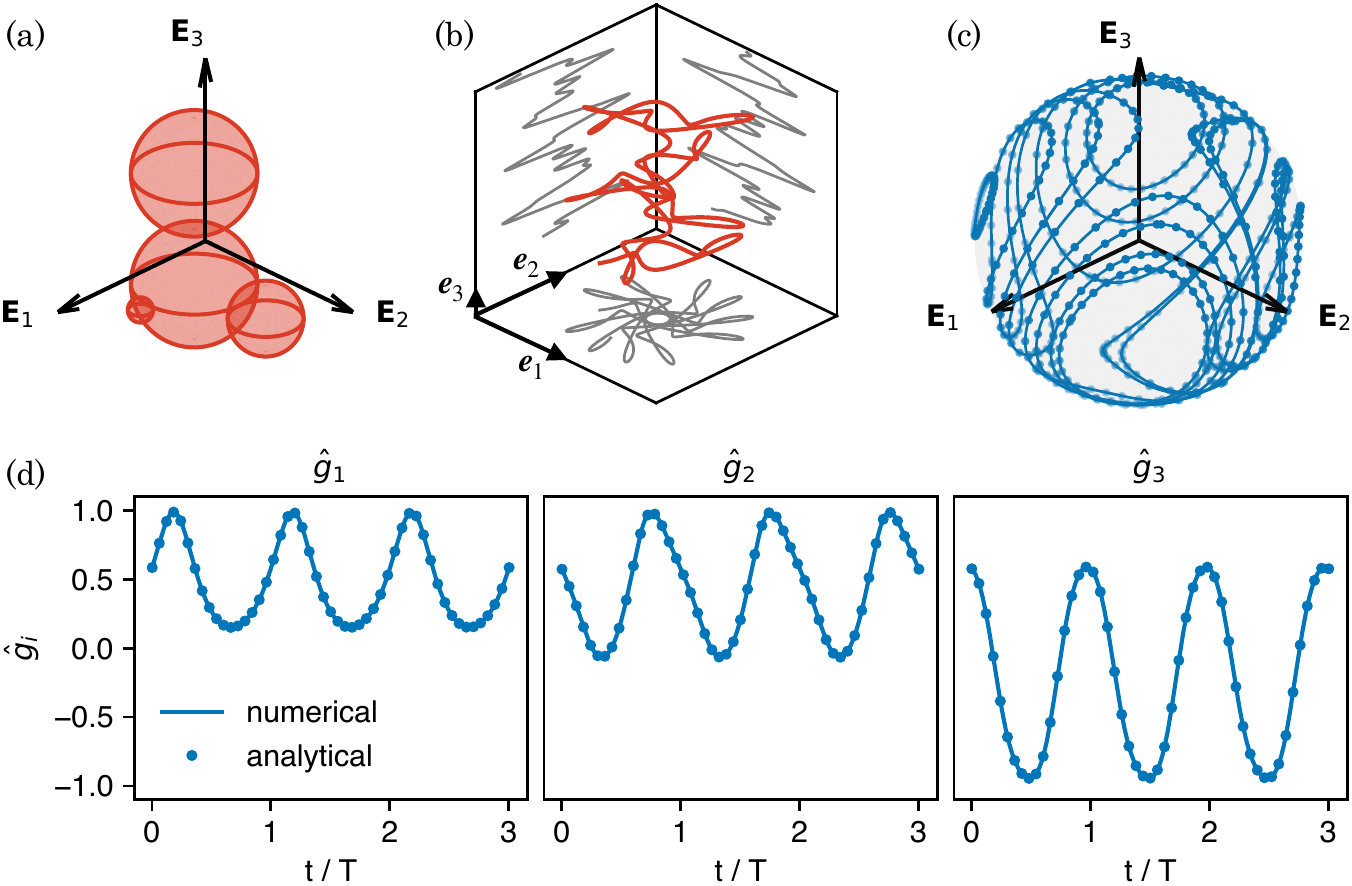}
\caption{
	Sinking dynamics of a chiral rigid body.
	(a) Geometry of the body, with body frame $(\bE_1, \bE_2, \bE_3)$ centered on the center of mass (identical to the hydrodynamic center of mobility).
	(b) Example of trajectory $\br_0$ of the center of mass in the laboratory frame, of basis $(\be_1,\be_2,\be_3)$ (the vertical axis $\be_3$ is compressed by a factor 300). 
	(c) Periodic orbits of the unit gravity vector $\hg$ in the body basis (lines: simulation, dots: analytical solution). 
	(d) Example of a periodic orbit: the three components of $\hg$ are shown as a function of time.  
}
\label{fig:sinking_rigid}
\end{center}
\end{figure}

The body's geometry is made of four spheres touching each other. It is chiral so that the rotation--translation coupling block of the mobility tensor is non-trivial. In practice, we compute the hydrodynamic center of mobility and then distribute the masses of the four spheres so that the center of mass coincides with the hydrodynamic mobility center (Fig.~\ref{fig:sinking_rigid}a). 
We integrate the equation of motion (\ref{eq:rigid_mobility_body}) with our code, which implements a fourth-order Runge--Kutta scheme, with 400 timesteps per period (Fig.~\ref{fig:sinking_rigid}b). The numerical and
analytical trajectories of the unit vector $\hg$ agree to better than numerical precision (Fig.~\ref{fig:sinking_rigid}c--d).

\subsection{Flexible fiber\label{sec:fibers}}

We now want to test the code implementation of elasticity. The most common deformable object in the literature is a flexible fiber~\cite{Du-Roure2019}.
In our case, the fiber is modeled as $N$ touching spheres joined by linear torsional springs of stiffness $k=B/2a$. The parameter $B$ is equivalent to the bending rigidity, such that $B = (\pi/4) E a^4$ for a fiber of radius $a$, with $E$ the Young's modulus. 

We first consider the problem of a negatively buoyant fiber sinking in a fluid at rest. The fiber is modeled with $N=10$ spheres.
The problem is governed by the dimensionless rigidity $\hat B = B / F_\perp L^2$, with $F_\perp = N m g$ the total weight of the fiber, $m$ being the mass of each sphere~\cite{Li2013,Delmotte2015}. 

\begin{figure}[t]
\begin{center}
    \includegraphics[scale=0.60]{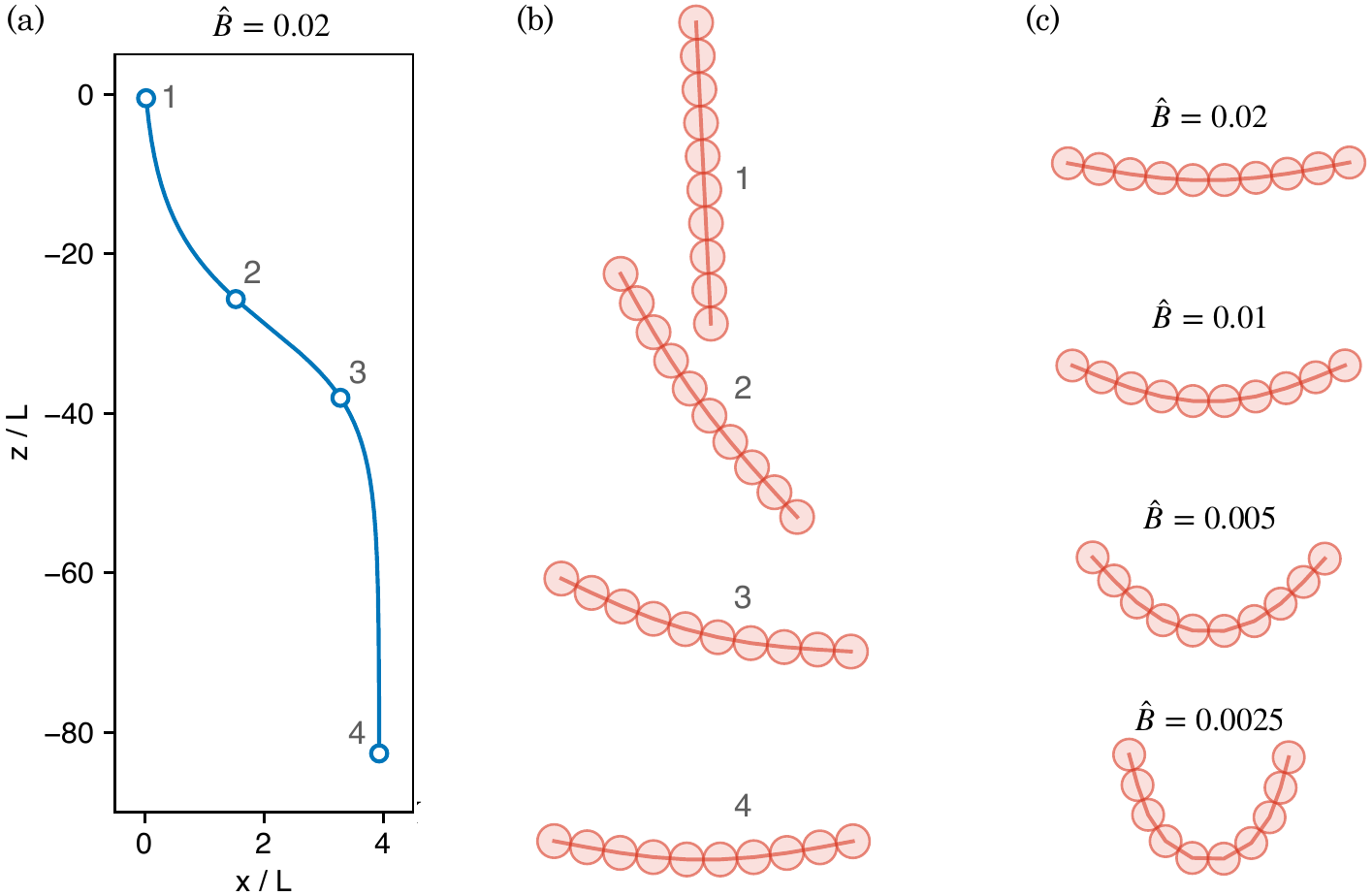}
\caption{
	Sinking dynamics of a flexible fiber ($N=10$ beads of identical radius $a$).
	(a) For an initial angle of $\pi/32$ with respect to the vertical, the graph shows the position of the center of mass of the fiber in the laboratory frame.  
	(b) Snapshots showing the deformation of the fiber along the trajectory shown in (a). 
	(c) Final shape for fibers of decreasing rigidity as labeled. 
}
\label{fig:sinking_fiber}
\end{center}
\end{figure}

As the fiber settles under gravity in a quiescent fluid, the problem activates the $N_Q=N-1$ soft degrees of freedom $\bQ$ of the planar fiber and the stiffness mobility $\tM_K$ for the first time (Fig.~\ref{fig:sinking_fiber}a). 
Starting from a straight configuration, the fiber relaxes to a steady horseshoe shape (Fig.~\ref{fig:sinking_fiber}b), whose curvature
increases as $\hat B$ decreases (Fig.~\ref{fig:sinking_fiber}c). 
We recover the progression from a nearly straight rod
to a tightly curled U reported in Refs~\cite{Li2013,Delmotte2015} with slight differences due to the different implementations: Li et al.~\cite{Li2013} consider a fiber with elliptic shape, and Delmotte et al.~\cite{Delmotte2015} consider resistive force theory only, which does not include hydrodynamic interactions between the different parts of the fiber. 

The next case breaks planarity and introduces prescribed kinematics. 
We consider a flexible fiber made of $N=10$ spheres, clamped at one end to an off-axis support
that is forced to rotate around a fixed axis~\cite{Wiggins1998,Coq2008,Coq2009}. 
Assuming infinitely stiff twisting rigidity, this soft body has $N_Q=18$ degrees of freedom (two bending degrees per adjacent sphere pair). 
The clamp acts as a kinematic constraint and generates an unknown reaction force and torque. 
It is handled through a different soft mobility system derived in Appendix~\ref{app:clamped_anchor},
which takes as unknown $\bQd$ and $\bff_0=[\bF_0,\bT_0]$, the force and torque acting on the clamped sphere 0. 

\begin{figure}[t]
\begin{center}
    \includegraphics[scale=0.60]{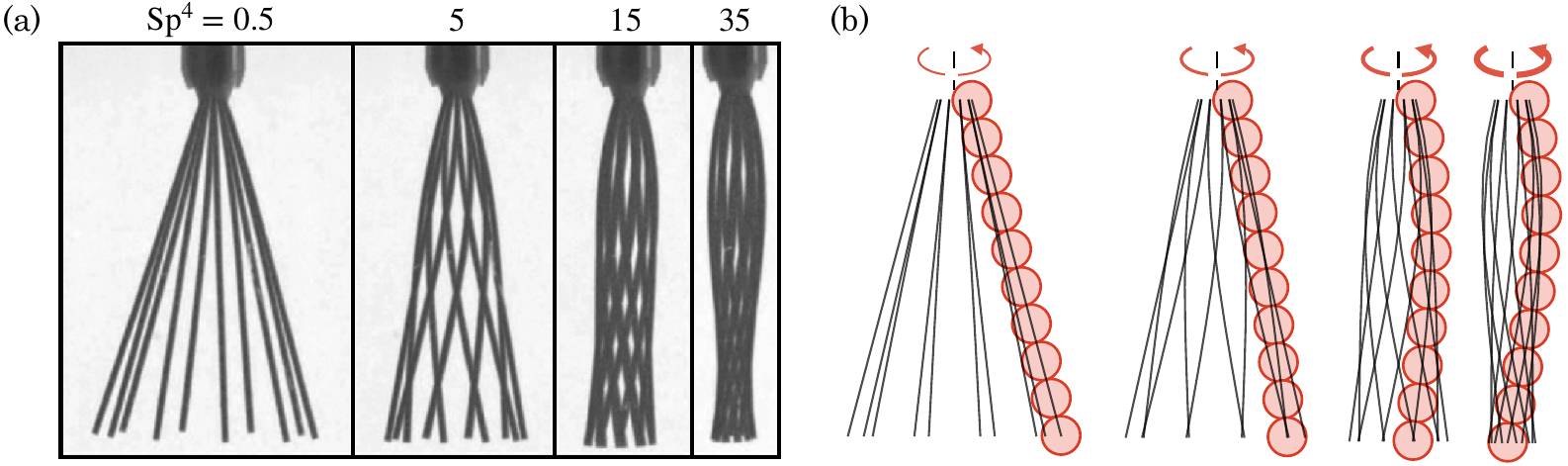}
\caption{
	Shape of a rotating flexible fiber. 
	(a) Experiments with increasing angular velocity, as measured by the Sperm number Sp. The fiber is anchored at an angle $\psi=15^\circ$ with an offset of $\delta_0/L=0.05$ with respect to the rotation axis (reproduced with permission from Ref.~\cite{Coq2009} \copyright 2009 IOP Publishing).
	(b) The fiber is made of $N=10$ touching beads, such that $L/a=20$, with matching Sperm number.  
}
\label{fig:rotating_fiber}
\end{center}
\end{figure}

In this problem, the dimensionless control parameter is the Sperm number $\mathrm{Sp} = {L}/{L_\omega}$, with $L_\omega = \left({B}/{\mu_\perp\,\omega}\right)^{1/4}$ and $\mu_\perp = 4\pi\mu/[\ln(L/a) + 1/2]$, the slender-body transverse drag. 
Our aim is to reproduce the experimental results of Refs~\cite{Coq2008,Coq2009}, in which the fiber is set at an angle $\psi=15^\circ$ with respect to the rotating axis. In these references, the Sperm number is defined as the fourth power of the standard definition used here.
As $\mathrm{Sp}^4$ is increased from $0.5$ to $35$, the shape progresses from a nearly straight rotating fiber through a steady three-dimensional shape that progressively wraps onto the rotation axis, reproducing the regimes documented experimentally (Fig.~\ref{fig:rotating_fiber}). 
Steady state is assessed by the body-frame condition $\max\|\bQd\|\,\tau_\mathrm{el}\ll 1$, where $\tau_\mathrm{el} = (2a)^4\mu / B$ is the typical elastic relaxation time.

\begin{figure}[t]
\begin{center}
   \includegraphics[scale=0.60]{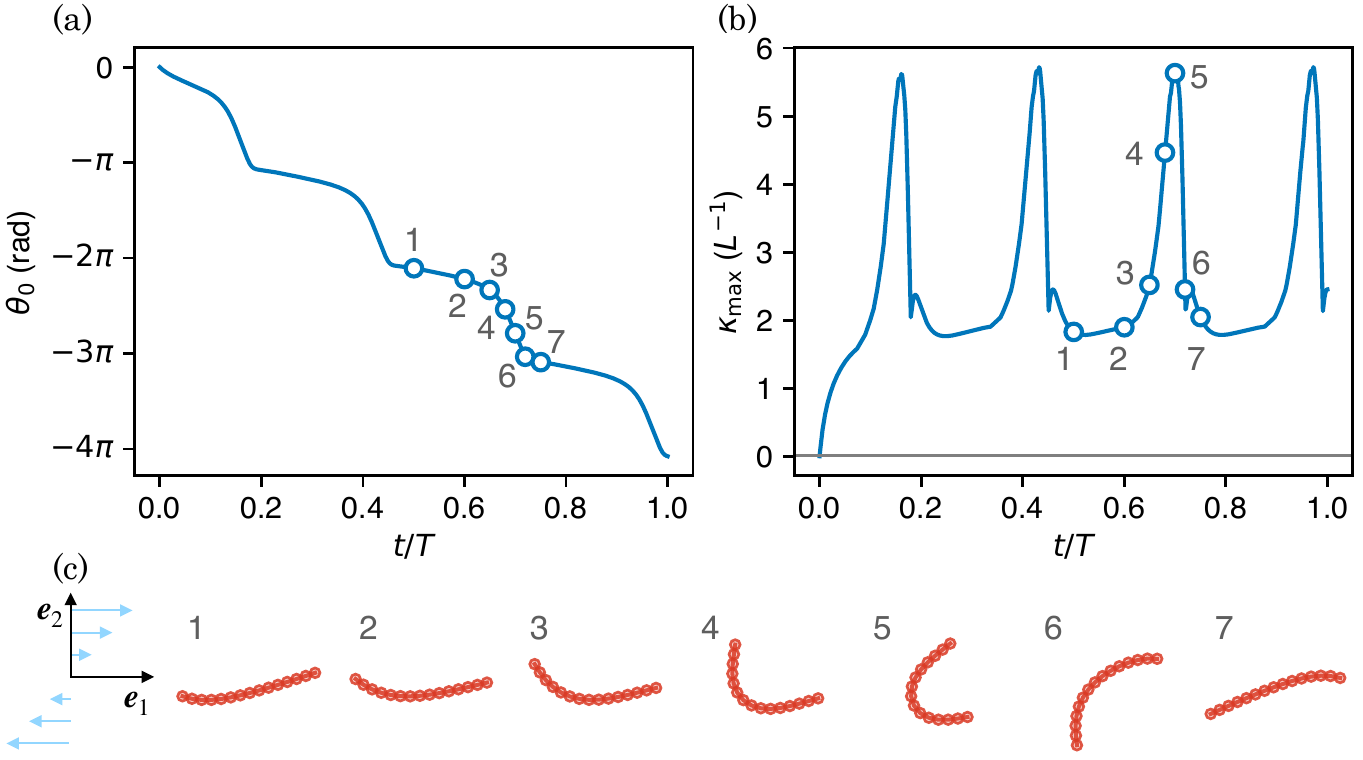}
\caption{
	Flexible fiber tumbling in pure shear flow for a dimensionless bending rigidity $\bar B = B / (8\pi\mu\dot\gamma a^4)=200$. 
	(a) Orientation angle $\theta_0$ around $\be_3$ of the leftmost sphere as a function of time, normalized by Jeffery's period of the rigid assembly $T$.
	(b) Maximum curvature of the fiber in units of length as a function of time. The horizontal gray line shows the intrinsic curvature of the fiber $\kappa_0 = 0.01 / L$.
	(c) Snapshots of the fiber deformation as labeled in (b).  
}
\label{fig:tumbling_fiber}
\end{center}
\end{figure}

The third case considers a flexible fiber made of $N=15$ spheres, in a pure shear flow $\bu^\infty = \dot\gamma\,y\,\be_1$. 
To break the symmetry, the fiber has an intrinsic curvature of $\kappa_0=0.01/L$. 
The problem is governed by the dimensionless rigidity $\bar B = B / (8\pi\mu\dot\gamma a^4)$, which compares elastic and viscous forces.
Figure~\ref{fig:tumbling_fiber} shows the tumbling dynamics when $\bar B = 200$: after a short transient, the fiber periodically rotates in the shear plane. Its maximum curvature and its angular velocity both reach a peak when the fiber is normal to the flow direction (C shape numbered 5 in Fig.~\ref{fig:tumbling_fiber}c). 
This tumbling dynamics is similar to what has been reported in experiments~\cite{Liu2018} and simulations~\cite{Schmid2000,Delmotte2015}.

Together, these three cases test the implementation of the soft degrees of freedom and of the elastic mobility $\tM_K$ (sinking), of the anchor problem of Appendix~\ref{app:clamped_anchor} together with prescribed kinematics (rotation), and of the coupling to a background flow (tumbling).

\subsection{Jeffery's orbits\label{sec.jeffery}}

We now place a rigid dumbbell made of two identical spheres in a pure shear flow. 
This axisymmetric body follows periodic trajectories, known as Jeffery's orbits~\cite{Jeffery1922, Bretherton1962, Thorp2019, Ishimoto2023}. 
The period and shape of the orbits are set by a single shape parameter $\beta$. 
For an ellipsoid of aspect ratio $c=a_\parallel / a_\perp$, the shape parameter is $\beta = (c^2 - 1) / (c^2 + 1)$. 
It was later generalized by Bretherton~\cite{Bretherton1962} to arbitrary axisymmetric bodies through an equivalent aspect ratio.
In our case, the Bretherton parameter $\beta$ is the component of the tensor $\tC_E$ coupling the component $\bE_3$ of the angular velocity $\bomega_0$ to the component $\bE_1 \otimes \bE_2$ of the rate-of-strain tensor $\tE_0^\infty$. 

Writing $\bE_1$ for the dumbbell's axis of symmetry, the rigid mobility equation reduces to
\begin{equation}
\bEd_1 = \bomega_0^\infty \times \bE_{1} +
	\beta\left(\tE_0^\infty\cdot\bE_{1} - (\bE_{1}\cdot\tE_0^\infty\cdot\bE_{1})\bE_{1}\right).
\end{equation}
In a shear flow $\bu^\infty = \dot\gamma\,y\,\be_1$, the angular velocity and rate-of-strain tensor reduce to $\bomega_0^\infty =- \dot\gamma\be_3/2 $ and $\tE_0^\infty = \dot\gamma(\be_1 \otimes \be_2 + \be_2 \otimes \be_1)/2$. 
In spherical coordinates, $\bE_{1} = (\sin\theta\cos\varphi,\sin\theta\sin\varphi,\cos\theta)$ and Jeffery's orbit admits an analytical solution such that
\begin{equation}\label{eq:Jeffery_solution}
\tan\varphi(t) = -\frac{1}{c} \tan\left(\frac{\dot\gamma t}{c+1/c} + \varphi_0 \right), \quad
\tan^2\theta \left(\cos^2\varphi + c^2\sin^2\varphi \right)=K^2,
\end{equation}
with  equivalent aspect ratio $c=\sqrt{(1+\beta)/(1-\beta)}$, period $T=2\pi(c+1/c)/\dot\gamma$, and
 constant of motion $K$, which is set by the initial condition.  
The above solution is now used to test the implementation of the flow-coupling tensor $\tC_E$.

\begin{figure}[t]
\begin{center}
    \includegraphics[scale=0.60]{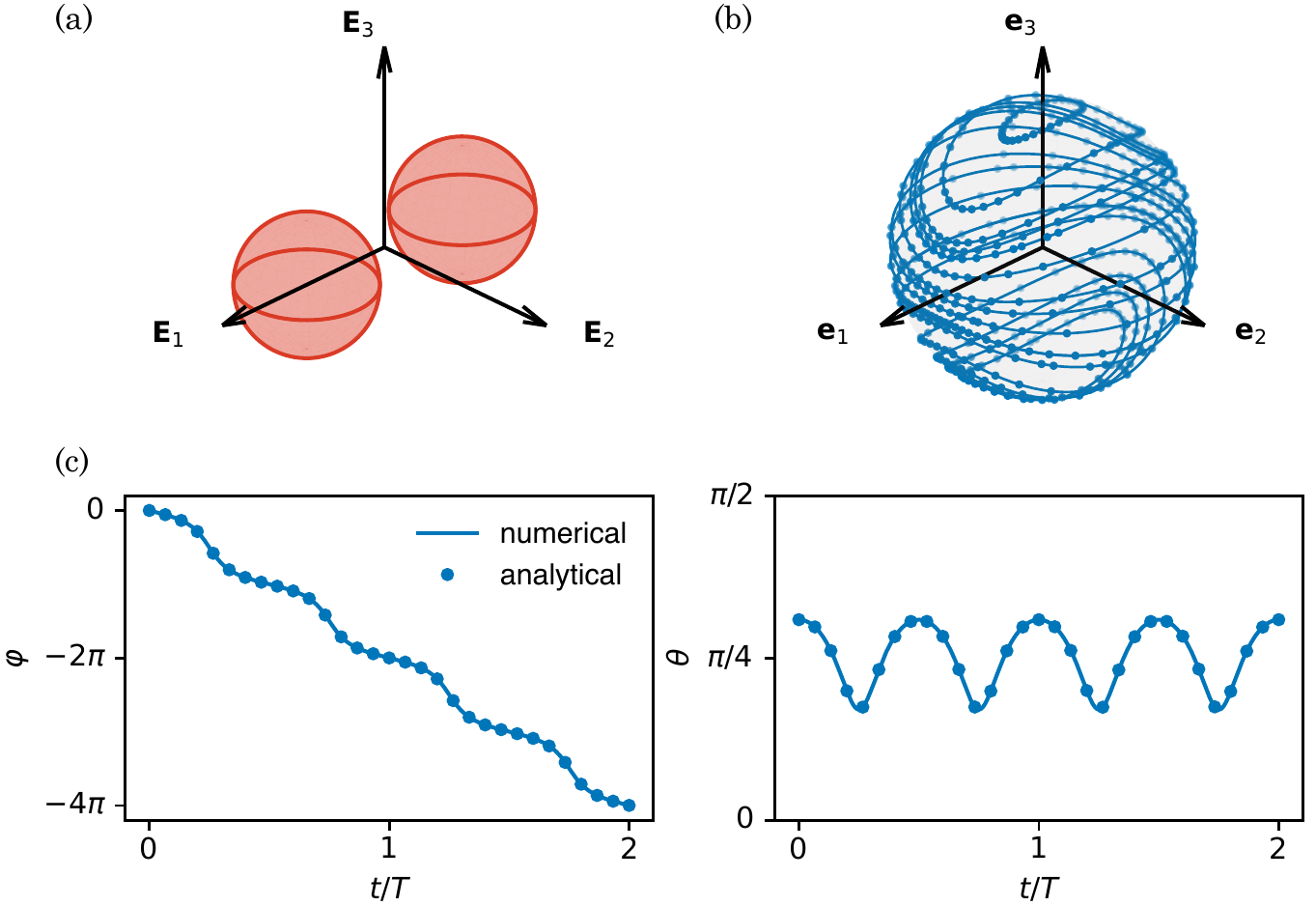}
\caption{
	Jeffery orbits of a rigid dumbbell in a pure shear flow of unit shear rate in the $\be_1$-$\be_2$ plane: $\bu^\infty = [y, 0, 0]$.
	(a) Geometry of the rigid body considered in the body frame.
	(b) Jeffery periodic orbits of $\bE_1$ in the laboratory frame for different initial conditions (lines: simulation, dots: analytical solution).
	(c) Example of an orbit of $\bE_1$ as a function of time in the spherical coordinate $(r, \theta, \varphi)$.	
}
\label{fig:Jeffery_rigid}
\end{center}
\end{figure}

The rigid body geometry is shown in Fig.~\ref{fig:Jeffery_rigid}a: it is made of two spheres of radius $a=1$ separated by a gap $\delta = 1$. 
The dumbbell's Bretherton parameter $\beta=0.717$ is recovered from the rigid mobility tensor and substituted into the analytical orbit.
In Figure~\ref{fig:Jeffery_rigid}, we compare the orbits computed from the library to the analytical solution given by Eq.~(\ref{eq:Jeffery_solution}), which agree within numerical precision. 

\begin{figure}[bht]
\begin{center}
    \includegraphics[scale=0.60]{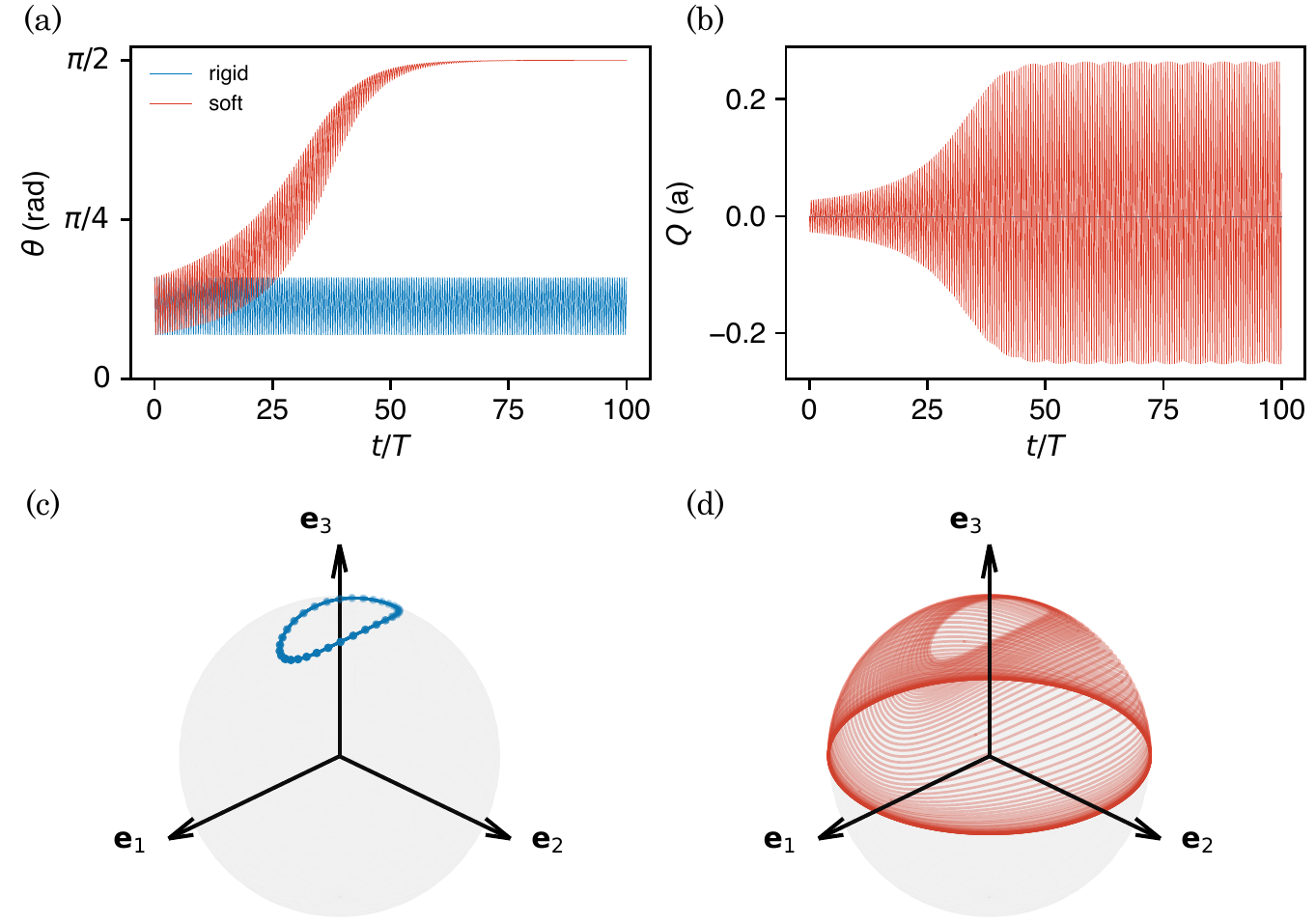}
\caption{
	Jeffery's orbits of a soft dumbbell. 
	(a) Polar angle as a function of time for a rigid ($k=\infty$) and a soft ($k=100\mu \dot\gamma a$) dumbbell.
	(b) Elongation of the spring $Q$ in units of the sphere radius $a$. 
	(c) Similar to Fig.~\ref{fig:Jeffery_rigid} for the initial condition of (a). 
	(d) Orbit of $\bE_1$ the dumbbell's axis of symmetry, showing that the orbit is attracted to the shear plane $\be_1$--$\be_2$, corresponding to a polar angle $\theta=\pi /2$.
	}
\label{fig:jeffery_soft}
\end{center}
\end{figure}

Jeffery's orbits are degenerate, but the degeneracy can be lifted by a small amount of inertia, or a non-Newtonian rheology~\cite{Ishimoto2023}.
Another way to lift degeneracy is to add flexibility~\cite{Zhang2019}. This can be easily tested in the proposed library by replacing the rigid bond between the two spheres of the dumbbell by a linear spring of stiffness $k=100\mu \dot\gamma a$.
In that case, the orbits migrate toward the shear plane, as shown in Fig.~\ref{fig:jeffery_soft}.

\subsection{Three-sphere swimmer}

\begin{figure}[t]
\begin{center}
    \includegraphics[scale=0.60]{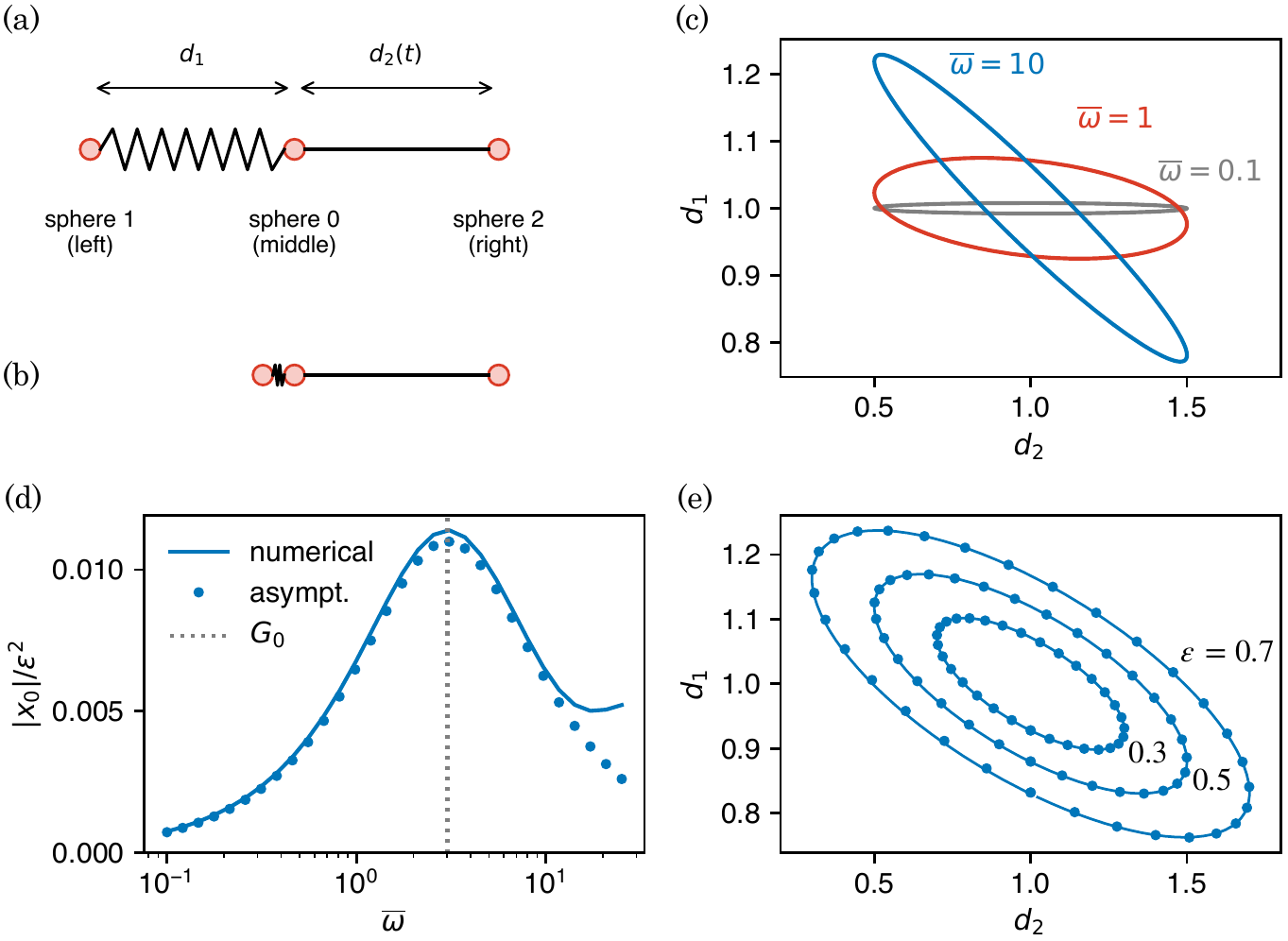}
\caption{
	Optimization of the three-sphere swimmer.
	(a) Default geometry with $d_2 = 1+\varepsilon\sin(t)$, spring rest length $l_1=1$, and sphere radius $a_i=0.05$.
	(b) Optimized geometry for free $k$, $a_1$ and $l_1$.
	(c) Orbits of the swimmer in the $d_2$--$d_1$ plane for different values of the dimensionless frequency $\overline{\omega} = 2 / k$, with $k$ the spring stiffness. 
	(d) Swimming displacement over one period $|x_0|$ as a function of $\overline{\omega}$, for the default geometry and $\varepsilon = 0.1$ (line: simulation, dots: analytical asymptotics in the limit of small $\varepsilon$).
	(e) Orbits of the swimmer at optimal frequency $\overline{\omega} = G_0$ for different values of the forcing amplitude $\varepsilon$, as labeled. 
}
\label{fig:three_sphere}
\end{center}
\end{figure}

We next turn to active swimming and optimization. 
Najafi and Golestanian proposed a simple three-sphere swimmer~\cite{Najafi2004}, which we use here for validation. 
This swimmer is made of three spheres of radii $a_i$ connected by deformable links, breaking time-reversal symmetry through a non-reciprocal actuation cycle (Fig.~\ref{fig:three_sphere}a). 
Here we drive the swimmer by a sinusoidal elongation $d_2 = l_2(1+\varepsilon\sin(\omega t))$ between the right and middle spheres. The left sphere is connected to the middle sphere by a passive spring of stiffness $k$ and rest length $l_1$. 
The problem is entirely determined by the dimensionless parameters
\begin{equation}
\overline{\omega} = \omega \mu (l_1 + l_2) / k, \quad
\varepsilon, \quad
l_1/l_2, \quad
a_i/l_2. 
\end{equation}

The goal is to find the rigidity, or equivalently $\overline{\omega}$, such that the displacement $x_0$ of the middle sphere over one period (after a transient) is maximum. Montino \& DeSimone~\cite{Montino2015} showed that the system admits a well-defined optimum $\overline{\omega}^* = G_0$ in the asymptotic limit of small $\varepsilon$, for fixed $l_1/l_2$ and $a_i / l_2$. 
We will use their prediction as a benchmark of our optimization method. 

We set $\omega=l_2=\mu=1$ in the simulation and start with spheres of identical radii $a_i=0.05$ and $l_1=l_2$ (Fig.~\ref{fig:three_sphere}a). Varying the spring stiffness $k$ varies the dimensionless frequency $\overline{\omega}$. In  Fig.~\ref{fig:three_sphere}c, we show the periodic shape changes in the $d_2$--$d_1$ plane after a transient of four periods. For an optimal $\overline{\omega}^* = G_0$, the displacement over one period is maximized (Fig.~\ref{fig:three_sphere}d). We test that our numerical optimization predicts $G_0$ within less than 1\% for $\varepsilon=0.1$. Figure~\ref{fig:three_sphere}e shows that the simulation at this value of $\overline{\omega}$ reproduces the asymptotic predictions valid for small $\varepsilon$~\cite{Montino2015}.

To test the optimization method further, we optimize three design parameters: $k$, $a_1$, and $l_1$ (for fixed $a_0=a_2=0.05$, and $\varepsilon=0.5$). We show that the final shape (Fig.~\ref{fig:three_sphere}b) displaces the swimmer $7.48$ times farther per period than the shape with $a_1=0.05$ and $l_1=1$ (optimal values: $k^*=1.19$, $a_1^*=0.049$, $l_1^*=0.155$). 

\subsection{Soft surfer\label{sec.surfer}}

\subsubsection{Setting}

We now consider the optimization of a navigation problem: a gyrotactic swimmer in a Taylor--Green  flow~\cite{Colabrese2017}.
We will examine whether elastic deformations may help the swimmer to orient in the flow and ascend more rapidly.
The flow velocity of the Taylor--Green vortices is $\bu^\infty = V_\mathrm{max} [0, \sin(y/\ell)\cos(z/\ell), -\cos(y/\ell)\sin(z/\ell)]$, with typical length $\ell$, such that the maximum angular flow velocity is $\omega_\mathrm{max}=V_\mathrm{max}/\ell$.

This flow is steady and single-scale, but features both elliptic points (pure rotation) and hyperbolic points (pure strain). 
It has been used as a ``model'' of turbulence in the literature on navigation and preferential sampling~\cite{Colabrese2017}. 
We choose it because it remains cheap enough to allow an exhaustive scan over initial conditions, which a direct numerical simulation of turbulence would not. 
Consistently with \S\ref{sec.assumptions}, the swimmer is assumed small compared with $\ell$, so that it experiences an approximately linear flow, given by Eq.~(\ref{eq:bg_flow}), at each instant.

Gravity is directed along $-\be_3$, so that ascending means moving along $+\be_3$. The flow has no component along $\be_1$ and does not depend on $x$. The dynamics is therefore two-dimensional in the $(y,z)$ plane, as long as the initial swimming axis is in the same plane.

\subsubsection{Rigid gyrotactic swimmer}

\begin{figure}[t]
\begin{center}
    \includegraphics[scale=0.60]{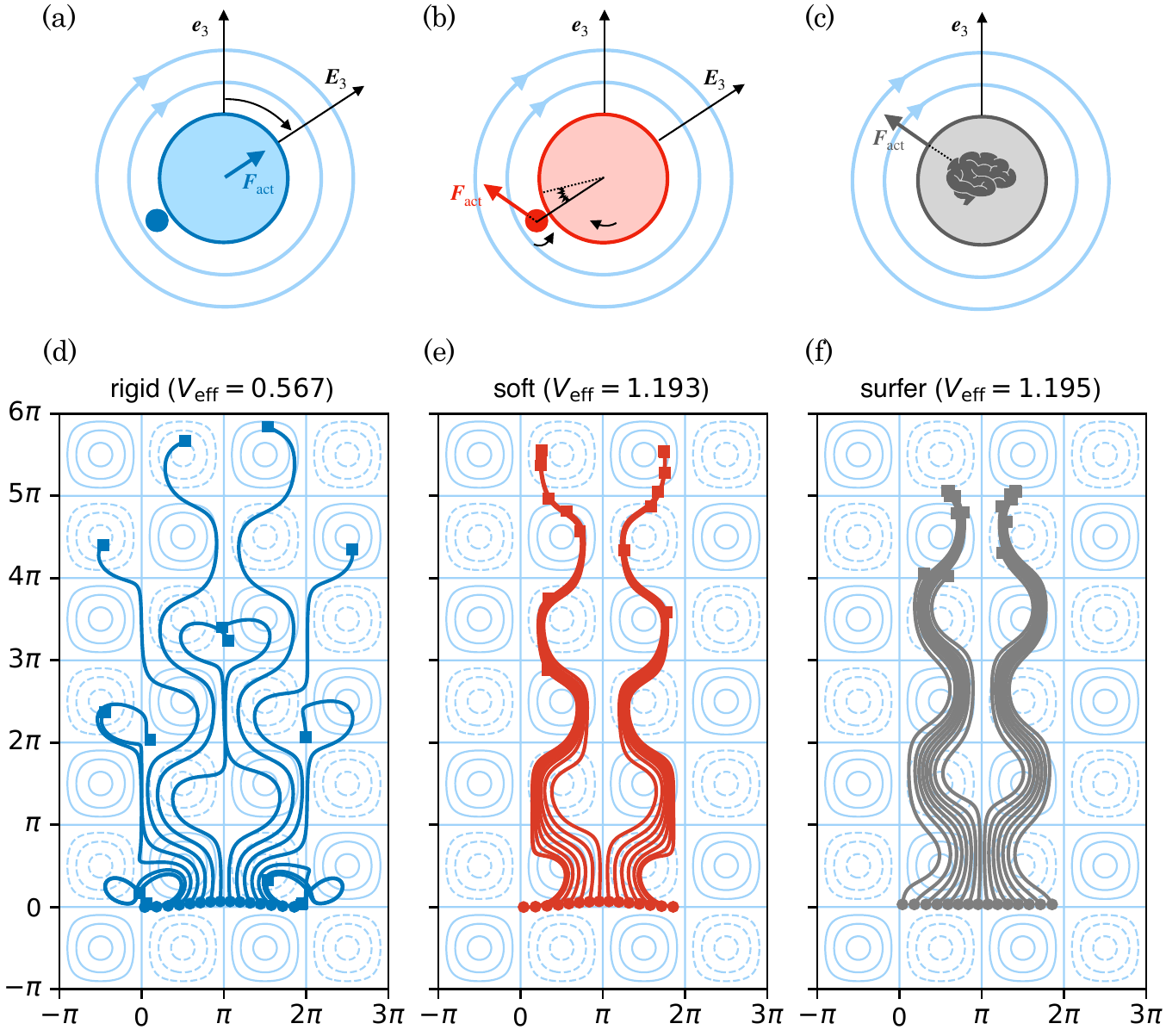}
\caption{
	Vertical migration by gyrotactic swimmers in a flow.
	We consider three types of swimmers:
	(a) a rigid asymmetric dumbbell with the small sphere heavier than the large one, of radii $a_1 \leq a_0$ and equivalent masses $m_0 = -1$, $m_1 = 1$, whose axis $\bE_3$ carries both the bottom-heaviness and the swimming direction;
	(b) a soft swimmer with the same shape as (a) but with the two spheres able to roll on each other, with a torsional pull-back spring of stiffness $k$. The rolling angle $\Theta_1$ of the heavy sphere is the single degree of freedom $Q$, and the swimming direction, fixed in the frame of the heavy sphere, is tilted by $\Theta_1$ from $\bE_3$;
	(c) a surfer, which chooses the direction of swimming according to the surfing policy aligned with $\be_3\cdot \exp(\tau \nabla \bu^\infty)$.
	These three swimmers with identical swimming speed $V_\mathrm{swim}$ have different trajectories in a Taylor--Green vortical flow, shown in the $(y,z)$ plane with gravity along $-\be_3$:
	(d) rigid swimmers can get trapped in vortical cells and their effective vertical velocity is $V_\mathrm{eff} = 0.567 V_\mathrm{swim}$; 
	(e) soft swimmers ``slalom'' the vortical cells and enhance their effective vertical velocity to $V_\mathrm{eff} = 1.193 V_\mathrm{swim}$;
	(f) surfers reach a similar effective vertical velocity of $V_\mathrm{eff} = 1.195 V_\mathrm{swim}$ with a more conservative strategy (more vertical).
	Light blue lines represent the flow streamlines (dashed for negative vorticity, solid for positive vorticity). 
	}
\label{fig:surfers}
\end{center}
\end{figure}

We first consider a rigid bottom-heavy swimmer made of a light sphere, numbered 0, of radius $a_0$ and equivalent mass $m_0=-1$, and a heavy sphere of radius $a_1\leq a_0$ and opposite mass $m_1=1$ (Fig.~\ref{fig:surfers}a).
Its axis $\bE_3$ joins the two sphere centers and points from the heavy to the light sphere. It is both the bottom-heavy axis and, for this rigid swimmer, the swimming direction.

The orientation dynamics follows from the angular velocity block of the soft mobility equation~(\ref{eq:soft_mobility_body}). With $N_Q = 0$ and no prescribed kinematics, only the field $\bH = -g \be_3$ and rate-of-strain terms survive,
\begin{equation}\label{eq:omega_surfer}
\bomega_0 = \bomega_0^\infty + \left[\tM_H \cdot \bH\right]_\omega 
+ \left[\tC_E : \tE_0^\infty\right]_\omega,
\end{equation}
and the swimming axis is transported by $\bEd_3 = \bomega_0 \times \bE_3$. For an axisymmetric body, the rate-of-strain term reduces to the Bretherton form with shape factor $\beta$, and the gravitational term to an alignment toward the vertical, so that Eq.~(\ref{eq:omega_surfer}) becomes the Pedley--Kessler equation~\cite{Pedley1992}
\begin{equation}\label{eq:Pedley-Kessler}
\bEd_{3} = \bomega_0^\infty \times \bE_{3} + \beta\left(\tE_0^\infty\cdot\bE_{3} - (\bE_{3}\cdot\tE_0^\infty\cdot\bE_{3})\bE_{3}\right) + \frac{1}{2 \tau_\mathrm{align}} \left(\be_3 - (\be_3\cdot\bE_{3} ) \bE_{3}\right),
\end{equation}
where the first two terms on the right-hand side are due to the viscous torques and the third term to the gravitational torque.
In practice, for a rigid body with axis of symmetry $\bE_3$, the alignment timescale is $\tau_\mathrm{align} = 1/(2 m_1 g M_H)$, where $M_H$ is the entry of the mobility tensor $\tM_H$ that couples the $\bE_1$ component of the angular velocity $\bomega_0$ to the $\bE_2$ component of the gravitational force.
For the rigid body represented in Fig.~\ref{fig:surfers}a, we find $\tau_\mathrm{align} = 0.075 / \omega_\mathrm{max}$ for $m_1 g = 50 \mu a_0^2 \omega_\mathrm{max}$.

The propulsion is modeled by an effective force, fixed in the body frame and directed along $\bE_3$, adjusted so that the swimming velocity in a fluid at rest is $V_\mathrm{swim} = V_\mathrm{max}$. This is a standard device for describing the kinematics of a force-free swimmer through the mobility matrix~\cite{Kummel2014}: here it is equivalent to integrating Eq.~(\ref{eq:Pedley-Kessler}) for the orientation together with $\bu_0 = \bu_0^\infty + V_\mathrm{swim}\bE_3$ for the translation, which is the gyrotactic swimmer of Pedley and Kessler. 
The trajectories for different initial positions are depicted in Fig.~\ref{fig:surfers}d. The swimmer tends to orient upwards, thanks to its bottom-heaviness, when it is near a separatrix where the vorticity is small. Otherwise, it tends to orient at an angle that can be obtained by balancing the viscous and gravitational torques in Eq.~(\ref{eq:Pedley-Kessler}). This equilibrium direction is illustrated in Fig.~\ref{fig:surfers}a for a steady vortical flow. It tends to move the swimmer in regions of smaller vertical velocity, which results in a preferential sampling of negative vertical flow velocities. This slows down the vertical migration and the averaged effective vertical velocity is $V_\mathrm{eff} = 0.567 V_\mathrm{swim}$ for the 15 initial conditions tested in Fig.~\ref{fig:surfers}d and a simulation time of $T=4\pi/ \omega_\mathrm{max}$.  

\subsubsection{Soft gyrotactic swimmer}

We now consider the same geometry, but instead of having a rigid body, the two spheres are able to roll onto each other with a torsional spring of stiffness $k$ (Fig.~\ref{fig:surfers}b). Rolling without slip ties the two sphere orientations, $a_1 \bTheta_1 = -a_0\bTheta_0$, leaving a single degree of freedom $\bQ = [\Theta_1]$, the rolling angle of the heavy sphere about $\bE_1$. The effective force is now fixed in the frame of the heavy sphere rather than in the body frame, so that the swimming direction is tilted by $\Theta_1$ away from $\bE_3$: the axis $\bE_3$ still joins the two sphere centers and still carries the bottom-heaviness, but the swimmer no longer swims along it. Because the effective force is applied to the heavy sphere, it also enters the deformation block of Eq.~(\ref{eq:soft_mobility_body}) and therefore contributes, alongside the viscous and elastic torques, to setting $\Theta_1$.
Flexibility adds a dimensionless rigidity to the problem: $\bar k = k/(\mu \omega_\mathrm{max} a_0^3)$.
Through the optimization, we found the design parameters of this soft swimmer that maximize the average effective vertical velocity $V_\mathrm{eff}$ for the same 15 initial conditions as the rigid gyrotactic swimmer and the same simulation time of $T=4\pi/ \omega_\mathrm{max}$.
These optimal parameters are $a_1/a_0=0.169$ and $\bar k = 18.2$, and the effective velocity is $V_\mathrm{eff} = 1.193 V_\mathrm{swim}$, which means that this soft swimmer swims 19\% faster than without background flow (Fig.~\ref{fig:surfers}e).

The mechanism behind this speed-up is the following. In a quiescent fluid, the torsional spring relaxes both sphere orientations to zero, the swimming direction then coincides with the bottom-heavy axis $\bE_3$, and the soft swimmer behaves like the rigid one. 
In a vortical flow, the viscous torques on the two spheres have the same sign, and the light sphere being larger, its torque is larger. The rolling constraint $a_1\bTheta_1 = -a_0\bTheta_0$ forces the two rotations into opposition, so that, balanced against the spring, the torques set an equilibrium with a small angular offset between the heavy sphere and $\bE_3$, whose amplitude is controlled by $a_1/a_0$ and $\bar k$. 
The optimum $(a_1/a_0, \bar k) = (0.169, 18.2)$ found by the optimization lies in between, where the offset is just large enough to tilt the swimming axis in the opposite sense to that of the rigid swimmer: a mirror swimming direction with the same bottom-heavy axis (Fig.~\ref{fig:surfers}b). The bottom-heavy axis still points toward regions of negative vertical flow, but the swimming axis now points toward regions of positive vertical flow. This reversal of the gyrotactic bias produces the preferential sampling of upward flow and the 19\% speed-up of $V_\mathrm{eff}$ over $V_\mathrm{swim}$.

Although this ``soft surfer'' would be difficult to fabricate, it has a computational advantage: the rolling constraint $a_1\bTheta_1 = -a_0\bTheta_0$ ties the two sphere orientations to each other, and the center-to-center vector remains aligned with $\bE_3$. The  positions of the two spheres are thus frozen in the body frame $(\bE_1, \bE_2, \bE_3)$, so the soft mobility tensors in Eq.~(\ref{eq:soft_mobility_body}) do not depend on the degree of freedom $\bQ$ and can be evaluated once instead of at every timestep.

\subsubsection{Comparison with the surfing strategy}

We compare this performance with that of a surfer that chooses its swimming direction according to the surfing strategy of Ref.~\cite{Monthiller2022}, which aligns the swimming direction along $\be_3\cdot\exp(\tau\nabla\bu^\infty)$ with a time delay $\tau_\mathrm{align}$ to make a fair comparison with the rigid or soft gyrotactic swimmers considered before. 
This surfing strategy is based on a linear approximation of the background flow integrated during a typical correlation time $\tau$. We have shown that this strategy is quasi-optimal for turbulent flows~\cite{Mecanna2025}.
Here $\tau$ can be viewed as a parameter that has to be optimized for maximal $V_\mathrm{eff}$: we find $\tau = 0.547 / \omega_\mathrm{max}$ and $V_\mathrm{eff}= 1.195 V_\mathrm{swim}$. So the performance is similar to that of the optimal soft swimmer, with trajectories that are slightly more vertical  (Fig.~\ref{fig:surfers}f).
The soft swimmer thus achieves, through passive elasticity alone, what the surfing strategy achieves through an explicit measurement of the velocity gradient: the deformation performs the computation that the policy would otherwise have to carry out, an instance of morphological computation~\cite{Pfeifer2014,Hauser2014}.

A real microswimmer is force-free, its thrust being balanced by the drag on an unresolved propulsive appendage, so the effective force reproduces the swimmer's kinematics but not its flow~\cite{Felderhof2014,Kummel2014}. 
This has no consequence here, because we consider an isolated swimmer in an externally imposed flow, but it would matter for interactions between swimmers or with boundaries.

\section{Discussion\label{sec.discussion}}

We have presented a soft mobility framework that extends classical rigid-body mobility theory to deformable bodies in Stokes flow. 
Considering a soft body immersed in a linearized background flow, we showed that its idealization as an assembly of spheres connected by rods and springs obeys a saddle-point problem, whose explicit solution is the soft mobility equation.
This equation is an ordinary differential equation for the position, orientation, and deformation, in which the soft mobility tensors are explicit functions of the deformation state. The framework was validated on canonical problems and exploited as an inverse-design tool through end-to-end automatic differentiation, illustrated by the soft surfer: a passive gyrotactic swimmer able to exploit the background flow to ascend faster than without flow. 

The soft mobility equation~(\ref{eq:soft_mobility}) sits between two established families. 
Bead and multiblob methods~\cite{Delmotte2015,Balboa-Usabiaga2022,Fuchter2023} discretize a deformable body into rigid sub-bodies and enforce the internal constraints through Lagrange multipliers, solving the resulting saddle-point system iteratively at every timestep, and, as \S\ref{sec.saddle} makes explicit, our formulation shares that structure. 
What differs is that we eliminate the constraint loads rather than iterate on them, at a cost in $N$ that buys an interpretable and differentiable reduced system. 
At the other end, the generalized-coordinate framework of Solovev and Friedrich~\cite{Solovev2021} projects the deformation onto a small set of Lagrangian coordinates and balances generalized forces in their conjugate space, so that the constraints never appear at all. 
Our reduction is the same in spirit, and both have the structure of Onsager's variational principle for dissipative systems~\cite{Doi2011}.
It differs in two ways: their mobility matrix is tabulated on the generalized coordinates and interpolated rather than given in closed form, which limits differentiability, and their formulation carries no coupling to the rate-of-strain of the background flow.

The present framework therefore combines the projection idea of Ref.~\cite{Solovev2021} with three additions: a closed-form differentiable mobility (here Rotne--Prager--Yamakawa, though any differentiable mobility model is compatible), explicit elastic and flow-gradient coupling tensors, and a JAX implementation that makes every tensor differentiable with respect to the design parameters. 
The cost of one forward simulation exceeds that of an explicit bead-spring or multiblob simulation, but the gradient of any scalar objective with respect to the design parameters comes essentially for free through automatic differentiation.

The soft mobility theory is tailored for systems with few degrees of freedom, typically $N_Q = \mathcal{O}(1)$, in contrast with standard Stokesian dynamics approaches that handle suspensions with $N_Q = \mathcal{O}(N)$, where $N = \mathcal{O}(10^3)$ rigid bodies or more~\cite{Brady1988,Fiore2019,Swan2016}. Those methods invert the saddle-point system arising from the constraints using preconditioned iterative solvers (typically GMRES~\cite{Saad1986}, with a cost of $\mathcal{O}(N^2)$ per iteration, or $\mathcal{O}(N)$ when combined with fast multipole methods~\cite{Greengard1987,Tornberg2008,Liang2013}).
These iterative procedures are efficient for forward simulation but difficult to differentiate. 
Our framework instead inverts the dense $6N\times 6N$ grand mobility matrix directly in Eq.~(\ref{eq:schur_solution}). This $\mathcal{O}(N^3)$ inversion is computationally costly, but it is differentiable. 
For a soft body with small $N_Q$ and no prescribed kinematics, the tensors depend on $\bQ$ alone and the inversion can be tabulated on a discrete grid in $\bQ$ and interpolated during simulation, making the cost competitive with iterative approaches. When part of the kinematics is prescribed, the grid must be extended over the gait phase as well.
The framework is therefore most useful when one wants to optimize the design of a single deformable body, and least competitive when one wants to simulate large suspensions for which Stokesian dynamics remains the right tool.

Three modeling assumptions bound the domain of validity. 
First, the background flow is linearized around the body, which requires $L \ll \ell$, where $L$ is the body size and $\ell$ the smallest scale over which the flow varies (the Kolmogorov scale $\eta$ in turbulence). 
Second, the Rotne--Prager--Yamakawa approximation is a truncated multipole expansion that does not resolve the lubrication forces between close surfaces. 
This is a genuine approximation here, since the spheres of our fiber assemblies or soft gyrotactic swimmer are in contact by construction. 
Rotne--Prager--Yamakawa nonetheless remains symmetric positive definite for overlapping spheres, which is what keeps $\tM_G$ invertible and differentiable, whereas lubrication corrections are built from tabulated resistance functions~\cite{Jeffrey1984} that are difficult to 
differentiate properly. 
The third assumption is structural rather than physical: the degrees of freedom $\bQ$ are required to parameterize the deformation minimally, so that all internal constraints are holonomic and eliminated by the choice of coordinates. Some architectures cannot be parameterized in this way and fall outside the present formulation: for instance, statically indeterminate assemblies such as tensegrities, unilateral constraints such as contact between non-adjacent spheres.

Soft mobility theory could be extended in three directions: first, thermal fluctuations; second, interactions between soft bodies; and third, arbitrary geometries. 

First, Brownian fluctuations could be added by including a stochastic force term in the soft mobility equation, calibrated to satisfy the fluctuation--dissipation theorem on the reduced generalized coordinates. The spurious drift term this requires, proportional to the divergence of the generalized mobility, is itself a derivative of a tensor we already compute, and is therefore available at no additional analytical cost through automatic differentiation. 
This is the strategy adopted in Brownian dynamics codes built on the same tools~\cite{Waszkiewicz2023}. 
The extension would bring soft macromolecules and microswimmers in thermal baths within reach~\cite{Cichocki2021,Funkenbusch2024}. 

Second, far-field hydrodynamic interactions between soft bodies could be incorporated by treating each soft body as a deformable multipole and coupling them through a truncated Fax\'en expansion. Pairs and dilute suspensions of soft bodies in flow would then become tractable, opening the soft-mobility analogue of Stokesian dynamics for dilute suspensions of deformable bodies, complementing existing Stokesian dynamics for rigid suspensions~\cite{Brady1988,Fiore2019}.

Third, the continuum formulation of Appendix~\ref{app:continuum} invites an implementation of its own. Nothing in the derivation requires the body to be an assembly of spheres.
The natural choice would combine finite-element modes for the elastic operator with a boundary-element or regularized-Stokeslet evaluation of the Stokes forces. 
A compact basis could be built from the least attenuated viscous eigenmodes of the body in a quiescent fluid, retaining the $N_Q$ slowest deformation modes. 
The obstacle is differentiability: the reduced tensors would follow from large linear solves rather than closed-form expressions.
In our view, the natural route to the inverse design of soft bodies of arbitrary geometry would be an adjoint formulation of the soft mobility equation. 

\begin{acknowledgments}

Most of this work has been developed during a sabbatical stay in Cambridge, funded by the French Embassy Churchill College Fellowship.
This project has received funding from the European Research Council (ERC) under the European Union's Horizon 2020 research and innovation program (grant agreement No 834238).

\end{acknowledgments}

\appendix

\section{From generalized coordinates to sphere velocities\label{app:kinematics}}

Orientations are parameterized by Rodrigues vectors $\bTheta\in\mathbb{R}^3$, whose 
direction $\hTheta = \bTheta/\Theta$ (with $\Theta = \|\bTheta\|$) is the rotation 
axis and whose magnitude $\Theta$ is the rotation angle. The corresponding rotation 
matrix follows from the Euler--Rodrigues formula
\begin{equation}\label{eq:Rodrigues_formula}
\mRot(\bTheta) = \cos\Theta\,\tI + \sin\Theta\,[\hTheta]_\times 
+ (1-\cos\Theta)\,\hTheta\otimes\hTheta,
\end{equation}
with $\otimes$ the outer product, so that $\bx = \mRot(\btheta_0)\cdot\bX$ converts 
body-basis into laboratory-basis components. This parameterization is 
singularity-free for $\Theta\in[0,2\pi)$, unlike Euler angles.

The angular velocity is not the time derivative of the Rodrigues vector. 
Differentiating Eq.~(\ref{eq:Rodrigues_formula}) gives the Jacobian $\tB_3$ of the exponential map $\mathfrak{so}(3)\to SO(3)$, whose inverse is known as the Bortz equation in the attitude-estimation literature~\cite{Bortz1971},
\begin{equation}\label{eq.Bortz_kinematics}
\begin{aligned}
&\bOmega = \tB_3(\bTheta) \cdot \bThetad,\qquad
\tB_3(\bTheta) = \frac{\sin\Theta}{\Theta}\,\tI
	+ \frac{1-\cos\Theta}{\Theta}\left[\hTheta\right]_\times
	+ \left(1 - \frac{\sin\Theta}{\Theta}\right)\hTheta\otimes\hTheta,\\
&\bThetad = \tB_3^{-1}(\bTheta) \cdot \bOmega,\qquad
\tB_3^{-1}(\bTheta) = \frac{\Theta}{2}\cotan\!\left(\frac{\Theta}{2}\right)\tI
	- \frac{\Theta}{2}\left[\hTheta\right]_\times
	+ \left(1 - \frac{\Theta}{2}\cotan\!\left(\frac{\Theta}{2}\right)\right)\hTheta\otimes\hTheta.
\end{aligned}
\end{equation}
Here the angular velocity $\bOmega$, defined by $[\bOmega]_\times = \dot{\mRot}\cdot\mRot^\top$, has its components in the basis of the parent frame: the body basis for a sphere, and the laboratory basis for the body frame in Eq.~(\ref{eq:generalized_velocity}). In the basis of the rotated frame, $\tB_3$ is replaced by its transpose. Both matrices tend to $\tI$ as $\Theta\to 0$, and $\det\tB_3 = [\sin(\Theta/2)/(\Theta/2)]^2$. Since the translational part is unaffected, 
$\bU_i = \bRd_i$, the six-component velocity of sphere $i$ follows from 
$\bXd_i = [\bRd_i, \bThetad_i]$ through the block-diagonal Bortz operator
\begin{equation}\label{eq:B6}
\bV_i = \tB_6(\bTheta_i)\cdot\bXd_i,
\qquad
\tB_6(\bTheta_i) =
\begin{bmatrix} \tI & \mathbf{0} \\ \mathbf{0} & \tB_3(\bTheta_i) \end{bmatrix}.
\end{equation}
The same construction applied to the body frame gives the $N_q\times N_q$ matrix 
$\tB(\btheta_0)$ of Eq.~(\ref{eq:generalized_velocity}), block-diagonal with 
$\tB_6(\btheta_0)$ in the upper-left block and $\tI$ in the lower-right.

The six-component positions $\bX_i(\bQ,t)$ are functions of the degrees 
of freedom and, where the kinematics is prescribed, of time. Applying the chain rule 
to Eq.~(\ref{eq:B6}) splits the body-frame velocity into a deformation and an 
active part,
\begin{equation}\label{eq:Jacobian_sphere}
\bV_i = \tJ_i \cdot \bQd + (\bV_\mathrm{act})_i,
\qquad
\tJ_i = \tB_6 \cdot \frac{\partial \bX_i}{\partial \bQ},
\qquad
(\bV_\mathrm{act})_i = \tB_6 \cdot \frac{\partial\bX_i}{\partial t},
\end{equation}
which stack into $\bV = \tJ_\mathrm{sph} \cdot \bQd + \bV_\mathrm{act}$, with 
$\tJ_\mathrm{sph}$ of size $6N \times N_Q$ and $\bV_\mathrm{act}$ of size $6N$.

The laboratory-frame velocity of sphere $i$ adds the rigid-body transport due to 
$\bv_0 = [\bu_0,\bomega_0]$,
\begin{equation}\label{eq:composition_per_sphere}
\bu_i = \bU_i + \bu_0 + \bomega_0 \times \bR_i, 
\qquad \bomega_i = \bOmega_i + \bomega_0,
\end{equation}
so that $\bv = \bV + \tC_U \cdot \bv_0$, where $\tC_U$ is the $6N\times 6$ stack of 
blocks
\begin{equation}\label{eq:CU_blocks}
(\tC_U)_i = 
\begin{bmatrix} \tI & -[\bR_i]_\times \\ \mathbf{0} & \tI \end{bmatrix}.
\end{equation}
Together, Eqs.~(\ref{eq:Jacobian_sphere})--(\ref{eq:CU_blocks}) give the kinematic 
relation~(\ref{eq.U_is_J_Qdot}) of the main text, with the kinematic Jacobian 
assembled by concatenation, $\tJ = [\,\tC_U,\; \tJ_\mathrm{sph}\,]$. Because every 
entry is evaluated from the sphere positions and orientations in the body frame, 
$\tJ$ is a function of $\bQ$, and of $t$ where the kinematics is prescribed, but 
never of $\bx_0$.

The same transport applies to the background flow. Evaluating the 
linearization~(\ref{eq:bg_flow}) at the sphere centers and using 
$\bomega_i^\infty = \bomega_0^\infty$ gives Eq.~(\ref{eq:grandflowvelocity}), with 
the stresslet coupling tensor acting as
\begin{equation}\label{eq:CS_blocks}
(\tC_S : \tE_0^\infty)_i = \left[\,\tE_0^\infty\cdot\bR_i,\; \mathbf{0}\,\right],
\end{equation}
the rate of strain being uniform at the scale of the body.

One numerical precaution is needed: $\tB_3$ is singular as $\Theta\to 2\pi$, where its determinant vanishes and $\cotan(\Theta/2)$ in its inverse diverges. The Rodrigues vectors are therefore remapped after each 
timestep, $\bTheta \leftarrow \bTheta - 2\pi\,\hTheta$ whenever 
$\|\bTheta\|\geq\pi$. This subtracts a full turn about the rotation axis, leaving 
the orientation unchanged while keeping $\|\bTheta\|\leq\pi$.

\section{The anchor problem\label{app:clamped_anchor}}

In several applications, one prescribes the velocity of a single sphere of the assembly, the anchor, rather than letting the assembly translate and rotate freely under external forces. Examples include clamped fibers in \S\ref{sec:fibers}. We extend here the soft mobility equation~(\ref{eq:soft_mobility_body}) to such an anchor problem.

We choose the body frame so that its origin coincides with the center of the anchored sphere, numbered $0$. So the subscript $0$ refers to both the body frame and the $0$th sphere's frame, which coincide. Its axis $\bE_1$ aligns with sphere $0$'s orientation, such that it is fixed in the body frame: $\bR_0 = \bm{0}$ and $\bTheta_0 = \bm{0}$ for all time. 

The body's lab-frame position $\br_0$ and orientation $\btheta_0$ are then identified with sphere $0$'s lab-frame position and orientation, and the actuation imposes them through two prescribed time-functions, $\bx_0(t)=[\br_0,\btheta_0]$. 
The body's six-component velocity $\bv_0 = [\bu_0,\bomega_0]$ in the laboratory frame follows from the prescription via the Bortz Jacobian~(\ref{eq.Bortz_kinematics}) of sphere~$0$,
$\bv_0(t) = \tB_6(\btheta_0)\cdot\bxd_0$, with $\tB_6$ given by Eq.~(\ref{eq:B6}).

The anchor problem is a mixed mobility problem: $\bv_0$ is now a known input rather than a velocity solved for, while the six-component reaction load $\bff_0 = [\bF_0,\bT_0]$ that the anchor exerts on sphere $0$ becomes an unknown. We reintroduce $\bff_0$ as an additional grand external load concentrated on sphere $0$. The associated contribution is added to the right-hand side of the soft mobility equation~(\ref{eq:soft_mobility_body}) as 
\begin{equation}\label{eq:soft_mobility_clamped}
\bp - \bp_0^\infty
=
\tM_K\cdot\bQ + \tM_H\cdot\bH + \tC_E:\tE_0^\infty - \tPi\cdot\bV_\mathrm{act}
+ \tM_0\cdot\bff_0,
\end{equation}
where $\tM_0\cdot\bff_0 = \tM\cdot\tJ^\top\cdot[\bff_0, \bm{0}] = \tM\cdot[\bff_0, \bm{0}]$, sphere~$0$ being fixed at the origin of the body frame, so that $\tM_0$ is made of the first six columns of $\tM$.

Splitting $\tM_0$ in a $6\times 6$ velocity block and an $N_Q\times 6$ deformation block,
\begin{equation}\label{eq:M_anchor_partition}
\tM_0 =
\begin{bmatrix} \tM_0^{v} \\[0.5ex] \tM_0^{Q} \end{bmatrix},
\end{equation}
the top six rows of Eq.~(\ref{eq:soft_mobility_clamped}) form a $6\times 6$ linear system for the unknown $\bff_0$,
\begin{equation}\label{eq:f_anchor_solve}
\tM_0^{v}\cdot\bff_0
\;=\;
\bv_0 - \bv_0^\infty
- \left[\tM_K\cdot\bQ + \tM_H\cdot\bH + \tC_E:\tE_0^\infty - \tPi\cdot\bV_\mathrm{act}\right]_{1{:}6},
\end{equation}
inverted at every integrator substep to produce $\bff_0$. The remaining $N_Q$ rows then yield the deformation rate
\begin{equation}\label{eq:Qdot_clamped}
\bQd
\;=\;
\left[\tM_K\cdot\bQ + \tM_H\cdot\bH + \tC_E:\tE_0^\infty - \tPi\cdot\bV_\mathrm{act}\right]_{7{:}N_q}
+ \tM_0^{Q}\cdot\bff_0.
\end{equation}
Equations~(\ref{eq:f_anchor_solve}--\ref{eq:Qdot_clamped}) replace the soft mobility equations for a clamped system. This system has the same number of unknowns as the unconstrained system and uses the same integration method in time (fourth-order Runge--Kutta).  

\section{Continuum origin of the soft mobility equation\label{app:continuum}}

The soft mobility equation is not specific to assemblies of spheres. We sketch here 
its continuum counterpart and show that the discrete system of \S\ref{sec.dynamics} 
is a Galerkin restriction of it.

A hyperelastic body occupies a reference volume $\Omega_0$ and is carried to its 
current configuration $\Omega_t$ by the deformation map $\bchi(\bX,t)$, with deformation 
gradient $\tF = \partial\bchi/\partial\bX$ and Green--Lagrange strain 
$\tE = (\tF^\top\cdot\tF - \tI)/2$. The second Piola--Kirchhoff stress derives from 
a strain energy density, $\tS = \partial W/\partial\tE$, the Saint 
Venant--Kirchhoff model being the simplest choice, reducing to linear elasticity at 
small strain while remaining invariant under large rotations. Body forces 
$\bF_v = \rho\,\tC_H\cdot\bH$ are taken linear in a field $\bH$ uniform at the 
scale of the body. Inertia being 
negligible, the body is in quasi-static equilibrium, whose weak form states that 
for any kinematically admissible virtual velocity $\bV^*$ the internal virtual 
power balances the power of the external forces,
\begin{equation}\label{eq:virtual_power}
\int_{\Omega_0} \tS : \tEd^* \, \rmd V =
	\int_{\Omega_0} \bF_v \cdot \bV^* \, \rmd V +
	\int_{\partial \Omega_0} \bF_s \cdot \bV^* \, \rmd S,
\end{equation}
with $\tEd^* = \sym(\tF^\top \cdot \nabla \bV^*)$ and $\bF_s$ the traction exerted 
by the fluid, referred to the reference configuration.

The virtual velocity $\bV^*$ is a field on $\Omega_0$. 
Being a velocity, it is a spatial vector, identical at the material point $\bX$ in $\Omega_0$ and at its image $\bx = \bchi(\bX,t)$ in $\Omega_t$. The deformation enters through the surface measure and the traction density instead. 
Writing $\bff_s = \tsigma\cdot\bn$ for the Cauchy traction on the deformed surface $\partial\Omega_t$, with $\tsigma$ the fluid stress tensor and $\bn$ the outward normal, the two are related by $\bF_s = J_s\,\bff_s$, where
\begin{equation}\label{eq:surface_Jacobian}
J_s = \|\tF^{-\top}\cdot \bN\|\,\det\tF
\end{equation}
is the surface Jacobian mapping a reference surface element to its deformed image, with $\bN$ the normal in the reference configuration. 
The virtual power of the surface forces can therefore be computed in either 
configuration,
\begin{equation}\label{eq:surface_power}
\int_{\partial \Omega_0} \bF_s \cdot \bV^* \, \rmd S = 
\int_{\partial \Omega_t} \bff_s \cdot \bV^* \, \rmd s.
\end{equation}

\begin{figure}[t]
\begin{center}
    \includegraphics[scale=0.60]{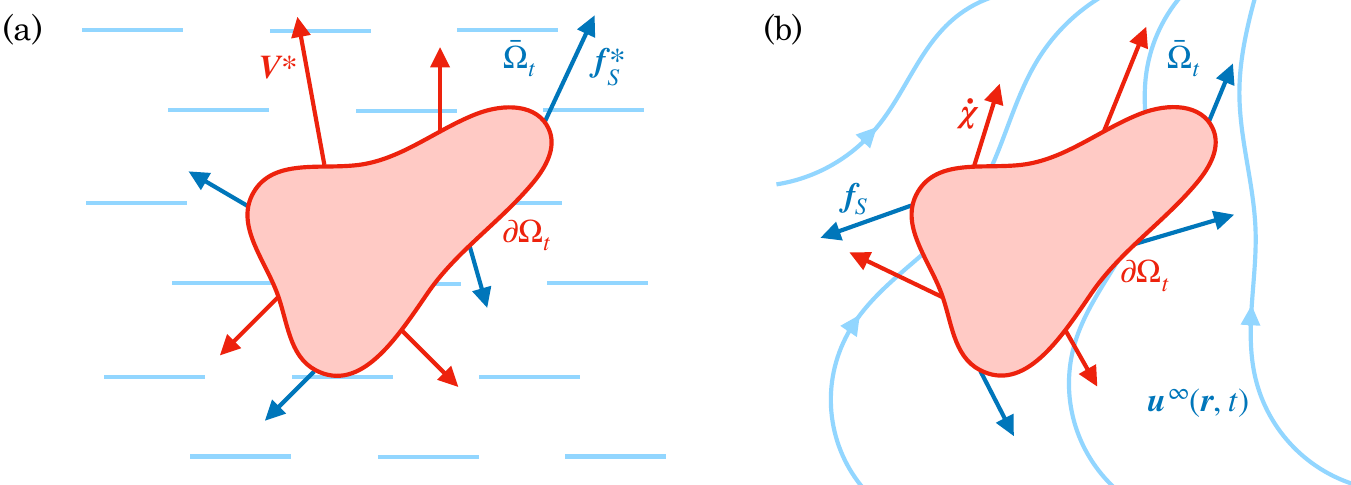}
\caption{Two solutions of the Stokes equations used in the reciprocal theorem: (a) the auxiliary problem, driven by virtual surface velocity $\bV^*$ in an otherwise quiescent fluid, producing traction $\bff_s^*$ on the right-hand side of Eq.~(\ref{eq:reciprocity_deformed}); (b) the physical problem, the body moving with velocity $\bxd$ in background flow $\bu^\infty$, producing traction $\bff_s$ on the left-hand side of Eq.~(\ref{eq:reciprocity_deformed}).}
\label{fig:two_stokes}
\end{center}
\end{figure}

The surface term is the only one coupling the body to the fluid, and the Lorentz reciprocal theorem~\cite{Happel2012} converts it into a form linear in the unknown velocity. 
We compare two Stokes flows sharing the instantaneous geometry $\partial\Omega_t$ (Fig.~\ref{fig:two_stokes}): an auxiliary flow driven by the virtual velocity $\bV^*$ in an otherwise quiescent fluid and producing $\bff_s^*$, and the physical flow, driven by the body moving at $\bxd$ in the background flow $\bu^\infty$ and producing the traction $\bff_s$.
The reciprocal theorem gives
\begin{equation}\label{eq:reciprocity_deformed}
\int_{\partial \Omega_t} \bff_s \cdot \bV^* \, \rmd s =
\int_{\partial \Omega_t} \bff_s^* \cdot \left(\bxd - \bu^\infty\right) \, \rmd s,
\end{equation}
which, transferred back through Eq.~(\ref{eq:surface_power}), is the reference-configuration 
identity
\begin{equation}\label{eq:reciprocity}
\int_{\partial \Omega_0} \bF_s \cdot \bV^* \, \rmd S =
\int_{\partial \Omega_0} \bF^*_s \cdot \left(\bxd - \bu^\infty\right) \, \rmd S,
\end{equation}
with $\bF_s^* = J_s\,\bff_s^*$. The dependence of $\bF_s^*$ on $\bV^*$ is linear, by 
linearity of the Stokes equations, and nonlinear in the current shape. It is the 
continuum counterpart of the grand mobility $\tM_G$, and may be evaluated by any 
Stokes solver.

Substituting Eq.~(\ref{eq:reciprocity}) into 
Eq.~(\ref{eq:virtual_power}) gives the weak form of the problem: one seeks $\bxd$ in 
the space $V$ of kinematically admissible velocity fields such that
\begin{equation}\label{eq:weak_form}
\int_{\partial \Omega_0} \bxd \cdot \bF_s^*\, \rmd S =
	\int_{\partial \Omega_0} \bu^\infty \cdot \bF_s^* \, \rmd S +
	\int_{\Omega_0} \tS : \tEd^* \, \rmd V -
	\int_{\Omega_0} \bF_v \cdot \bV^* \, \rmd V
\end{equation}
for all $\bV^*\in V$. The left-hand side is linear in the unknown $\bxd$ and nonlinear in the shape. 
The three terms on the right are the virtual powers of the background flow, of the elastic stresses, and of the body forces. Two structural properties of the soft mobility equation are already visible here. 
First, rigid-body virtual velocities make the elastic term vanish, since $\tF^\top\cdot\nabla\bV^*$ is then skew-symmetric. 
Second, the same operator acts on the rigid-body part of $\bxd$ on the left and on the 
linearized background flow~(\ref{eq:bg_flow}) on the right, so that only relative 
velocities couple, as Galilean invariance requires.

Using a Galerkin discretization, we restrict $\bxd$ to a finite-dimensional subspace spanned by $n$ velocity fields $\{\bV_j\}$.
Writing $\bxd = \sum_j \dot a_j \bV_j$ and testing Eq.~(\ref{eq:weak_form}) against each $\bV_i$ turns the weak form into an $n\times n$ system,
\begin{equation}\label{eq:galerkin_system}
\hat{\tR} \cdot \bad = \hat{\tK}(\ba) + \hat{\tC}_H\cdot\bH
+ \hat{\tC}_V \cdot \bv_0^\infty + \hat{\tC}_E : \tE_0^\infty,
\qquad
\hat R_{ij} = -\int_{\partial\Omega_0} \bV_j\cdot\bF_{s,i}^*\,\rmd S,
\end{equation}
where $\hat{\tK}$, $\hat{\tC}_H$, $\hat{\tC}_V$ and $\hat{\tC}_E$ are the corresponding
projections of the three right-hand-side terms of Eq.~(\ref{eq:weak_form}), the
background-flow term splitting into a rigid-body part and a rate-of-strain part.
The right-hand side of Eq.~\eqref{eq:galerkin_system} is linear in $\bv_0^\infty$ and $\tE_0^\infty$ because the 
background flow enters Eq.~(\ref{eq:weak_form}) only through its 
linearization~(\ref{eq:bg_flow}). The matrix $\hat{\tR}$ is symmetric positive 
definite by the reciprocal theorem, so that Eq.~(\ref{eq:galerkin_system}) inverts 
to an explicit ordinary differential equation for $\ba$ with the structure of the soft mobility equation~(\ref{eq:soft_mobility}).

For the sphere assembly of \S\ref{sec.dynamics}, the reduction is more direct. 
Neglecting the volume of the springs and rods, the reference
volume is the union of the spheres, $\Omega_0 = \cup_k \Omega^{(k)}$, and the natural basis is the
one generated by the generalized coordinates: the $j$-th generalized coordinate produces the velocity field $\bV_j$, rigid on each sphere and fixed there by the $(k,j)$ six-component block $\tJ_{kj}$ of the kinematic Jacobian. 
Pairing a traction with a field that is rigid on $\partial\Omega^{(k)}$ leaves only its resultants,
\begin{equation}\label{eq:rigid_pairing}
\int_{\partial\Omega^{(k)}} \bV_j\cdot\bF_s\,\rmd S
= \bU_{kj}\cdot\!\!\int_{\partial\Omega^{(k)}}\!\!\bF_s\,\rmd S
+ \bOmega_{kj}\cdot\!\!\int_{\partial\Omega^{(k)}}\!\!(\bX-\bR_k)\times\bF_s\,\rmd S
= \tJ_{kj}\cdot\bff_k,
\end{equation}
with $\tJ_{kj} = [\bU_{kj},\,\bOmega_{kj}]$.
Summed over the spheres, the surface integral of Eq.~(\ref{eq:virtual_power}) is therefore the
generalized hydrodynamic load $\tJ^\top\cdot\bff_\mathrm{hydro}$, and the body-force integral
reduces in the same way to $\tJ^\top\cdot\bff_\mathrm{ext}$. The elastic integral vanishes on the
spheres, where $\tEd^* = \bm{0}$, and is carried entirely by the springs, which contribute
$-\tJ^\top\cdot\bff_K$.
Restricted to this basis, Eq.~(\ref{eq:virtual_power}) is thus
$\tJ^\top\cdot(\bff_\mathrm{hydro}+\bff_\mathrm{ext}+\bff_K)=0$, which is
Eq.~(\ref{eq.generalized_force_Q}) combined with the load balance~(\ref{eq.force_balance}),
precisely the point from which \S\ref{sec.dynamics} starts.
Closing it with the grand mobility relation~(\ref{eq.mobility_problem_all_spheres}) returns the
saddle-point problem~(\ref{eq:saddle_point}), whose Schur complement is
$\hat{\tR} = \tJ^\top\cdot\tM_G^{-1}\cdot\tJ$.

The discrete theory is thus the Galerkin restriction of the continuum problem to
piecewise-rigid velocity fields, exact for that basis, with the approximation entering at one
place only: the exact relation between sphere velocities and sphere loads is replaced by the
Rotne--Prager--Yamakawa mobility, a pairwise far-field expansion between sphere centers.

\bibliography{softmobility.bib}

\begin{thebibliography}{102}%
\makeatletter
\providecommand \@ifxundefined [1]{%
 \@ifx{#1\undefined}
}%
\providecommand \@ifnum [1]{%
 \ifnum #1\expandafter \@firstoftwo
 \else \expandafter \@secondoftwo
 \fi
}%
\providecommand \@ifx [1]{%
 \ifx #1\expandafter \@firstoftwo
 \else \expandafter \@secondoftwo
 \fi
}%
\providecommand \natexlab [1]{#1}%
\providecommand \enquote  [1]{``#1''}%
\providecommand \bibnamefont  [1]{#1}%
\providecommand \bibfnamefont [1]{#1}%
\providecommand \citenamefont [1]{#1}%
\providecommand \href@noop [0]{\@secondoftwo}%
\providecommand \href [0]{\begingroup \@sanitize@url \@href}%
\providecommand \@href[1]{\@@startlink{#1}\@@href}%
\providecommand \@@href[1]{\endgroup#1\@@endlink}%
\providecommand \@sanitize@url [0]{\catcode `\\12\catcode `\$12\catcode
  `\&12\catcode `\#12\catcode `\^12\catcode `\_12\catcode `\%12\relax}%
\providecommand \@@startlink[1]{}%
\providecommand \@@endlink[0]{}%
\providecommand \url  [0]{\begingroup\@sanitize@url \@url }%
\providecommand \@url [1]{\endgroup\@href {#1}{\urlprefix }}%
\providecommand \urlprefix  [0]{URL }%
\providecommand \Eprint [0]{\href }%
\providecommand \doibase [0]{https://doi.org/}%
\providecommand \selectlanguage [0]{\@gobble}%
\providecommand \bibinfo  [0]{\@secondoftwo}%
\providecommand \bibfield  [0]{\@secondoftwo}%
\providecommand \translation [1]{[#1]}%
\providecommand \BibitemOpen [0]{}%
\providecommand \bibitemStop [0]{}%
\providecommand \bibitemNoStop [0]{.\EOS\space}%
\providecommand \EOS [0]{\spacefactor3000\relax}%
\providecommand \BibitemShut  [1]{\csname bibitem#1\endcsname}%
\let\auto@bib@innerbib\@empty
%</preamble>
\bibitem [{\citenamefont {Visser}(2001)}]{Visser2001}%
  \BibitemOpen
  \bibfield  {author} {\bibinfo {author} {\bibfnamefont {A.~W.}\ \bibnamefont
  {Visser}},\ }\bibfield  {title} {\bibinfo {title} {Hydromechanical signals in
  the plankton},\ }\href@noop {} {\bibfield  {journal} {\bibinfo  {journal}
  {Mar. Ecol. Prog. Ser.}\ }\textbf {\bibinfo {volume} {222}},\ \bibinfo
  {pages} {1} (\bibinfo {year} {2001})}\BibitemShut {NoStop}%
\bibitem [{\citenamefont {Wheeler}\ \emph {et~al.}(2019)\citenamefont
  {Wheeler}, \citenamefont {Secchi}, \citenamefont {Rusconi},\ and\
  \citenamefont {Stocker}}]{Wheeler2019}%
  \BibitemOpen
  \bibfield  {author} {\bibinfo {author} {\bibfnamefont {J.~D.}\ \bibnamefont
  {Wheeler}}, \bibinfo {author} {\bibfnamefont {E.}~\bibnamefont {Secchi}},
  \bibinfo {author} {\bibfnamefont {R.}~\bibnamefont {Rusconi}},\ and\ \bibinfo
  {author} {\bibfnamefont {R.}~\bibnamefont {Stocker}},\ }\bibfield  {title}
  {\bibinfo {title} {Not just going with the flow: the effects of fluid flow on
  bacteria and plankton},\ }\href@noop {} {\bibfield  {journal} {\bibinfo
  {journal} {Annu. Rev. Cell Dev. Biol.}\ }\textbf {\bibinfo {volume} {35}},\
  \bibinfo {pages} {213} (\bibinfo {year} {2019})}\BibitemShut {NoStop}%
\bibitem [{\citenamefont {Lauga}\ and\ \citenamefont
  {Powers}(2009)}]{Lauga2009}%
  \BibitemOpen
  \bibfield  {author} {\bibinfo {author} {\bibfnamefont {E.}~\bibnamefont
  {Lauga}}\ and\ \bibinfo {author} {\bibfnamefont {T.~R.}\ \bibnamefont
  {Powers}},\ }\bibfield  {title} {\bibinfo {title} {The hydrodynamics of
  swimming microorganisms},\ }\href@noop {} {\bibfield  {journal} {\bibinfo
  {journal} {Rep. Prog. Phys.}\ }\textbf {\bibinfo {volume} {72}},\ \bibinfo
  {pages} {096601} (\bibinfo {year} {2009})}\BibitemShut {NoStop}%
\bibitem [{\citenamefont {Lauga}(2016)}]{Lauga2016}%
  \BibitemOpen
  \bibfield  {author} {\bibinfo {author} {\bibfnamefont {E.}~\bibnamefont
  {Lauga}},\ }\bibfield  {title} {\bibinfo {title} {Bacterial hydrodynamics},\
  }\href@noop {} {\bibfield  {journal} {\bibinfo  {journal} {Annu. Rev. Fluid
  Mech.}\ }\textbf {\bibinfo {volume} {48}},\ \bibinfo {pages} {105} (\bibinfo
  {year} {2016})}\BibitemShut {NoStop}%
\bibitem [{\citenamefont {Lauga}(2020)}]{Lauga2020}%
  \BibitemOpen
  \bibfield  {author} {\bibinfo {author} {\bibfnamefont {E.}~\bibnamefont
  {Lauga}},\ }\href@noop {} {\emph {\bibinfo {title} {The fluid dynamics of
  cell motility}}},\ Vol.~\bibinfo {volume} {62}\ (\bibinfo  {publisher}
  {Cambridge University Press},\ \bibinfo {year} {2020})\BibitemShut {NoStop}%
\bibitem [{\citenamefont {Guasto}\ \emph {et~al.}(2012)\citenamefont {Guasto},
  \citenamefont {Rusconi},\ and\ \citenamefont {Stocker}}]{Guasto2012}%
  \BibitemOpen
  \bibfield  {author} {\bibinfo {author} {\bibfnamefont {J.~S.}\ \bibnamefont
  {Guasto}}, \bibinfo {author} {\bibfnamefont {R.}~\bibnamefont {Rusconi}},\
  and\ \bibinfo {author} {\bibfnamefont {R.}~\bibnamefont {Stocker}},\
  }\bibfield  {title} {\bibinfo {title} {Fluid mechanics of planktonic
  microorganisms},\ }\href@noop {} {\bibfield  {journal} {\bibinfo  {journal}
  {Annu. Rev. Fluid Mech.}\ }\textbf {\bibinfo {volume} {44}},\ \bibinfo
  {pages} {373} (\bibinfo {year} {2012})}\BibitemShut {NoStop}%
\bibitem [{\citenamefont {Barth\`es-Biesel}(2016)}]{BarthesBiesel2016}%
  \BibitemOpen
  \bibfield  {author} {\bibinfo {author} {\bibfnamefont {D.}~\bibnamefont
  {Barth\`es-Biesel}},\ }\bibfield  {title} {\bibinfo {title} {Motion and
  deformation of elastic capsules and vesicles in flow},\ }\href
  {https://doi.org/10.1146/annurev-fluid-122414-034345} {\bibfield  {journal}
  {\bibinfo  {journal} {Annu. Rev. Fluid Mech.}\ }\textbf {\bibinfo {volume}
  {48}},\ \bibinfo {pages} {25} (\bibinfo {year} {2016})}\BibitemShut {NoStop}%
\bibitem [{\citenamefont {Farutin}\ \emph {et~al.}(2013)\citenamefont
  {Farutin}, \citenamefont {Rafa{\"\i}}, \citenamefont {Dysthe}, \citenamefont
  {Duperray}, \citenamefont {Peyla},\ and\ \citenamefont
  {Misbah}}]{Farutin2013}%
  \BibitemOpen
  \bibfield  {author} {\bibinfo {author} {\bibfnamefont {A.}~\bibnamefont
  {Farutin}}, \bibinfo {author} {\bibfnamefont {S.}~\bibnamefont {Rafa{\"\i}}},
  \bibinfo {author} {\bibfnamefont {D.~K.}\ \bibnamefont {Dysthe}}, \bibinfo
  {author} {\bibfnamefont {A.}~\bibnamefont {Duperray}}, \bibinfo {author}
  {\bibfnamefont {P.}~\bibnamefont {Peyla}},\ and\ \bibinfo {author}
  {\bibfnamefont {C.}~\bibnamefont {Misbah}},\ }\bibfield  {title} {\bibinfo
  {title} {Amoeboid swimming: A generic self-propulsion of cells in fluids by
  means of membrane deformations},\ }\href@noop {} {\bibfield  {journal}
  {\bibinfo  {journal} {Phys. Rev. Lett.}\ }\textbf {\bibinfo {volume} {111}},\
  \bibinfo {pages} {228102} (\bibinfo {year} {2013})}\BibitemShut {NoStop}%
\bibitem [{\citenamefont {Smayda}(1974)}]{Smayda1974}%
  \BibitemOpen
  \bibfield  {author} {\bibinfo {author} {\bibfnamefont {T.~J.}\ \bibnamefont
  {Smayda}},\ }\bibfield  {title} {\bibinfo {title} {Some experiments on the
  sinking characteristics of two freshwater diatoms},\ }\href@noop {}
  {\bibfield  {journal} {\bibinfo  {journal} {Limnol. Oceanogr.}\ }\textbf
  {\bibinfo {volume} {19}},\ \bibinfo {pages} {628} (\bibinfo {year}
  {1974})}\BibitemShut {NoStop}%
\bibitem [{\citenamefont {Wiseman}\ and\ \citenamefont
  {Reynolds}(1981)}]{Wiseman1981}%
  \BibitemOpen
  \bibfield  {author} {\bibinfo {author} {\bibfnamefont {S.~W.}\ \bibnamefont
  {Wiseman}}\ and\ \bibinfo {author} {\bibfnamefont {C.~S.}\ \bibnamefont
  {Reynolds}},\ }\bibfield  {title} {\bibinfo {title} {Sinking rate and
  electrophoretic mobility of the freshwater diatom asterionella formosa: an
  experimental investigation},\ }\href@noop {} {\bibfield  {journal} {\bibinfo
  {journal} {Br. Phycol. J.}\ }\textbf {\bibinfo {volume} {16}},\ \bibinfo
  {pages} {357} (\bibinfo {year} {1981})}\BibitemShut {NoStop}%
\bibitem [{\citenamefont {Padis{\'a}k}\ \emph {et~al.}(2003)\citenamefont
  {Padis{\'a}k}, \citenamefont {Sor{\'o}czki-Pint{\'e}r},\ and\ \citenamefont
  {Rezner}}]{Padisak2003}%
  \BibitemOpen
  \bibfield  {author} {\bibinfo {author} {\bibfnamefont {J.}~\bibnamefont
  {Padis{\'a}k}}, \bibinfo {author} {\bibfnamefont {{\'E}.}~\bibnamefont
  {Sor{\'o}czki-Pint{\'e}r}},\ and\ \bibinfo {author} {\bibfnamefont
  {Z.}~\bibnamefont {Rezner}},\ }\bibfield  {title} {\bibinfo {title} {Sinking
  properties of some phytoplankton shapes and the relation of form resistance
  to morphological diversity of plankton---an experimental study},\ }in\
  \href@noop {} {\emph {\bibinfo {booktitle} {Aquatic Biodiversity: A
  Celebratory Volume in Honour of {H}enri {J}. {D}umont}}}\ (\bibinfo
  {publisher} {Springer},\ \bibinfo {year} {2003})\ pp.\ \bibinfo {pages}
  {243--257}\BibitemShut {NoStop}%
\bibitem [{\citenamefont {Wiggins}\ \emph {et~al.}(1998)\citenamefont
  {Wiggins}, \citenamefont {Riveline}, \citenamefont {Ott},\ and\ \citenamefont
  {Goldstein}}]{Wiggins1998}%
  \BibitemOpen
  \bibfield  {author} {\bibinfo {author} {\bibfnamefont {C.~H.}\ \bibnamefont
  {Wiggins}}, \bibinfo {author} {\bibfnamefont {D.}~\bibnamefont {Riveline}},
  \bibinfo {author} {\bibfnamefont {A.}~\bibnamefont {Ott}},\ and\ \bibinfo
  {author} {\bibfnamefont {R.~E.}\ \bibnamefont {Goldstein}},\ }\bibfield
  {title} {\bibinfo {title} {Trapping and wiggling: elastohydrodynamics of
  driven microfilaments},\ }\href@noop {} {\bibfield  {journal} {\bibinfo
  {journal} {Biophys. J.}\ }\textbf {\bibinfo {volume} {74}},\ \bibinfo {pages}
  {1043} (\bibinfo {year} {1998})}\BibitemShut {NoStop}%
\bibitem [{\citenamefont {Brouzet}\ \emph {et~al.}(2014)\citenamefont
  {Brouzet}, \citenamefont {Verhille},\ and\ \citenamefont
  {Le~Gal}}]{Brouzet2014}%
  \BibitemOpen
  \bibfield  {author} {\bibinfo {author} {\bibfnamefont {C.}~\bibnamefont
  {Brouzet}}, \bibinfo {author} {\bibfnamefont {G.}~\bibnamefont {Verhille}},\
  and\ \bibinfo {author} {\bibfnamefont {P.}~\bibnamefont {Le~Gal}},\
  }\bibfield  {title} {\bibinfo {title} {Flexible fiber in a turbulent flow: A
  macroscopic polymer},\ }\href@noop {} {\bibfield  {journal} {\bibinfo
  {journal} {Phys. Rev. Lett.}\ }\textbf {\bibinfo {volume} {112}},\ \bibinfo
  {pages} {074501} (\bibinfo {year} {2014})}\BibitemShut {NoStop}%
\bibitem [{\citenamefont {Du~Roure}\ \emph {et~al.}(2019)\citenamefont
  {Du~Roure}, \citenamefont {Lindner}, \citenamefont {Nazockdast},\ and\
  \citenamefont {Shelley}}]{Du-Roure2019}%
  \BibitemOpen
  \bibfield  {author} {\bibinfo {author} {\bibfnamefont {O.}~\bibnamefont
  {Du~Roure}}, \bibinfo {author} {\bibfnamefont {A.}~\bibnamefont {Lindner}},
  \bibinfo {author} {\bibfnamefont {E.~N.}\ \bibnamefont {Nazockdast}},\ and\
  \bibinfo {author} {\bibfnamefont {M.~J.}\ \bibnamefont {Shelley}},\
  }\bibfield  {title} {\bibinfo {title} {Dynamics of flexible fibers in viscous
  flows and fluids},\ }\href@noop {} {\bibfield  {journal} {\bibinfo  {journal}
  {Annu. Rev. Fluid Mech.}\ }\textbf {\bibinfo {volume} {51}},\ \bibinfo
  {pages} {539} (\bibinfo {year} {2019})}\BibitemShut {NoStop}%
\bibitem [{\citenamefont {{\.Z}uk}\ \emph {et~al.}(2021)\citenamefont
  {{\.Z}uk}, \citenamefont {S{\l}owicka}, \citenamefont {Ekiel-Je{\.z}ewska},\
  and\ \citenamefont {Stone}}]{Zuk2021}%
  \BibitemOpen
  \bibfield  {author} {\bibinfo {author} {\bibfnamefont {P.~J.}\ \bibnamefont
  {{\.Z}uk}}, \bibinfo {author} {\bibfnamefont {A.~M.}\ \bibnamefont
  {S{\l}owicka}}, \bibinfo {author} {\bibfnamefont {M.~L.}\ \bibnamefont
  {Ekiel-Je{\.z}ewska}},\ and\ \bibinfo {author} {\bibfnamefont {H.~A.}\
  \bibnamefont {Stone}},\ }\bibfield  {title} {\bibinfo {title} {Universal
  features of the shape of elastic fibres in shear flow},\ }\href@noop {}
  {\bibfield  {journal} {\bibinfo  {journal} {J. Fluid Mech.}\ }\textbf
  {\bibinfo {volume} {914}},\ \bibinfo {pages} {A31} (\bibinfo {year}
  {2021})}\BibitemShut {NoStop}%
\bibitem [{\citenamefont {Sitti}(2021)}]{Sitti2021}%
  \BibitemOpen
  \bibfield  {author} {\bibinfo {author} {\bibfnamefont {M.}~\bibnamefont
  {Sitti}},\ }\bibfield  {title} {\bibinfo {title} {Physical intelligence as a
  new paradigm},\ }\href@noop {} {\bibfield  {journal} {\bibinfo  {journal}
  {Extreme Mech. Lett.}\ }\textbf {\bibinfo {volume} {46}},\ \bibinfo {pages}
  {101340} (\bibinfo {year} {2021})}\BibitemShut {NoStop}%
\bibitem [{\citenamefont {Yasa}\ \emph {et~al.}(2023)\citenamefont {Yasa},
  \citenamefont {Toshimitsu}, \citenamefont {Michelis}, \citenamefont {Jones},
  \citenamefont {Filippi}, \citenamefont {Buchner},\ and\ \citenamefont
  {Katzschmann}}]{Yasa2023}%
  \BibitemOpen
  \bibfield  {author} {\bibinfo {author} {\bibfnamefont {O.}~\bibnamefont
  {Yasa}}, \bibinfo {author} {\bibfnamefont {Y.}~\bibnamefont {Toshimitsu}},
  \bibinfo {author} {\bibfnamefont {M.~Y.}\ \bibnamefont {Michelis}}, \bibinfo
  {author} {\bibfnamefont {L.~S.}\ \bibnamefont {Jones}}, \bibinfo {author}
  {\bibfnamefont {M.}~\bibnamefont {Filippi}}, \bibinfo {author} {\bibfnamefont
  {T.}~\bibnamefont {Buchner}},\ and\ \bibinfo {author} {\bibfnamefont {R.~K.}\
  \bibnamefont {Katzschmann}},\ }\bibfield  {title} {\bibinfo {title} {An
  overview of soft robotics},\ }\href@noop {} {\bibfield  {journal} {\bibinfo
  {journal} {Annu. Rev. Control Robot. Auton. Syst.}\ }\textbf {\bibinfo
  {volume} {6}},\ \bibinfo {pages} {1} (\bibinfo {year} {2023})}\BibitemShut
  {NoStop}%
\bibitem [{\citenamefont {Tsang}\ \emph {et~al.}(2020)\citenamefont {Tsang},
  \citenamefont {Demir}, \citenamefont {Ding},\ and\ \citenamefont
  {Pak}}]{Tsang2020b}%
  \BibitemOpen
  \bibfield  {author} {\bibinfo {author} {\bibfnamefont {A.~C.}\ \bibnamefont
  {Tsang}}, \bibinfo {author} {\bibfnamefont {E.}~\bibnamefont {Demir}},
  \bibinfo {author} {\bibfnamefont {Y.}~\bibnamefont {Ding}},\ and\ \bibinfo
  {author} {\bibfnamefont {O.~S.}\ \bibnamefont {Pak}},\ }\bibfield  {title}
  {\bibinfo {title} {Roads to smart artificial microswimmers},\ }\href@noop {}
  {\bibfield  {journal} {\bibinfo  {journal} {Adv. Intell. Syst.}\ }\textbf
  {\bibinfo {volume} {2}},\ \bibinfo {pages} {1900137} (\bibinfo {year}
  {2020})}\BibitemShut {NoStop}%
\bibitem [{\citenamefont {Elnaggar}\ \emph {et~al.}(2024)\citenamefont
  {Elnaggar}, \citenamefont {Kang}, \citenamefont {Tian}, \citenamefont {Han},\
  and\ \citenamefont {Keshavarz}}]{Elnaggar2024}%
  \BibitemOpen
  \bibfield  {author} {\bibinfo {author} {\bibfnamefont {A.}~\bibnamefont
  {Elnaggar}}, \bibinfo {author} {\bibfnamefont {S.}~\bibnamefont {Kang}},
  \bibinfo {author} {\bibfnamefont {M.}~\bibnamefont {Tian}}, \bibinfo {author}
  {\bibfnamefont {B.}~\bibnamefont {Han}},\ and\ \bibinfo {author}
  {\bibfnamefont {M.}~\bibnamefont {Keshavarz}},\ }\bibfield  {title} {\bibinfo
  {title} {State of the art in actuation of micro/nanorobots for biomedical
  applications},\ }\href@noop {} {\bibfield  {journal} {\bibinfo  {journal}
  {Small Sci.}\ }\textbf {\bibinfo {volume} {4}},\ \bibinfo {pages} {2300211}
  (\bibinfo {year} {2024})}\BibitemShut {NoStop}%
\bibitem [{\citenamefont {Dabbagh}\ \emph {et~al.}(2022)\citenamefont
  {Dabbagh}, \citenamefont {Sarabi}, \citenamefont {Birtek}, \citenamefont
  {Seyfi}, \citenamefont {Sitti},\ and\ \citenamefont {Tasoglu}}]{Dabbagh2022}%
  \BibitemOpen
  \bibfield  {author} {\bibinfo {author} {\bibfnamefont {S.~R.}\ \bibnamefont
  {Dabbagh}}, \bibinfo {author} {\bibfnamefont {M.~R.}\ \bibnamefont {Sarabi}},
  \bibinfo {author} {\bibfnamefont {M.~T.}\ \bibnamefont {Birtek}}, \bibinfo
  {author} {\bibfnamefont {S.}~\bibnamefont {Seyfi}}, \bibinfo {author}
  {\bibfnamefont {M.}~\bibnamefont {Sitti}},\ and\ \bibinfo {author}
  {\bibfnamefont {S.}~\bibnamefont {Tasoglu}},\ }\bibfield  {title} {\bibinfo
  {title} {3d-printed microrobots from design to translation},\ }\href@noop {}
  {\bibfield  {journal} {\bibinfo  {journal} {Nat. Commun.}\ }\textbf {\bibinfo
  {volume} {13}},\ \bibinfo {pages} {5875} (\bibinfo {year}
  {2022})}\BibitemShut {NoStop}%
\bibitem [{\citenamefont {Mo}\ \emph {et~al.}(2023)\citenamefont {Mo},
  \citenamefont {Li},\ and\ \citenamefont {Bian}}]{Mo2023}%
  \BibitemOpen
  \bibfield  {author} {\bibinfo {author} {\bibfnamefont {C.}~\bibnamefont
  {Mo}}, \bibinfo {author} {\bibfnamefont {G.}~\bibnamefont {Li}},\ and\
  \bibinfo {author} {\bibfnamefont {X.}~\bibnamefont {Bian}},\ }\bibfield
  {title} {\bibinfo {title} {Challenges and attempts to make intelligent
  microswimmers},\ }\href@noop {} {\bibfield  {journal} {\bibinfo  {journal}
  {Front. Phys.}\ }\textbf {\bibinfo {volume} {11}},\ \bibinfo {pages}
  {1279883} (\bibinfo {year} {2023})}\BibitemShut {NoStop}%
\bibitem [{\citenamefont {Jiang}\ \emph {et~al.}(2022)\citenamefont {Jiang},
  \citenamefont {Yang}, \citenamefont {Ferreira},\ and\ \citenamefont
  {Zhang}}]{Jiang2022}%
  \BibitemOpen
  \bibfield  {author} {\bibinfo {author} {\bibfnamefont {J.}~\bibnamefont
  {Jiang}}, \bibinfo {author} {\bibfnamefont {Z.}~\bibnamefont {Yang}},
  \bibinfo {author} {\bibfnamefont {A.}~\bibnamefont {Ferreira}},\ and\
  \bibinfo {author} {\bibfnamefont {L.}~\bibnamefont {Zhang}},\ }\bibfield
  {title} {\bibinfo {title} {Control and autonomy of microrobots: Recent
  progress and perspective},\ }\href@noop {} {\bibfield  {journal} {\bibinfo
  {journal} {Adv. Intell. Syst.}\ }\textbf {\bibinfo {volume} {4}},\ \bibinfo
  {pages} {2100279} (\bibinfo {year} {2022})}\BibitemShut {NoStop}%
\bibitem [{\citenamefont {Sutherland}\ \emph {et~al.}(2023)\citenamefont
  {Sutherland}, \citenamefont {DiBenedetto}, \citenamefont {Kaminski},\ and\
  \citenamefont {Van Den~Bremer}}]{Sutherland2023}%
  \BibitemOpen
  \bibfield  {author} {\bibinfo {author} {\bibfnamefont {B.~R.}\ \bibnamefont
  {Sutherland}}, \bibinfo {author} {\bibfnamefont {M.}~\bibnamefont
  {DiBenedetto}}, \bibinfo {author} {\bibfnamefont {A.}~\bibnamefont
  {Kaminski}},\ and\ \bibinfo {author} {\bibfnamefont {T.}~\bibnamefont {Van
  Den~Bremer}},\ }\bibfield  {title} {\bibinfo {title} {Fluid dynamics
  challenges in predicting plastic pollution transport in the ocean: A
  perspective},\ }\href@noop {} {\bibfield  {journal} {\bibinfo  {journal}
  {Phys. Rev. Fluids}\ }\textbf {\bibinfo {volume} {8}},\ \bibinfo {pages}
  {070701} (\bibinfo {year} {2023})}\BibitemShut {NoStop}%
\bibitem [{\citenamefont {Brenner}(1963)}]{Brenner1963}%
  \BibitemOpen
  \bibfield  {author} {\bibinfo {author} {\bibfnamefont {H.}~\bibnamefont
  {Brenner}},\ }\bibfield  {title} {\bibinfo {title} {The stokes resistance of
  an arbitrary particle},\ }\href@noop {} {\bibfield  {journal} {\bibinfo
  {journal} {Chem. Eng. Sci.}\ }\textbf {\bibinfo {volume} {18}},\ \bibinfo
  {pages} {1} (\bibinfo {year} {1963})}\BibitemShut {NoStop}%
\bibitem [{\citenamefont {Happel}\ and\ \citenamefont
  {Brenner}(1983)}]{Happel2012}%
  \BibitemOpen
  \bibfield  {author} {\bibinfo {author} {\bibfnamefont {J.}~\bibnamefont
  {Happel}}\ and\ \bibinfo {author} {\bibfnamefont {H.}~\bibnamefont
  {Brenner}},\ }\href@noop {} {\emph {\bibinfo {title} {Low Reynolds number
  hydrodynamics: with special applications to particulate media}}},\ \bibinfo
  {edition} {first paperback edition}\ ed.\ (\bibinfo  {publisher} {Martinus
  Nijhoff Publishers, The Hague},\ \bibinfo {year} {1983})\BibitemShut
  {NoStop}%
\bibitem [{\citenamefont {Kim}\ and\ \citenamefont {Karrila}(2013)}]{Kim2013}%
  \BibitemOpen
  \bibfield  {author} {\bibinfo {author} {\bibfnamefont {S.}~\bibnamefont
  {Kim}}\ and\ \bibinfo {author} {\bibfnamefont {S.~J.}\ \bibnamefont
  {Karrila}},\ }\href@noop {} {\emph {\bibinfo {title} {Microhydrodynamics:
  principles and selected applications}}}\ (\bibinfo  {publisher}
  {Butterworth-Heinemann},\ \bibinfo {year} {2013})\BibitemShut {NoStop}%
\bibitem [{\citenamefont {Kim}\ \emph {et~al.}(2013)\citenamefont {Kim},
  \citenamefont {Laschi},\ and\ \citenamefont {Trimmer}}]{Kim2013a}%
  \BibitemOpen
  \bibfield  {author} {\bibinfo {author} {\bibfnamefont {S.}~\bibnamefont
  {Kim}}, \bibinfo {author} {\bibfnamefont {C.}~\bibnamefont {Laschi}},\ and\
  \bibinfo {author} {\bibfnamefont {B.}~\bibnamefont {Trimmer}},\ }\bibfield
  {title} {\bibinfo {title} {Soft robotics: a bioinspired evolution in
  robotics},\ }\href@noop {} {\bibfield  {journal} {\bibinfo  {journal} {Trends
  Biotechnol.}\ }\textbf {\bibinfo {volume} {31}},\ \bibinfo {pages} {287}
  (\bibinfo {year} {2013})}\BibitemShut {NoStop}%
\bibitem [{\citenamefont {Pfeifer}\ and\ \citenamefont
  {Bongard}(2006)}]{Pfeifer2006}%
  \BibitemOpen
  \bibfield  {author} {\bibinfo {author} {\bibfnamefont {R.}~\bibnamefont
  {Pfeifer}}\ and\ \bibinfo {author} {\bibfnamefont {J.}~\bibnamefont
  {Bongard}},\ }\href@noop {} {\emph {\bibinfo {title} {How the body shapes the
  way we think: a new view of intelligence}}}\ (\bibinfo  {publisher} {MIT
  press},\ \bibinfo {year} {2006})\BibitemShut {NoStop}%
\bibitem [{\citenamefont {Pfeifer}\ \emph {et~al.}(2014)\citenamefont
  {Pfeifer}, \citenamefont {Iida},\ and\ \citenamefont
  {Lungarella}}]{Pfeifer2014}%
  \BibitemOpen
  \bibfield  {author} {\bibinfo {author} {\bibfnamefont {R.}~\bibnamefont
  {Pfeifer}}, \bibinfo {author} {\bibfnamefont {F.}~\bibnamefont {Iida}},\ and\
  \bibinfo {author} {\bibfnamefont {M.}~\bibnamefont {Lungarella}},\ }\bibfield
   {title} {\bibinfo {title} {Cognition from the bottom up: on biological
  inspiration, body morphology, and soft materials},\ }\href@noop {} {\bibfield
   {journal} {\bibinfo  {journal} {Trends Cogn. Sci.}\ }\textbf {\bibinfo
  {volume} {18}},\ \bibinfo {pages} {404} (\bibinfo {year} {2014})}\BibitemShut
  {NoStop}%
\bibitem [{\citenamefont {Hauser}\ \emph {et~al.}(2011)\citenamefont {Hauser},
  \citenamefont {Ijspeert}, \citenamefont {F{\"u}chslin}, \citenamefont
  {Pfeifer},\ and\ \citenamefont {Maass}}]{Hauser2011}%
  \BibitemOpen
  \bibfield  {author} {\bibinfo {author} {\bibfnamefont {H.}~\bibnamefont
  {Hauser}}, \bibinfo {author} {\bibfnamefont {A.~J.}\ \bibnamefont
  {Ijspeert}}, \bibinfo {author} {\bibfnamefont {R.~M.}\ \bibnamefont
  {F{\"u}chslin}}, \bibinfo {author} {\bibfnamefont {R.}~\bibnamefont
  {Pfeifer}},\ and\ \bibinfo {author} {\bibfnamefont {W.}~\bibnamefont
  {Maass}},\ }\bibfield  {title} {\bibinfo {title} {Towards a theoretical
  foundation for morphological computation with compliant bodies},\ }\href@noop
  {} {\bibfield  {journal} {\bibinfo  {journal} {Biol. Cybern.}\ }\textbf
  {\bibinfo {volume} {105}},\ \bibinfo {pages} {355} (\bibinfo {year}
  {2011})}\BibitemShut {NoStop}%
\bibitem [{\citenamefont {Hauser}\ \emph {et~al.}(2014)\citenamefont {Hauser},
  \citenamefont {F{\"u}chslin},\ and\ \citenamefont {Pfeifer}}]{Hauser2014}%
  \BibitemOpen
  \bibinfo {editor} {\bibfnamefont {H.}~\bibnamefont {Hauser}}, \bibinfo
  {editor} {\bibfnamefont {R.~M.}\ \bibnamefont {F{\"u}chslin}},\ and\ \bibinfo
  {editor} {\bibfnamefont {R.}~\bibnamefont {Pfeifer}},\ eds.,\ \href@noop {}
  {\emph {\bibinfo {title} {Opinions and outlooks on morphological
  computation}}}\ (\bibinfo {year} {2014})\BibitemShut {NoStop}%
\bibitem [{\citenamefont {Nakajima}\ \emph {et~al.}(2015)\citenamefont
  {Nakajima}, \citenamefont {Hauser}, \citenamefont {Li},\ and\ \citenamefont
  {Pfeifer}}]{Nakajima2015}%
  \BibitemOpen
  \bibfield  {author} {\bibinfo {author} {\bibfnamefont {K.}~\bibnamefont
  {Nakajima}}, \bibinfo {author} {\bibfnamefont {H.}~\bibnamefont {Hauser}},
  \bibinfo {author} {\bibfnamefont {T.}~\bibnamefont {Li}},\ and\ \bibinfo
  {author} {\bibfnamefont {R.}~\bibnamefont {Pfeifer}},\ }\bibfield  {title}
  {\bibinfo {title} {Information processing via physical soft body},\
  }\href@noop {} {\bibfield  {journal} {\bibinfo  {journal} {Sci. Rep.}\
  }\textbf {\bibinfo {volume} {5}},\ \bibinfo {pages} {10487} (\bibinfo {year}
  {2015})}\BibitemShut {NoStop}%
\bibitem [{\citenamefont {Hauser}\ \emph {et~al.}(2023)\citenamefont {Hauser},
  \citenamefont {Nanayakkara},\ and\ \citenamefont {Forni}}]{Hauser2023}%
  \BibitemOpen
  \bibfield  {author} {\bibinfo {author} {\bibfnamefont {H.}~\bibnamefont
  {Hauser}}, \bibinfo {author} {\bibfnamefont {T.}~\bibnamefont
  {Nanayakkara}},\ and\ \bibinfo {author} {\bibfnamefont {F.}~\bibnamefont
  {Forni}},\ }\bibfield  {title} {\bibinfo {title} {Leveraging morphological
  computation for controlling soft robots: Learning from nature to control soft
  robots},\ }\href@noop {} {\bibfield  {journal} {\bibinfo  {journal} {IEEE
  Control Syst. Mag.}\ }\textbf {\bibinfo {volume} {43}},\ \bibinfo {pages}
  {114} (\bibinfo {year} {2023})}\BibitemShut {NoStop}%
\bibitem [{\citenamefont {Pozrikidis}(1992)}]{Pozrikidis1992}%
  \BibitemOpen
  \bibfield  {author} {\bibinfo {author} {\bibfnamefont {C.}~\bibnamefont
  {Pozrikidis}},\ }\href@noop {} {\emph {\bibinfo {title} {Boundary integral
  and singularity methods for linearized viscous flow}}},\ Cambridge Texts in
  Applied Mathematics\ (\bibinfo  {publisher} {Cambridge University Press},\
  \bibinfo {year} {1992})\BibitemShut {NoStop}%
\bibitem [{\citenamefont {Balboa~Usabiaga}\ \emph {et~al.}(2017)\citenamefont
  {Balboa~Usabiaga}, \citenamefont {Kallemov}, \citenamefont {Delmotte},
  \citenamefont {Bhalla}, \citenamefont {Griffith},\ and\ \citenamefont
  {Donev}}]{Balboa-Usabiaga2017}%
  \BibitemOpen
  \bibfield  {author} {\bibinfo {author} {\bibfnamefont {F.}~\bibnamefont
  {Balboa~Usabiaga}}, \bibinfo {author} {\bibfnamefont {B.}~\bibnamefont
  {Kallemov}}, \bibinfo {author} {\bibfnamefont {B.}~\bibnamefont {Delmotte}},
  \bibinfo {author} {\bibfnamefont {A.}~\bibnamefont {Bhalla}}, \bibinfo
  {author} {\bibfnamefont {B.}~\bibnamefont {Griffith}},\ and\ \bibinfo
  {author} {\bibfnamefont {A.}~\bibnamefont {Donev}},\ }\bibfield  {title}
  {\bibinfo {title} {Hydrodynamics of suspensions of passive and active rigid
  particles: a rigid multiblob approach},\ }\href@noop {} {\bibfield  {journal}
  {\bibinfo  {journal} {Commun. Appl. Math. Comput. Sci.}\ }\textbf {\bibinfo
  {volume} {11}},\ \bibinfo {pages} {217} (\bibinfo {year} {2017})}\BibitemShut
  {NoStop}%
\bibitem [{\citenamefont {Balboa~Usabiaga}\ and\ \citenamefont
  {Delmotte}(2022)}]{Balboa-Usabiaga2022}%
  \BibitemOpen
  \bibfield  {author} {\bibinfo {author} {\bibfnamefont {F.}~\bibnamefont
  {Balboa~Usabiaga}}\ and\ \bibinfo {author} {\bibfnamefont {B.}~\bibnamefont
  {Delmotte}},\ }\bibfield  {title} {\bibinfo {title} {A numerical method for
  suspensions of articulated bodies in viscous flows},\ }\href@noop {}
  {\bibfield  {journal} {\bibinfo  {journal} {J. Comput. Phys.}\ }\textbf
  {\bibinfo {volume} {464}},\ \bibinfo {pages} {111365} (\bibinfo {year}
  {2022})}\BibitemShut {NoStop}%
\bibitem [{\citenamefont {Broms}\ \emph {et~al.}(2023)\citenamefont {Broms},
  \citenamefont {Sandberg},\ and\ \citenamefont {Tornberg}}]{Broms2023}%
  \BibitemOpen
  \bibfield  {author} {\bibinfo {author} {\bibfnamefont {A.}~\bibnamefont
  {Broms}}, \bibinfo {author} {\bibfnamefont {M.}~\bibnamefont {Sandberg}},\
  and\ \bibinfo {author} {\bibfnamefont {A.-K.}\ \bibnamefont {Tornberg}},\
  }\bibfield  {title} {\bibinfo {title} {A locally corrected multiblob method
  with hydrodynamically matched grids for the stokes mobility problem},\
  }\href@noop {} {\bibfield  {journal} {\bibinfo  {journal} {J. Comput. Phys.}\
  }\textbf {\bibinfo {volume} {487}},\ \bibinfo {pages} {112172} (\bibinfo
  {year} {2023})}\BibitemShut {NoStop}%
\bibitem [{\citenamefont {Tornberg}\ and\ \citenamefont
  {Shelley}(2004)}]{Tornberg2004}%
  \BibitemOpen
  \bibfield  {author} {\bibinfo {author} {\bibfnamefont {A.-K.}\ \bibnamefont
  {Tornberg}}\ and\ \bibinfo {author} {\bibfnamefont {M.~J.}\ \bibnamefont
  {Shelley}},\ }\bibfield  {title} {\bibinfo {title} {Simulating the dynamics
  and interactions of flexible fibers in stokes flows},\ }\href@noop {}
  {\bibfield  {journal} {\bibinfo  {journal} {J. Comput. Phys.}\ }\textbf
  {\bibinfo {volume} {196}},\ \bibinfo {pages} {8} (\bibinfo {year}
  {2004})}\BibitemShut {NoStop}%
\bibitem [{\citenamefont {Nazockdast}\ \emph {et~al.}(2017)\citenamefont
  {Nazockdast}, \citenamefont {Rahimian}, \citenamefont {Zorin},\ and\
  \citenamefont {Shelley}}]{Nazockdast2017}%
  \BibitemOpen
  \bibfield  {author} {\bibinfo {author} {\bibfnamefont {E.}~\bibnamefont
  {Nazockdast}}, \bibinfo {author} {\bibfnamefont {A.}~\bibnamefont
  {Rahimian}}, \bibinfo {author} {\bibfnamefont {D.}~\bibnamefont {Zorin}},\
  and\ \bibinfo {author} {\bibfnamefont {M.}~\bibnamefont {Shelley}},\
  }\bibfield  {title} {\bibinfo {title} {A fast platform for simulating
  semi-flexible fiber suspensions applied to cell mechanics},\ }\href@noop {}
  {\bibfield  {journal} {\bibinfo  {journal} {J. Comput. Phys.}\ }\textbf
  {\bibinfo {volume} {329}},\ \bibinfo {pages} {173} (\bibinfo {year}
  {2017})}\BibitemShut {NoStop}%
\bibitem [{\citenamefont {Maxian}\ \emph {et~al.}(2022)\citenamefont {Maxian},
  \citenamefont {Sprinkle}, \citenamefont {Peskin},\ and\ \citenamefont
  {Donev}}]{Maxian2022}%
  \BibitemOpen
  \bibfield  {author} {\bibinfo {author} {\bibfnamefont {O.}~\bibnamefont
  {Maxian}}, \bibinfo {author} {\bibfnamefont {B.}~\bibnamefont {Sprinkle}},
  \bibinfo {author} {\bibfnamefont {C.~S.}\ \bibnamefont {Peskin}},\ and\
  \bibinfo {author} {\bibfnamefont {A.}~\bibnamefont {Donev}},\ }\bibfield
  {title} {\bibinfo {title} {Hydrodynamics of a twisting, bending, inextensible
  fiber in stokes flow},\ }\href@noop {} {\bibfield  {journal} {\bibinfo
  {journal} {Phys. Rev. Fluids}\ }\textbf {\bibinfo {volume} {7}},\ \bibinfo
  {pages} {074101} (\bibinfo {year} {2022})}\BibitemShut {NoStop}%
\bibitem [{\citenamefont {Garc{\'\i}a de~la Torre}\ and\ \citenamefont
  {Bloomfield}(1981)}]{Garcia-de-la-Torre1981}%
  \BibitemOpen
  \bibfield  {author} {\bibinfo {author} {\bibfnamefont {J.}~\bibnamefont
  {Garc{\'\i}a de~la Torre}}\ and\ \bibinfo {author} {\bibfnamefont {V.~A.}\
  \bibnamefont {Bloomfield}},\ }\bibfield  {title} {\bibinfo {title}
  {Hydrodynamic properties of complex, rigid, biological macromolecules: theory
  and applications},\ }\href@noop {} {\bibfield  {journal} {\bibinfo  {journal}
  {Q. Rev. Biophys.}\ }\textbf {\bibinfo {volume} {14}},\ \bibinfo {pages} {81}
  (\bibinfo {year} {1981})}\BibitemShut {NoStop}%
\bibitem [{\citenamefont {Zuk}\ \emph {et~al.}(2014)\citenamefont {Zuk},
  \citenamefont {Wajnryb}, \citenamefont {Mizerski},\ and\ \citenamefont
  {Szymczak}}]{Zuk2014}%
  \BibitemOpen
  \bibfield  {author} {\bibinfo {author} {\bibfnamefont {P.~J.}\ \bibnamefont
  {Zuk}}, \bibinfo {author} {\bibfnamefont {E.}~\bibnamefont {Wajnryb}},
  \bibinfo {author} {\bibfnamefont {K.}~\bibnamefont {Mizerski}},\ and\
  \bibinfo {author} {\bibfnamefont {P.}~\bibnamefont {Szymczak}},\ }\bibfield
  {title} {\bibinfo {title} {{R}otne--{P}rager--{Y}amakawa approximation for
  different-sized particles in application to macromolecular bead models},\
  }\href@noop {} {\bibfield  {journal} {\bibinfo  {journal} {J. Fluid Mech.}\
  }\textbf {\bibinfo {volume} {741}},\ \bibinfo {pages} {R5} (\bibinfo {year}
  {2014})}\BibitemShut {NoStop}%
\bibitem [{\citenamefont {Zuk}\ \emph {et~al.}(2018)\citenamefont {Zuk},
  \citenamefont {Cichocki},\ and\ \citenamefont {Szymczak}}]{Zuk2018}%
  \BibitemOpen
  \bibfield  {author} {\bibinfo {author} {\bibfnamefont {P.~J.}\ \bibnamefont
  {Zuk}}, \bibinfo {author} {\bibfnamefont {B.}~\bibnamefont {Cichocki}},\ and\
  \bibinfo {author} {\bibfnamefont {P.}~\bibnamefont {Szymczak}},\ }\bibfield
  {title} {\bibinfo {title} {Grpy: an accurate bead method for calculation of
  hydrodynamic properties of rigid biomacromolecules},\ }\href@noop {}
  {\bibfield  {journal} {\bibinfo  {journal} {Biophys. J.}\ }\textbf {\bibinfo
  {volume} {115}},\ \bibinfo {pages} {782} (\bibinfo {year}
  {2018})}\BibitemShut {NoStop}%
\bibitem [{\citenamefont {Gonzalez}\ \emph {et~al.}(2004)\citenamefont
  {Gonzalez}, \citenamefont {Graf},\ and\ \citenamefont
  {Maddocks}}]{Gonzalez2004}%
  \BibitemOpen
  \bibfield  {author} {\bibinfo {author} {\bibfnamefont {O.}~\bibnamefont
  {Gonzalez}}, \bibinfo {author} {\bibfnamefont {A.}~\bibnamefont {Graf}},\
  and\ \bibinfo {author} {\bibfnamefont {J.}~\bibnamefont {Maddocks}},\
  }\bibfield  {title} {\bibinfo {title} {Dynamics of a rigid body in a {S}tokes
  fluid},\ }\href@noop {} {\bibfield  {journal} {\bibinfo  {journal} {J. Fluid
  Mech.}\ }\textbf {\bibinfo {volume} {519}},\ \bibinfo {pages} {133} (\bibinfo
  {year} {2004})}\BibitemShut {NoStop}%
\bibitem [{\citenamefont {Witten}\ and\ \citenamefont
  {Diamant}(2020)}]{Witten2020}%
  \BibitemOpen
  \bibfield  {author} {\bibinfo {author} {\bibfnamefont {T.~A.}\ \bibnamefont
  {Witten}}\ and\ \bibinfo {author} {\bibfnamefont {H.}~\bibnamefont
  {Diamant}},\ }\bibfield  {title} {\bibinfo {title} {A review of shaped
  colloidal particles in fluids: anisotropy and chirality},\ }\href@noop {}
  {\bibfield  {journal} {\bibinfo  {journal} {Rep. Prog. Phys.}\ }\textbf
  {\bibinfo {volume} {83}},\ \bibinfo {pages} {116601} (\bibinfo {year}
  {2020})}\BibitemShut {NoStop}%
\bibitem [{\citenamefont {Moreau}\ \emph {et~al.}(2022)\citenamefont {Moreau},
  \citenamefont {Ishimoto},\ and\ \citenamefont {Privat}}]{Moreau2022}%
  \BibitemOpen
  \bibfield  {author} {\bibinfo {author} {\bibfnamefont {C.}~\bibnamefont
  {Moreau}}, \bibinfo {author} {\bibfnamefont {K.}~\bibnamefont {Ishimoto}},\
  and\ \bibinfo {author} {\bibfnamefont {Y.}~\bibnamefont {Privat}},\
  }\bibfield  {title} {\bibinfo {title} {Shapes optimising grand resistance
  tensor entries for a rigid body in a {S}tokes flow},\ }\href@noop {}
  {\bibfield  {journal} {\bibinfo  {journal} {arXiv:2207.06023}\ } (\bibinfo
  {year} {2022})}\BibitemShut {NoStop}%
\bibitem [{\citenamefont {Brady}\ and\ \citenamefont
  {Bossis}(1988)}]{Brady1988}%
  \BibitemOpen
  \bibfield  {author} {\bibinfo {author} {\bibfnamefont {J.~F.}\ \bibnamefont
  {Brady}}\ and\ \bibinfo {author} {\bibfnamefont {G.}~\bibnamefont {Bossis}},\
  }\bibfield  {title} {\bibinfo {title} {Stokesian dynamics},\ }\href
  {https://doi.org/10.1146/annurev.fl.20.010188.000551} {\bibfield  {journal}
  {\bibinfo  {journal} {Annu. Rev. Fluid Mech.}\ }\textbf {\bibinfo {volume}
  {20}},\ \bibinfo {pages} {111} (\bibinfo {year} {1988})}\BibitemShut
  {NoStop}%
\bibitem [{\citenamefont {Ichiki}(2002)}]{Ichiki2002}%
  \BibitemOpen
  \bibfield  {author} {\bibinfo {author} {\bibfnamefont {K.}~\bibnamefont
  {Ichiki}},\ }\bibfield  {title} {\bibinfo {title} {Improvement of the
  stokesian dynamics method for systems with a finite number of particles},\
  }\href@noop {} {\bibfield  {journal} {\bibinfo  {journal} {J. Fluid Mech.}\
  }\textbf {\bibinfo {volume} {452}},\ \bibinfo {pages} {231} (\bibinfo {year}
  {2002})}\BibitemShut {NoStop}%
\bibitem [{\citenamefont {Fiore}\ and\ \citenamefont {Swan}(2019)}]{Fiore2019}%
  \BibitemOpen
  \bibfield  {author} {\bibinfo {author} {\bibfnamefont {A.~M.}\ \bibnamefont
  {Fiore}}\ and\ \bibinfo {author} {\bibfnamefont {J.~W.}\ \bibnamefont
  {Swan}},\ }\bibfield  {title} {\bibinfo {title} {Fast stokesian dynamics},\
  }\href@noop {} {\bibfield  {journal} {\bibinfo  {journal} {J. Fluid Mech.}\
  }\textbf {\bibinfo {volume} {878}},\ \bibinfo {pages} {544} (\bibinfo {year}
  {2019})}\BibitemShut {NoStop}%
\bibitem [{\citenamefont {Swan}\ and\ \citenamefont {Wang}(2016)}]{Swan2016}%
  \BibitemOpen
  \bibfield  {author} {\bibinfo {author} {\bibfnamefont {J.~W.}\ \bibnamefont
  {Swan}}\ and\ \bibinfo {author} {\bibfnamefont {G.}~\bibnamefont {Wang}},\
  }\bibfield  {title} {\bibinfo {title} {Rapid calculation of hydrodynamic and
  transport properties in concentrated solutions of colloidal particles and
  macromolecules},\ }\href@noop {} {\bibfield  {journal} {\bibinfo  {journal}
  {Phys. Fluids}\ }\textbf {\bibinfo {volume} {28}} (\bibinfo {year}
  {2016})}\BibitemShut {NoStop}%
\bibitem [{\citenamefont {Rotne}\ and\ \citenamefont
  {Prager}(1969)}]{Rotne1969}%
  \BibitemOpen
  \bibfield  {author} {\bibinfo {author} {\bibfnamefont {J.}~\bibnamefont
  {Rotne}}\ and\ \bibinfo {author} {\bibfnamefont {S.}~\bibnamefont {Prager}},\
  }\bibfield  {title} {\bibinfo {title} {Variational treatment of hydrodynamic
  interaction in polymers},\ }\href@noop {} {\bibfield  {journal} {\bibinfo
  {journal} {J. Chem. Phys.}\ }\textbf {\bibinfo {volume} {50}},\ \bibinfo
  {pages} {4831} (\bibinfo {year} {1969})}\BibitemShut {NoStop}%
\bibitem [{\citenamefont {Yamakawa}(1970)}]{Yamakawa1970}%
  \BibitemOpen
  \bibfield  {author} {\bibinfo {author} {\bibfnamefont {H.}~\bibnamefont
  {Yamakawa}},\ }\bibfield  {title} {\bibinfo {title} {Transport properties of
  polymer chains in dilute solution: hydrodynamic interaction},\ }\href@noop {}
  {\bibfield  {journal} {\bibinfo  {journal} {J. Chem. Phys.}\ }\textbf
  {\bibinfo {volume} {53}},\ \bibinfo {pages} {436} (\bibinfo {year}
  {1970})}\BibitemShut {NoStop}%
\bibitem [{\citenamefont {Wajnryb}\ \emph {et~al.}(2013)\citenamefont
  {Wajnryb}, \citenamefont {Mizerski}, \citenamefont {Zuk},\ and\ \citenamefont
  {Szymczak}}]{Wajnryb2013}%
  \BibitemOpen
  \bibfield  {author} {\bibinfo {author} {\bibfnamefont {E.}~\bibnamefont
  {Wajnryb}}, \bibinfo {author} {\bibfnamefont {K.~A.}\ \bibnamefont
  {Mizerski}}, \bibinfo {author} {\bibfnamefont {P.~J.}\ \bibnamefont {Zuk}},\
  and\ \bibinfo {author} {\bibfnamefont {P.}~\bibnamefont {Szymczak}},\
  }\bibfield  {title} {\bibinfo {title} {Generalization of the
  rotne--prager--yamakawa mobility and shear disturbance tensors},\ }\href@noop
  {} {\bibfield  {journal} {\bibinfo  {journal} {J. Fluid Mech.}\ }\textbf
  {\bibinfo {volume} {731}},\ \bibinfo {pages} {R3} (\bibinfo {year}
  {2013})}\BibitemShut {NoStop}%
\bibitem [{\citenamefont {Cichocki}\ \emph {et~al.}(2021)\citenamefont
  {Cichocki}, \citenamefont {Szymczak},\ and\ \citenamefont
  {{\.Z}uk}}]{Cichocki2021}%
  \BibitemOpen
  \bibfield  {author} {\bibinfo {author} {\bibfnamefont {B.}~\bibnamefont
  {Cichocki}}, \bibinfo {author} {\bibfnamefont {P.}~\bibnamefont {Szymczak}},\
  and\ \bibinfo {author} {\bibfnamefont {P.~J.}\ \bibnamefont {{\.Z}uk}},\
  }\bibfield  {title} {\bibinfo {title} {Generalized rotne--prager--yamakawa
  approximation for brownian dynamics in shear flow in bounded, unbounded, and
  periodic domains},\ }\href@noop {} {\bibfield  {journal} {\bibinfo  {journal}
  {J. Chem. Phys.}\ }\textbf {\bibinfo {volume} {154}} (\bibinfo {year}
  {2021})}\BibitemShut {NoStop}%
\bibitem [{\citenamefont {Waszkiewicz}\ \emph {et~al.}(2024)\citenamefont
  {Waszkiewicz}, \citenamefont {Micha\'s}, \citenamefont {Bia{\l}obrzewski},
  \citenamefont {Klepka}, \citenamefont {Cieplak-Rotowska}, \citenamefont
  {Stasza{\l}ek}, \citenamefont {Cichocki}, \citenamefont {Lisicki},
  \citenamefont {Szymczak},\ and\ \citenamefont
  {Niedzwiecka}}]{Waszkiewicz2024}%
  \BibitemOpen
  \bibfield  {author} {\bibinfo {author} {\bibfnamefont {R.}~\bibnamefont
  {Waszkiewicz}}, \bibinfo {author} {\bibfnamefont {A.}~\bibnamefont
  {Micha\'s}}, \bibinfo {author} {\bibfnamefont {M.~K.}\ \bibnamefont
  {Bia{\l}obrzewski}}, \bibinfo {author} {\bibfnamefont {B.~P.}\ \bibnamefont
  {Klepka}}, \bibinfo {author} {\bibfnamefont {M.~K.}\ \bibnamefont
  {Cieplak-Rotowska}}, \bibinfo {author} {\bibfnamefont {Z.}~\bibnamefont
  {Stasza{\l}ek}}, \bibinfo {author} {\bibfnamefont {B.}~\bibnamefont
  {Cichocki}}, \bibinfo {author} {\bibfnamefont {M.}~\bibnamefont {Lisicki}},
  \bibinfo {author} {\bibfnamefont {P.}~\bibnamefont {Szymczak}},\ and\
  \bibinfo {author} {\bibfnamefont {A.}~\bibnamefont {Niedzwiecka}},\
  }\bibfield  {title} {\bibinfo {title} {Hydrodynamic radii of intrinsically
  disordered proteins: Fast prediction by minimum dissipation approximation and
  experimental validation},\ }\href
  {https://doi.org/10.1021/acs.jpclett.4c00312} {\bibfield  {journal} {\bibinfo
   {journal} {J. Phys. Chem. Lett.}\ }\textbf {\bibinfo {volume} {15}},\
  \bibinfo {pages} {5024} (\bibinfo {year} {2024})}\BibitemShut {NoStop}%
\bibitem [{\citenamefont {Jeffrey}\ and\ \citenamefont
  {Onishi}(1984)}]{Jeffrey1984}%
  \BibitemOpen
  \bibfield  {author} {\bibinfo {author} {\bibfnamefont {D.~J.}\ \bibnamefont
  {Jeffrey}}\ and\ \bibinfo {author} {\bibfnamefont {Y.}~\bibnamefont
  {Onishi}},\ }\bibfield  {title} {\bibinfo {title} {Calculation of the
  resistance and mobility functions for two unequal rigid spheres in
  low-{R}eynolds-number flow},\ }\href
  {https://doi.org/10.1017/S0022112084000355} {\bibfield  {journal} {\bibinfo
  {journal} {J. Fluid Mech.}\ }\textbf {\bibinfo {volume} {139}},\ \bibinfo
  {pages} {261} (\bibinfo {year} {1984})}\BibitemShut {NoStop}%
\bibitem [{\citenamefont {Delmotte}\ \emph
  {et~al.}(2015{\natexlab{a}})\citenamefont {Delmotte}, \citenamefont
  {Keaveny}, \citenamefont {Plourabou{\'e}},\ and\ \citenamefont
  {Climent}}]{Delmotte2015b}%
  \BibitemOpen
  \bibfield  {author} {\bibinfo {author} {\bibfnamefont {B.}~\bibnamefont
  {Delmotte}}, \bibinfo {author} {\bibfnamefont {E.~E.}\ \bibnamefont
  {Keaveny}}, \bibinfo {author} {\bibfnamefont {F.}~\bibnamefont
  {Plourabou{\'e}}},\ and\ \bibinfo {author} {\bibfnamefont {E.}~\bibnamefont
  {Climent}},\ }\bibfield  {title} {\bibinfo {title} {Large-scale simulation of
  steady and time-dependent active suspensions with the force-coupling
  method},\ }\href@noop {} {\bibfield  {journal} {\bibinfo  {journal} {J.
  Comput. Phys.}\ }\textbf {\bibinfo {volume} {302}},\ \bibinfo {pages} {524}
  (\bibinfo {year} {2015}{\natexlab{a}})}\BibitemShut {NoStop}%
\bibitem [{\citenamefont {Delmotte}\ \emph
  {et~al.}(2015{\natexlab{b}})\citenamefont {Delmotte}, \citenamefont
  {Climent},\ and\ \citenamefont {Plourabou{\'e}}}]{Delmotte2015}%
  \BibitemOpen
  \bibfield  {author} {\bibinfo {author} {\bibfnamefont {B.}~\bibnamefont
  {Delmotte}}, \bibinfo {author} {\bibfnamefont {E.}~\bibnamefont {Climent}},\
  and\ \bibinfo {author} {\bibfnamefont {F.}~\bibnamefont {Plourabou{\'e}}},\
  }\bibfield  {title} {\bibinfo {title} {A general formulation of bead models
  applied to flexible fibers and active filaments at low {R}eynolds number},\
  }\href@noop {} {\bibfield  {journal} {\bibinfo  {journal} {J. Comput. Phys.}\
  }\textbf {\bibinfo {volume} {286}},\ \bibinfo {pages} {14} (\bibinfo {year}
  {2015}{\natexlab{b}})}\BibitemShut {NoStop}%
\bibitem [{\citenamefont {Fuchter}\ and\ \citenamefont
  {Bloomfield-Gad{\^e}lha}(2023)}]{Fuchter2023}%
  \BibitemOpen
  \bibfield  {author} {\bibinfo {author} {\bibfnamefont {P.}~\bibnamefont
  {Fuchter}}\ and\ \bibinfo {author} {\bibfnamefont {H.}~\bibnamefont
  {Bloomfield-Gad{\^e}lha}},\ }\bibfield  {title} {\bibinfo {title} {The
  three-dimensional coarse-graining formulation of interacting
  elastohydrodynamic filaments and multi-body microhydrodynamics},\ }\href@noop
  {} {\bibfield  {journal} {\bibinfo  {journal} {J. R. Soc. Interface}\
  }\textbf {\bibinfo {volume} {20}},\ \bibinfo {pages} {20230021} (\bibinfo
  {year} {2023})}\BibitemShut {NoStop}%
\bibitem [{\citenamefont {Liang}\ \emph {et~al.}(2013)\citenamefont {Liang},
  \citenamefont {Gimbutas}, \citenamefont {Greengard}, \citenamefont {Huang},\
  and\ \citenamefont {Jiang}}]{Liang2013}%
  \BibitemOpen
  \bibfield  {author} {\bibinfo {author} {\bibfnamefont {Z.}~\bibnamefont
  {Liang}}, \bibinfo {author} {\bibfnamefont {Z.}~\bibnamefont {Gimbutas}},
  \bibinfo {author} {\bibfnamefont {L.}~\bibnamefont {Greengard}}, \bibinfo
  {author} {\bibfnamefont {J.}~\bibnamefont {Huang}},\ and\ \bibinfo {author}
  {\bibfnamefont {S.}~\bibnamefont {Jiang}},\ }\bibfield  {title} {\bibinfo
  {title} {A fast multipole method for the {R}otne--{P}rager--{Y}amakawa tensor
  and its applications},\ }\href@noop {} {\bibfield  {journal} {\bibinfo
  {journal} {J. Comput. Phys.}\ }\textbf {\bibinfo {volume} {234}},\ \bibinfo
  {pages} {133} (\bibinfo {year} {2013})}\BibitemShut {NoStop}%
\bibitem [{\citenamefont {Solovev}\ and\ \citenamefont
  {Friedrich}(2021)}]{Solovev2021}%
  \BibitemOpen
  \bibfield  {author} {\bibinfo {author} {\bibfnamefont {A.}~\bibnamefont
  {Solovev}}\ and\ \bibinfo {author} {\bibfnamefont {B.~M.}\ \bibnamefont
  {Friedrich}},\ }\bibfield  {title} {\bibinfo {title} {Lagrangian mechanics of
  active systems},\ }\href@noop {} {\bibfield  {journal} {\bibinfo  {journal}
  {Eur. Phys. J. E}\ }\textbf {\bibinfo {volume} {44}},\ \bibinfo {pages} {1}
  (\bibinfo {year} {2021})}\BibitemShut {NoStop}%
\bibitem [{\citenamefont {Vilfan}\ and\ \citenamefont
  {Stark}(2009)}]{Vilfan2009}%
  \BibitemOpen
  \bibfield  {author} {\bibinfo {author} {\bibfnamefont {A.}~\bibnamefont
  {Vilfan}}\ and\ \bibinfo {author} {\bibfnamefont {H.}~\bibnamefont {Stark}},\
  }\bibfield  {title} {\bibinfo {title} {Comment on ``ferromagnetic
  microswimmers''},\ }\href {https://doi.org/10.1103/PhysRevLett.103.199801}
  {\bibfield  {journal} {\bibinfo  {journal} {Phys. Rev. Lett.}\ }\textbf
  {\bibinfo {volume} {103}},\ \bibinfo {pages} {199801} (\bibinfo {year}
  {2009})}\BibitemShut {NoStop}%
\bibitem [{\citenamefont {Klindt}\ \emph {et~al.}(2016)\citenamefont {Klindt},
  \citenamefont {Ruloff}, \citenamefont {Wagner},\ and\ \citenamefont
  {Friedrich}}]{Klindt2016}%
  \BibitemOpen
  \bibfield  {author} {\bibinfo {author} {\bibfnamefont {G.~S.}\ \bibnamefont
  {Klindt}}, \bibinfo {author} {\bibfnamefont {C.}~\bibnamefont {Ruloff}},
  \bibinfo {author} {\bibfnamefont {C.}~\bibnamefont {Wagner}},\ and\ \bibinfo
  {author} {\bibfnamefont {B.~M.}\ \bibnamefont {Friedrich}},\ }\bibfield
  {title} {\bibinfo {title} {Load response of the flagellar beat},\ }\href
  {https://doi.org/10.1103/PhysRevLett.117.258101} {\bibfield  {journal}
  {\bibinfo  {journal} {Phys. Rev. Lett.}\ }\textbf {\bibinfo {volume} {117}},\
  \bibinfo {pages} {258101} (\bibinfo {year} {2016})}\BibitemShut {NoStop}%
\bibitem [{\citenamefont {Klindt}\ \emph {et~al.}(2017)\citenamefont {Klindt},
  \citenamefont {Ruloff}, \citenamefont {Wagner},\ and\ \citenamefont
  {Friedrich}}]{Klindt2017}%
  \BibitemOpen
  \bibfield  {author} {\bibinfo {author} {\bibfnamefont {G.~S.}\ \bibnamefont
  {Klindt}}, \bibinfo {author} {\bibfnamefont {C.}~\bibnamefont {Ruloff}},
  \bibinfo {author} {\bibfnamefont {C.}~\bibnamefont {Wagner}},\ and\ \bibinfo
  {author} {\bibfnamefont {B.~M.}\ \bibnamefont {Friedrich}},\ }\bibfield
  {title} {\bibinfo {title} {In-phase and anti-phase flagellar synchronization
  by waveform compliance and basal coupling},\ }\href
  {https://doi.org/10.1088/1367-2630/aa9031} {\bibfield  {journal} {\bibinfo
  {journal} {New J. Phys.}\ }\textbf {\bibinfo {volume} {19}},\ \bibinfo
  {pages} {113052} (\bibinfo {year} {2017})}\BibitemShut {NoStop}%
\bibitem [{\citenamefont {Friedrich}\ and\ \citenamefont
  {J\"ulicher}(2012)}]{Friedrich2012}%
  \BibitemOpen
  \bibfield  {author} {\bibinfo {author} {\bibfnamefont {B.~M.}\ \bibnamefont
  {Friedrich}}\ and\ \bibinfo {author} {\bibfnamefont {F.}~\bibnamefont
  {J\"ulicher}},\ }\bibfield  {title} {\bibinfo {title} {Flagellar
  synchronization independent of hydrodynamic interactions},\ }\href
  {https://doi.org/10.1103/PhysRevLett.109.138102} {\bibfield  {journal}
  {\bibinfo  {journal} {Phys. Rev. Lett.}\ }\textbf {\bibinfo {volume} {109}},\
  \bibinfo {pages} {138102} (\bibinfo {year} {2012})}\BibitemShut {NoStop}%
\bibitem [{\citenamefont {Geyer}\ \emph {et~al.}(2013)\citenamefont {Geyer},
  \citenamefont {J\"ulicher}, \citenamefont {Howard},\ and\ \citenamefont
  {Friedrich}}]{Geyer2013}%
  \BibitemOpen
  \bibfield  {author} {\bibinfo {author} {\bibfnamefont {V.~F.}\ \bibnamefont
  {Geyer}}, \bibinfo {author} {\bibfnamefont {F.}~\bibnamefont {J\"ulicher}},
  \bibinfo {author} {\bibfnamefont {J.}~\bibnamefont {Howard}},\ and\ \bibinfo
  {author} {\bibfnamefont {B.~M.}\ \bibnamefont {Friedrich}},\ }\bibfield
  {title} {\bibinfo {title} {Cell-body rocking is a dominant mechanism for
  flagellar synchronization in a swimming alga},\ }\href
  {https://doi.org/10.1073/pnas.1300895110} {\bibfield  {journal} {\bibinfo
  {journal} {Proc. Natl. Acad. Sci. U.S.A.}\ }\textbf {\bibinfo {volume}
  {110}},\ \bibinfo {pages} {18058} (\bibinfo {year} {2013})}\BibitemShut
  {NoStop}%
\bibitem [{\citenamefont {Eloy}(2026)}]{softmobility}%
  \BibitemOpen
  \bibfield  {author} {\bibinfo {author} {\bibfnamefont {C.}~\bibnamefont
  {Eloy}},\ }\href@noop {} {\bibinfo {title} {Soft mobility {P}ython library}}
  (\bibinfo {year} {2026}),\ \bibinfo {note}
  {\url{https://github.com/C0PEP0D/SoftMobility}}\BibitemShut {NoStop}%
\bibitem [{\citenamefont {Bortz}(1971)}]{Bortz1971}%
  \BibitemOpen
  \bibfield  {author} {\bibinfo {author} {\bibfnamefont {J.~E.}\ \bibnamefont
  {Bortz}},\ }\bibfield  {title} {\bibinfo {title} {A new mathematical
  formulation for strapdown inertial navigation},\ }\href@noop {} {\bibfield
  {journal} {\bibinfo  {journal} {IEEE Trans. Aerosp. Electron. Syst.}\ ,\
  \bibinfo {pages} {61}} (\bibinfo {year} {1971})}\BibitemShut {NoStop}%
\bibitem [{\citenamefont {Saad}\ and\ \citenamefont
  {Schultz}(1986)}]{Saad1986}%
  \BibitemOpen
  \bibfield  {author} {\bibinfo {author} {\bibfnamefont {Y.}~\bibnamefont
  {Saad}}\ and\ \bibinfo {author} {\bibfnamefont {M.~H.}\ \bibnamefont
  {Schultz}},\ }\bibfield  {title} {\bibinfo {title} {Gmres: A generalized
  minimal residual algorithm for solving nonsymmetric linear systems},\
  }\href@noop {} {\bibfield  {journal} {\bibinfo  {journal} {SIAM J. Sci. Stat.
  Comput.}\ }\textbf {\bibinfo {volume} {7}},\ \bibinfo {pages} {856} (\bibinfo
  {year} {1986})}\BibitemShut {NoStop}%
\bibitem [{\citenamefont {Bradbury}\ \emph {et~al.}(2018)\citenamefont
  {Bradbury}, \citenamefont {Frostig}, \citenamefont {Hawkins}, \citenamefont
  {Johnson}, \citenamefont {Leary}, \citenamefont {Maclaurin}, \citenamefont
  {Necula}, \citenamefont {Paszke}, \citenamefont {Vander{P}las}, \citenamefont
  {Wanderman-{M}ilne},\ and\ \citenamefont {Zhang}}]{jax2018github}%
  \BibitemOpen
  \bibfield  {author} {\bibinfo {author} {\bibfnamefont {J.}~\bibnamefont
  {Bradbury}}, \bibinfo {author} {\bibfnamefont {R.}~\bibnamefont {Frostig}},
  \bibinfo {author} {\bibfnamefont {P.}~\bibnamefont {Hawkins}}, \bibinfo
  {author} {\bibfnamefont {M.~J.}\ \bibnamefont {Johnson}}, \bibinfo {author}
  {\bibfnamefont {C.}~\bibnamefont {Leary}}, \bibinfo {author} {\bibfnamefont
  {D.}~\bibnamefont {Maclaurin}}, \bibinfo {author} {\bibfnamefont
  {G.}~\bibnamefont {Necula}}, \bibinfo {author} {\bibfnamefont
  {A.}~\bibnamefont {Paszke}}, \bibinfo {author} {\bibfnamefont
  {J.}~\bibnamefont {Vander{P}las}}, \bibinfo {author} {\bibfnamefont
  {S.}~\bibnamefont {Wanderman-{M}ilne}},\ and\ \bibinfo {author}
  {\bibfnamefont {Q.}~\bibnamefont {Zhang}},\ }\href
  {http://github.com/google/jax} {\bibinfo {title} {{JAX}: composable
  transformations of {Python}+{NumPy} programs}} (\bibinfo {year}
  {2018})\BibitemShut {NoStop}%
\bibitem [{\citenamefont {DeepMind}\ \emph {et~al.}(2020)\citenamefont
  {DeepMind} \emph {et~al.}}]{optax2020github}%
  \BibitemOpen
  \bibfield  {author} {\bibinfo {author} {\bibnamefont {DeepMind}} \emph
  {et~al.},\ }\href {http://github.com/google-deepmind/optax} {\bibinfo {title}
  {The {D}eep{M}ind {JAX} {E}cosystem}} (\bibinfo {year} {2020})\BibitemShut
  {NoStop}%
\bibitem [{\citenamefont {Marsden}\ and\ \citenamefont
  {Ratiu}(2013)}]{Marsden2013}%
  \BibitemOpen
  \bibfield  {author} {\bibinfo {author} {\bibfnamefont {J.~E.}\ \bibnamefont
  {Marsden}}\ and\ \bibinfo {author} {\bibfnamefont {T.~S.}\ \bibnamefont
  {Ratiu}},\ }\href@noop {} {\emph {\bibinfo {title} {Introduction to mechanics
  and symmetry: a basic exposition of classical mechanical systems}}},\
  Vol.~\bibinfo {volume} {17}\ (\bibinfo  {publisher} {Springer Science \&
  Business Media},\ \bibinfo {year} {2013})\BibitemShut {NoStop}%
\bibitem [{\citenamefont {Weinberger}(1972)}]{Weinberger1972}%
  \BibitemOpen
  \bibfield  {author} {\bibinfo {author} {\bibfnamefont {H.}~\bibnamefont
  {Weinberger}},\ }\bibfield  {title} {\bibinfo {title} {Variational properties
  of steady fall in stokes flow},\ }\href@noop {} {\bibfield  {journal}
  {\bibinfo  {journal} {J. Fluid Mech.}\ }\textbf {\bibinfo {volume} {52}},\
  \bibinfo {pages} {321} (\bibinfo {year} {1972})}\BibitemShut {NoStop}%
\bibitem [{\citenamefont {Weber}\ \emph {et~al.}(2013)\citenamefont {Weber},
  \citenamefont {Carlen}, \citenamefont {Dietler}, \citenamefont {Rawdon},\
  and\ \citenamefont {Stasiak}}]{Weber2013}%
  \BibitemOpen
  \bibfield  {author} {\bibinfo {author} {\bibfnamefont {C.}~\bibnamefont
  {Weber}}, \bibinfo {author} {\bibfnamefont {M.}~\bibnamefont {Carlen}},
  \bibinfo {author} {\bibfnamefont {G.}~\bibnamefont {Dietler}}, \bibinfo
  {author} {\bibfnamefont {E.~J.}\ \bibnamefont {Rawdon}},\ and\ \bibinfo
  {author} {\bibfnamefont {A.}~\bibnamefont {Stasiak}},\ }\bibfield  {title}
  {\bibinfo {title} {Sedimentation of macroscopic rigid knots and its relation
  to gel electrophoretic mobility of dna knots},\ }\href@noop {} {\bibfield
  {journal} {\bibinfo  {journal} {Sci. Rep.}\ }\textbf {\bibinfo {volume}
  {3}},\ \bibinfo {pages} {1091} (\bibinfo {year} {2013})}\BibitemShut
  {NoStop}%
\bibitem [{\citenamefont {Palusa}\ \emph {et~al.}(2018)\citenamefont {Palusa},
  \citenamefont {De~Graaf}, \citenamefont {Brown},\ and\ \citenamefont
  {Morozov}}]{Palusa2018}%
  \BibitemOpen
  \bibfield  {author} {\bibinfo {author} {\bibfnamefont {M.}~\bibnamefont
  {Palusa}}, \bibinfo {author} {\bibfnamefont {J.}~\bibnamefont {De~Graaf}},
  \bibinfo {author} {\bibfnamefont {A.}~\bibnamefont {Brown}},\ and\ \bibinfo
  {author} {\bibfnamefont {A.}~\bibnamefont {Morozov}},\ }\bibfield  {title}
  {\bibinfo {title} {Sedimentation of a rigid helix in viscous media},\
  }\href@noop {} {\bibfield  {journal} {\bibinfo  {journal} {Phys. Rev.
  Fluids}\ }\textbf {\bibinfo {volume} {3}},\ \bibinfo {pages} {124301}
  (\bibinfo {year} {2018})}\BibitemShut {NoStop}%
\bibitem [{\citenamefont {Braverman}\ \emph {et~al.}(2020)\citenamefont
  {Braverman}, \citenamefont {Mowitz},\ and\ \citenamefont
  {Witten}}]{Braverman2020}%
  \BibitemOpen
  \bibfield  {author} {\bibinfo {author} {\bibfnamefont {L.}~\bibnamefont
  {Braverman}}, \bibinfo {author} {\bibfnamefont {A.}~\bibnamefont {Mowitz}},\
  and\ \bibinfo {author} {\bibfnamefont {T.~A.}\ \bibnamefont {Witten}},\
  }\bibfield  {title} {\bibinfo {title} {Chiral motion in colloidal
  electrophoresis},\ }\href@noop {} {\bibfield  {journal} {\bibinfo  {journal}
  {Phys. Rev. E}\ }\textbf {\bibinfo {volume} {101}},\ \bibinfo {pages}
  {062608} (\bibinfo {year} {2020})}\BibitemShut {NoStop}%
\bibitem [{\citenamefont {Miara}\ \emph {et~al.}(2024)\citenamefont {Miara},
  \citenamefont {Vaquero-Stainer}, \citenamefont {Pihler-Puzovi{\'c}},
  \citenamefont {Heil},\ and\ \citenamefont {Juel}}]{Miara2024}%
  \BibitemOpen
  \bibfield  {author} {\bibinfo {author} {\bibfnamefont {T.}~\bibnamefont
  {Miara}}, \bibinfo {author} {\bibfnamefont {C.}~\bibnamefont
  {Vaquero-Stainer}}, \bibinfo {author} {\bibfnamefont {D.}~\bibnamefont
  {Pihler-Puzovi{\'c}}}, \bibinfo {author} {\bibfnamefont {M.}~\bibnamefont
  {Heil}},\ and\ \bibinfo {author} {\bibfnamefont {A.}~\bibnamefont {Juel}},\
  }\bibfield  {title} {\bibinfo {title} {Dynamics of inertialess sedimentation
  of a rigid {U}-shaped disk},\ }\href@noop {} {\bibfield  {journal} {\bibinfo
  {journal} {Commun. Phys.}\ }\textbf {\bibinfo {volume} {7}},\ \bibinfo
  {pages} {47} (\bibinfo {year} {2024})}\BibitemShut {NoStop}%
\bibitem [{\citenamefont {Huseby}\ \emph {et~al.}(2025)\citenamefont {Huseby},
  \citenamefont {Gissinger}, \citenamefont {Candelier}, \citenamefont {Pujara},
  \citenamefont {Verhille}, \citenamefont {Mehlig},\ and\ \citenamefont
  {Voth}}]{Huseby2025}%
  \BibitemOpen
  \bibfield  {author} {\bibinfo {author} {\bibfnamefont {E.}~\bibnamefont
  {Huseby}}, \bibinfo {author} {\bibfnamefont {J.}~\bibnamefont {Gissinger}},
  \bibinfo {author} {\bibfnamefont {F.}~\bibnamefont {Candelier}}, \bibinfo
  {author} {\bibfnamefont {N.}~\bibnamefont {Pujara}}, \bibinfo {author}
  {\bibfnamefont {G.}~\bibnamefont {Verhille}}, \bibinfo {author}
  {\bibfnamefont {B.}~\bibnamefont {Mehlig}},\ and\ \bibinfo {author}
  {\bibfnamefont {G.}~\bibnamefont {Voth}},\ }\bibfield  {title} {\bibinfo
  {title} {Helical ribbons: Simple chiral sedimentation},\ }\href@noop {}
  {\bibfield  {journal} {\bibinfo  {journal} {Phys. Rev. Fluids}\ }\textbf
  {\bibinfo {volume} {10}},\ \bibinfo {pages} {024101} (\bibinfo {year}
  {2025})}\BibitemShut {NoStop}%
\bibitem [{\citenamefont {Joshi}\ and\ \citenamefont
  {Govindarajan}(2025)}]{Joshi2025}%
  \BibitemOpen
  \bibfield  {author} {\bibinfo {author} {\bibfnamefont {H.}~\bibnamefont
  {Joshi}}\ and\ \bibinfo {author} {\bibfnamefont {R.}~\bibnamefont
  {Govindarajan}},\ }\bibfield  {title} {\bibinfo {title} {Sedimentation
  dynamics of bodies with two planes of symmetry},\ }\href@noop {} {\bibfield
  {journal} {\bibinfo  {journal} {Phys. Rev. Lett.}\ }\textbf {\bibinfo
  {volume} {134}},\ \bibinfo {pages} {014002} (\bibinfo {year}
  {2025})}\BibitemShut {NoStop}%
\bibitem [{\citenamefont {Li}\ \emph {et~al.}(2013)\citenamefont {Li},
  \citenamefont {Manikantan}, \citenamefont {Saintillan},\ and\ \citenamefont
  {Spagnolie}}]{Li2013}%
  \BibitemOpen
  \bibfield  {author} {\bibinfo {author} {\bibfnamefont {L.}~\bibnamefont
  {Li}}, \bibinfo {author} {\bibfnamefont {H.}~\bibnamefont {Manikantan}},
  \bibinfo {author} {\bibfnamefont {D.}~\bibnamefont {Saintillan}},\ and\
  \bibinfo {author} {\bibfnamefont {S.~E.}\ \bibnamefont {Spagnolie}},\
  }\bibfield  {title} {\bibinfo {title} {The sedimentation of flexible
  filaments},\ }\href@noop {} {\bibfield  {journal} {\bibinfo  {journal} {J.
  Fluid Mech.}\ }\textbf {\bibinfo {volume} {735}},\ \bibinfo {pages} {705}
  (\bibinfo {year} {2013})}\BibitemShut {NoStop}%
\bibitem [{\citenamefont {Coq}\ \emph {et~al.}(2008)\citenamefont {Coq},
  \citenamefont {Du~Roure}, \citenamefont {Marthelot}, \citenamefont
  {Bartolo},\ and\ \citenamefont {Fermigier}}]{Coq2008}%
  \BibitemOpen
  \bibfield  {author} {\bibinfo {author} {\bibfnamefont {N.}~\bibnamefont
  {Coq}}, \bibinfo {author} {\bibfnamefont {O.}~\bibnamefont {Du~Roure}},
  \bibinfo {author} {\bibfnamefont {J.}~\bibnamefont {Marthelot}}, \bibinfo
  {author} {\bibfnamefont {D.}~\bibnamefont {Bartolo}},\ and\ \bibinfo {author}
  {\bibfnamefont {M.}~\bibnamefont {Fermigier}},\ }\bibfield  {title} {\bibinfo
  {title} {Rotational dynamics of a soft filament: Wrapping transition and
  propulsive forces},\ }\href@noop {} {\bibfield  {journal} {\bibinfo
  {journal} {Phys. Fluids}\ }\textbf {\bibinfo {volume} {20}} (\bibinfo {year}
  {2008})}\BibitemShut {NoStop}%
\bibitem [{\citenamefont {Coq}\ \emph {et~al.}(2009)\citenamefont {Coq},
  \citenamefont {Roure}, \citenamefont {Fermigier},\ and\ \citenamefont
  {Bartolo}}]{Coq2009}%
  \BibitemOpen
  \bibfield  {author} {\bibinfo {author} {\bibfnamefont {N.}~\bibnamefont
  {Coq}}, \bibinfo {author} {\bibfnamefont {O.~d.}\ \bibnamefont {Roure}},
  \bibinfo {author} {\bibfnamefont {M.}~\bibnamefont {Fermigier}},\ and\
  \bibinfo {author} {\bibfnamefont {D.}~\bibnamefont {Bartolo}},\ }\bibfield
  {title} {\bibinfo {title} {Helical beating of an actuated elastic filament},\
  }\href@noop {} {\bibfield  {journal} {\bibinfo  {journal} {J. Phys. Condens.
  Matter}\ }\textbf {\bibinfo {volume} {21}},\ \bibinfo {pages} {204109}
  (\bibinfo {year} {2009})}\BibitemShut {NoStop}%
\bibitem [{\citenamefont {Liu}\ \emph {et~al.}(2018)\citenamefont {Liu},
  \citenamefont {Chakrabarti}, \citenamefont {Saintillan}, \citenamefont
  {Lindner},\ and\ \citenamefont {Du~Roure}}]{Liu2018}%
  \BibitemOpen
  \bibfield  {author} {\bibinfo {author} {\bibfnamefont {Y.}~\bibnamefont
  {Liu}}, \bibinfo {author} {\bibfnamefont {B.}~\bibnamefont {Chakrabarti}},
  \bibinfo {author} {\bibfnamefont {D.}~\bibnamefont {Saintillan}}, \bibinfo
  {author} {\bibfnamefont {A.}~\bibnamefont {Lindner}},\ and\ \bibinfo {author}
  {\bibfnamefont {O.}~\bibnamefont {Du~Roure}},\ }\bibfield  {title} {\bibinfo
  {title} {Morphological transitions of elastic filaments in shear flow},\
  }\href@noop {} {\bibfield  {journal} {\bibinfo  {journal} {Proc. Natl. Acad.
  Sci. U.S.A.}\ }\textbf {\bibinfo {volume} {115}},\ \bibinfo {pages} {9438}
  (\bibinfo {year} {2018})}\BibitemShut {NoStop}%
\bibitem [{\citenamefont {Schmid}\ \emph {et~al.}(2000)\citenamefont {Schmid},
  \citenamefont {Switzer},\ and\ \citenamefont {Klingenberg}}]{Schmid2000}%
  \BibitemOpen
  \bibfield  {author} {\bibinfo {author} {\bibfnamefont {C.~F.}\ \bibnamefont
  {Schmid}}, \bibinfo {author} {\bibfnamefont {L.~H.}\ \bibnamefont
  {Switzer}},\ and\ \bibinfo {author} {\bibfnamefont {D.~J.}\ \bibnamefont
  {Klingenberg}},\ }\bibfield  {title} {\bibinfo {title} {Simulations of fiber
  flocculation: Effects of fiber properties and interfiber friction},\
  }\href@noop {} {\bibfield  {journal} {\bibinfo  {journal} {J. Rheol.}\
  }\textbf {\bibinfo {volume} {44}},\ \bibinfo {pages} {781} (\bibinfo {year}
  {2000})}\BibitemShut {NoStop}%
\bibitem [{\citenamefont {Jeffery}(1922)}]{Jeffery1922}%
  \BibitemOpen
  \bibfield  {author} {\bibinfo {author} {\bibfnamefont {G.~B.}\ \bibnamefont
  {Jeffery}},\ }\bibfield  {title} {\bibinfo {title} {The motion of ellipsoidal
  particles immersed in a viscous fluid},\ }\href@noop {} {\bibfield  {journal}
  {\bibinfo  {journal} {Proc. R. Soc. Lond. A}\ }\textbf {\bibinfo {volume}
  {102}},\ \bibinfo {pages} {161} (\bibinfo {year} {1922})}\BibitemShut
  {NoStop}%
\bibitem [{\citenamefont {Bretherton}(1962)}]{Bretherton1962}%
  \BibitemOpen
  \bibfield  {author} {\bibinfo {author} {\bibfnamefont {F.~P.}\ \bibnamefont
  {Bretherton}},\ }\bibfield  {title} {\bibinfo {title} {The motion of rigid
  particles in a shear flow at low {R}eynolds number},\ }\href@noop {}
  {\bibfield  {journal} {\bibinfo  {journal} {J. Fluid Mech.}\ }\textbf
  {\bibinfo {volume} {14}},\ \bibinfo {pages} {284} (\bibinfo {year}
  {1962})}\BibitemShut {NoStop}%
\bibitem [{\citenamefont {Thorp}\ and\ \citenamefont
  {Lister}(2019)}]{Thorp2019}%
  \BibitemOpen
  \bibfield  {author} {\bibinfo {author} {\bibfnamefont {I.~R.}\ \bibnamefont
  {Thorp}}\ and\ \bibinfo {author} {\bibfnamefont {J.~R.}\ \bibnamefont
  {Lister}},\ }\bibfield  {title} {\bibinfo {title} {Motion of a
  non-axisymmetric particle in viscous shear flow},\ }\href@noop {} {\bibfield
  {journal} {\bibinfo  {journal} {J. Fluid Mech.}\ }\textbf {\bibinfo {volume}
  {872}},\ \bibinfo {pages} {532} (\bibinfo {year} {2019})}\BibitemShut
  {NoStop}%
\bibitem [{\citenamefont {Ishimoto}(2023)}]{Ishimoto2023}%
  \BibitemOpen
  \bibfield  {author} {\bibinfo {author} {\bibfnamefont {K.}~\bibnamefont
  {Ishimoto}},\ }\bibfield  {title} {\bibinfo {title} {Jeffery's orbits and
  microswimmers in flows: A theoretical review},\ }\href@noop {} {\bibfield
  {journal} {\bibinfo  {journal} {J. Phys. Soc. Jpn.}\ }\textbf {\bibinfo
  {volume} {92}},\ \bibinfo {pages} {062001} (\bibinfo {year}
  {2023})}\BibitemShut {NoStop}%
\bibitem [{\citenamefont {Zhang}\ \emph {et~al.}(2019)\citenamefont {Zhang},
  \citenamefont {Lam},\ and\ \citenamefont {Graham}}]{Zhang2019}%
  \BibitemOpen
  \bibfield  {author} {\bibinfo {author} {\bibfnamefont {X.}~\bibnamefont
  {Zhang}}, \bibinfo {author} {\bibfnamefont {W.~A.}\ \bibnamefont {Lam}},\
  and\ \bibinfo {author} {\bibfnamefont {M.~D.}\ \bibnamefont {Graham}},\
  }\bibfield  {title} {\bibinfo {title} {Dynamics of deformable straight and
  curved prolate capsules in simple shear flow},\ }\href@noop {} {\bibfield
  {journal} {\bibinfo  {journal} {Phys. Rev. Fluids}\ }\textbf {\bibinfo
  {volume} {4}},\ \bibinfo {pages} {043103} (\bibinfo {year}
  {2019})}\BibitemShut {NoStop}%
\bibitem [{\citenamefont {Najafi}\ and\ \citenamefont
  {Golestanian}(2004)}]{Najafi2004}%
  \BibitemOpen
  \bibfield  {author} {\bibinfo {author} {\bibfnamefont {A.}~\bibnamefont
  {Najafi}}\ and\ \bibinfo {author} {\bibfnamefont {R.}~\bibnamefont
  {Golestanian}},\ }\bibfield  {title} {\bibinfo {title} {Simple swimmer at low
  reynolds number: Three linked spheres},\ }\href@noop {} {\bibfield  {journal}
  {\bibinfo  {journal} {Phys. Rev. E}\ }\textbf {\bibinfo {volume} {69}},\
  \bibinfo {pages} {062901} (\bibinfo {year} {2004})}\BibitemShut {NoStop}%
\bibitem [{\citenamefont {Montino}\ and\ \citenamefont
  {DeSimone}(2015)}]{Montino2015}%
  \BibitemOpen
  \bibfield  {author} {\bibinfo {author} {\bibfnamefont {A.}~\bibnamefont
  {Montino}}\ and\ \bibinfo {author} {\bibfnamefont {A.}~\bibnamefont
  {DeSimone}},\ }\bibfield  {title} {\bibinfo {title} {Three-sphere
  low-reynolds-number swimmer with a passive elastic arm},\ }\href@noop {}
  {\bibfield  {journal} {\bibinfo  {journal} {Eur. Phys. J. E}\ }\textbf
  {\bibinfo {volume} {38}},\ \bibinfo {pages} {1} (\bibinfo {year}
  {2015})}\BibitemShut {NoStop}%
\bibitem [{\citenamefont {Colabrese}\ \emph {et~al.}(2017)\citenamefont
  {Colabrese}, \citenamefont {Gustavsson}, \citenamefont {Celani},\ and\
  \citenamefont {Biferale}}]{Colabrese2017}%
  \BibitemOpen
  \bibfield  {author} {\bibinfo {author} {\bibfnamefont {S.}~\bibnamefont
  {Colabrese}}, \bibinfo {author} {\bibfnamefont {K.}~\bibnamefont
  {Gustavsson}}, \bibinfo {author} {\bibfnamefont {A.}~\bibnamefont {Celani}},\
  and\ \bibinfo {author} {\bibfnamefont {L.}~\bibnamefont {Biferale}},\
  }\bibfield  {title} {\bibinfo {title} {Flow navigation by smart microswimmers
  via reinforcement learning},\ }\href@noop {} {\bibfield  {journal} {\bibinfo
  {journal} {Phys. Rev. Lett.}\ }\textbf {\bibinfo {volume} {118}},\ \bibinfo
  {pages} {158004} (\bibinfo {year} {2017})}\BibitemShut {NoStop}%
\bibitem [{\citenamefont {Pedley}\ and\ \citenamefont
  {Kessler}(1992)}]{Pedley1992}%
  \BibitemOpen
  \bibfield  {author} {\bibinfo {author} {\bibfnamefont {T.}~\bibnamefont
  {Pedley}}\ and\ \bibinfo {author} {\bibfnamefont {J.~O.}\ \bibnamefont
  {Kessler}},\ }\bibfield  {title} {\bibinfo {title} {Hydrodynamic phenomena in
  suspensions of swimming microorganisms},\ }\href@noop {} {\bibfield
  {journal} {\bibinfo  {journal} {Annu. Rev. Fluid Mech.}\ }\textbf {\bibinfo
  {volume} {24}},\ \bibinfo {pages} {313} (\bibinfo {year} {1992})}\BibitemShut
  {NoStop}%
\bibitem [{\citenamefont {K\"ummel}\ \emph {et~al.}(2014)\citenamefont
  {K\"ummel}, \citenamefont {ten Hagen}, \citenamefont {Wittkowski},
  \citenamefont {Buttinoni}, \citenamefont {Eichhorn}, \citenamefont {Volpe},
  \citenamefont {L\"owen},\ and\ \citenamefont {Bechinger}}]{Kummel2014}%
  \BibitemOpen
  \bibfield  {author} {\bibinfo {author} {\bibfnamefont {F.}~\bibnamefont
  {K\"ummel}}, \bibinfo {author} {\bibfnamefont {B.}~\bibnamefont {ten Hagen}},
  \bibinfo {author} {\bibfnamefont {R.}~\bibnamefont {Wittkowski}}, \bibinfo
  {author} {\bibfnamefont {I.}~\bibnamefont {Buttinoni}}, \bibinfo {author}
  {\bibfnamefont {R.}~\bibnamefont {Eichhorn}}, \bibinfo {author}
  {\bibfnamefont {G.}~\bibnamefont {Volpe}}, \bibinfo {author} {\bibfnamefont
  {H.}~\bibnamefont {L\"owen}},\ and\ \bibinfo {author} {\bibfnamefont
  {C.}~\bibnamefont {Bechinger}},\ }\bibfield  {title} {\bibinfo {title}
  {K\"ummel et al. reply},\ }\href
  {https://doi.org/10.1103/PhysRevLett.113.029802} {\bibfield  {journal}
  {\bibinfo  {journal} {Phys. Rev. Lett.}\ }\textbf {\bibinfo {volume} {113}},\
  \bibinfo {pages} {029802} (\bibinfo {year} {2014})}\BibitemShut {NoStop}%
\bibitem [{\citenamefont {Monthiller}\ \emph {et~al.}(2022)\citenamefont
  {Monthiller}, \citenamefont {Loisy}, \citenamefont {Koehl}, \citenamefont
  {Favier},\ and\ \citenamefont {Eloy}}]{Monthiller2022}%
  \BibitemOpen
  \bibfield  {author} {\bibinfo {author} {\bibfnamefont {R.}~\bibnamefont
  {Monthiller}}, \bibinfo {author} {\bibfnamefont {A.}~\bibnamefont {Loisy}},
  \bibinfo {author} {\bibfnamefont {M.~A.}\ \bibnamefont {Koehl}}, \bibinfo
  {author} {\bibfnamefont {B.}~\bibnamefont {Favier}},\ and\ \bibinfo {author}
  {\bibfnamefont {C.}~\bibnamefont {Eloy}},\ }\bibfield  {title} {\bibinfo
  {title} {Surfing on turbulence: a strategy for planktonic navigation},\
  }\href@noop {} {\bibfield  {journal} {\bibinfo  {journal} {Phys. Rev. Lett.}\
  }\textbf {\bibinfo {volume} {129}},\ \bibinfo {pages} {064502} (\bibinfo
  {year} {2022})}\BibitemShut {NoStop}%
\bibitem [{\citenamefont {Mecanna}\ \emph {et~al.}(2025)\citenamefont
  {Mecanna}, \citenamefont {Loisy},\ and\ \citenamefont {Eloy}}]{Mecanna2025}%
  \BibitemOpen
  \bibfield  {author} {\bibinfo {author} {\bibfnamefont {S.}~\bibnamefont
  {Mecanna}}, \bibinfo {author} {\bibfnamefont {A.}~\bibnamefont {Loisy}},\
  and\ \bibinfo {author} {\bibfnamefont {C.}~\bibnamefont {Eloy}},\ }\bibfield
  {title} {\bibinfo {title} {A critical assessment of reinforcement learning
  methods for microswimmer navigation in complex flows},\ }\href@noop {}
  {\bibfield  {journal} {\bibinfo  {journal} {Eur. Phys. J. E}\ }\textbf
  {\bibinfo {volume} {48}},\ \bibinfo {pages} {58} (\bibinfo {year}
  {2025})}\BibitemShut {NoStop}%
\bibitem [{\citenamefont {Felderhof}(2014)}]{Felderhof2014}%
  \BibitemOpen
  \bibfield  {author} {\bibinfo {author} {\bibfnamefont {B.~U.}\ \bibnamefont
  {Felderhof}},\ }\bibfield  {title} {\bibinfo {title} {Comment on ``{C}ircular
  motion of asymmetric self-propelling particles''},\ }\href
  {https://doi.org/10.1103/PhysRevLett.113.029801} {\bibfield  {journal}
  {\bibinfo  {journal} {Phys. Rev. Lett.}\ }\textbf {\bibinfo {volume} {113}},\
  \bibinfo {pages} {029801} (\bibinfo {year} {2014})}\BibitemShut {NoStop}%
\bibitem [{\citenamefont {Doi}(2011)}]{Doi2011}%
  \BibitemOpen
  \bibfield  {author} {\bibinfo {author} {\bibfnamefont {M.}~\bibnamefont
  {Doi}},\ }\bibfield  {title} {\bibinfo {title} {Onsager's variational
  principle in soft matter},\ }\href
  {https://doi.org/10.1088/0953-8984/23/28/284118} {\bibfield  {journal}
  {\bibinfo  {journal} {J. Phys. Condens. Matter}\ }\textbf {\bibinfo {volume}
  {23}},\ \bibinfo {pages} {284118} (\bibinfo {year} {2011})}\BibitemShut
  {NoStop}%
\bibitem [{\citenamefont {Greengard}\ and\ \citenamefont
  {Rokhlin}(1987)}]{Greengard1987}%
  \BibitemOpen
  \bibfield  {author} {\bibinfo {author} {\bibfnamefont {L.}~\bibnamefont
  {Greengard}}\ and\ \bibinfo {author} {\bibfnamefont {V.}~\bibnamefont
  {Rokhlin}},\ }\bibfield  {title} {\bibinfo {title} {A fast algorithm for
  particle simulations},\ }\href@noop {} {\bibfield  {journal} {\bibinfo
  {journal} {J. Comput. Phys.}\ }\textbf {\bibinfo {volume} {73}},\ \bibinfo
  {pages} {325} (\bibinfo {year} {1987})}\BibitemShut {NoStop}%
\bibitem [{\citenamefont {Tornberg}\ and\ \citenamefont
  {Greengard}(2008)}]{Tornberg2008}%
  \BibitemOpen
  \bibfield  {author} {\bibinfo {author} {\bibfnamefont {A.-K.}\ \bibnamefont
  {Tornberg}}\ and\ \bibinfo {author} {\bibfnamefont {L.}~\bibnamefont
  {Greengard}},\ }\bibfield  {title} {\bibinfo {title} {A fast multipole method
  for the three-dimensional stokes equations},\ }\href@noop {} {\bibfield
  {journal} {\bibinfo  {journal} {J. Comput. Phys.}\ }\textbf {\bibinfo
  {volume} {227}},\ \bibinfo {pages} {1613} (\bibinfo {year}
  {2008})}\BibitemShut {NoStop}%
\bibitem [{\citenamefont {Waszkiewicz}\ \emph {et~al.}(2023)\citenamefont
  {Waszkiewicz}, \citenamefont {Bartczak}, \citenamefont {Kolasa},\ and\
  \citenamefont {Lisicki}}]{Waszkiewicz2023}%
  \BibitemOpen
  \bibfield  {author} {\bibinfo {author} {\bibfnamefont {R.}~\bibnamefont
  {Waszkiewicz}}, \bibinfo {author} {\bibfnamefont {M.}~\bibnamefont
  {Bartczak}}, \bibinfo {author} {\bibfnamefont {K.}~\bibnamefont {Kolasa}},\
  and\ \bibinfo {author} {\bibfnamefont {M.}~\bibnamefont {Lisicki}},\
  }\bibfield  {title} {\bibinfo {title} {Pychastic: Precise {B}rownian dynamics
  using {T}aylor--it{\=o} integrators in {P}ython},\ }\href
  {https://doi.org/10.21468/SciPostPhysCodeb.11} {\bibfield  {journal}
  {\bibinfo  {journal} {SciPost Phys. Codebases}\ ,\ \bibinfo {pages} {011}}
  (\bibinfo {year} {2023})}\BibitemShut {NoStop}%
\bibitem [{\citenamefont {Funkenbusch}\ \emph {et~al.}(2024)\citenamefont
  {Funkenbusch}, \citenamefont {Silmore},\ and\ \citenamefont
  {Swan}}]{Funkenbusch2024}%
  \BibitemOpen
  \bibfield  {author} {\bibinfo {author} {\bibfnamefont {W.~T.}\ \bibnamefont
  {Funkenbusch}}, \bibinfo {author} {\bibfnamefont {K.~S.}\ \bibnamefont
  {Silmore}},\ and\ \bibinfo {author} {\bibfnamefont {J.~W.}\ \bibnamefont
  {Swan}},\ }\bibfield  {title} {\bibinfo {title} {Approaches for fast brownian
  dynamics simulation with constraints},\ }\href@noop {} {\bibfield  {journal}
  {\bibinfo  {journal} {J. Comput. Phys.}\ }\textbf {\bibinfo {volume} {509}},\
  \bibinfo {pages} {113043} (\bibinfo {year} {2024})}\BibitemShut {NoStop}%
\end{thebibliography}%

\end{document}